\documentclass{aa}  

\usepackage{txfonts}
\usepackage{hyperref}
\usepackage{color}
\usepackage{subfig}
\usepackage{pdflscape}
\usepackage{geometry}
\usepackage{longtable}
\usepackage{threeparttable}
\usepackage{deluxetable}
\usepackage{epstopdf}
\usepackage{graphicx}
\usepackage{mathrsfs}  
\usepackage{float}
\usepackage{textcomp}
\usepackage{gensymb}
\usepackage{morefloats}
\usepackage{booktabs}
\usepackage{footmisc}
\usepackage{etex}

\begin{document} 

\title{ALMA observations of the early stages of substellar formation in the Lupus 1 and 3 molecular clouds}

\author{A. Santamar\'ia-Miranda \inst{1,2,3}
	\and
	I. de Gregorio-Monsalvo \inst{1}
	\and
	A.L. Plunkett \inst{4}
	\and
	N. Hu\'elamo \inst{5}
	\and
	C. L\'opez \inst{6}
	\and
	\'A. Ribas \inst{1}  
	\and
	M.R. Schreiber \inst{7,3}   
	\and
	K. Mu\v{z}i\'{c} \inst{8}
	\and
	A. Palau \inst{9}
	\and
	L.B.G. Knee \inst{10}
	\and
	A. Bayo \inst{2,3}
	\and
	F. Comer\'on \inst{11}
	\and
	A. Hales \inst{6}
	}

\institute{European Southern Observatory, Av. Alonso de C\'ordova 3107, Casilla 19001, Santiago, Chile\\
	\email{asantama@eso.org}
	\and
	Instituto de F\'isica y Astronom\'ia, Universidad de Valpara\'iso, Av. Gran Breta\~na 1111, Casilla 5030, Valpara\'iso, Chile
	\and
	N\'ucleo Milenio Formaci\'on Planetaria - NPF,  Valpara\'iso, Chile
	\and
	National Radio Astronomy Observatory, 520 Edgemont Rd., Charlottesville, VA 22903, U.S.A.
	\and
	Dpto. Astrof\'isica, Centro de Astrobiolog\'ia (INTA-CSIC), ESAC Campus, Camino Bajo del Castillo s/n, Urb. Villafranca del Castillo, 28692 Villanueva de la Ca\~nada, Spain
	\and
	Joint ALMA Observatory, Av. Alonso de C\'ordova 3107, Casilla 19001, Santiago, Chile
    \and
    Departamento de F\'isica, Universidad T\'ecnica Federico Santa Mar\'ia, Av. Espa\~na 1680, Valpara\'iso, Chile
	\and
	CENTRA, Faculdade de Ci\^{e}ncias, Universidade de Lisboa, Ed. C8, Campo Grande, 1749-016 Lisboa, Portugal
	\and
	Instituto de Radioastronom\'ia y Astrof\'isica, Universidad Nacional Aut\'onoma de M\'exico, P.O. Box 3-72, 58090, Morelia, Michoac\'an, M\'exico
	\and
	Herzberg Astronomy and Astrophysics Research Centre, 5071 West Saanich Rd., Victoria, BC, V9E 2E7, Canada
	\and
	European Southern Observatory, Karl-Schwarzschild-Strasse 2, 85748 Garching bei M\"unchen, Germany
	        }

\date{Received 14/09/2020 ; accepted 17/11/2020}

\abstract
    {The dominant mechanism leading to the formation of brown dwarfs (BDs) remains uncertain. While the census of Class II analogs in the substellar domain continues to grow, the most direct keys to formation, which are obtained from younger objects (pre-BD cores and proto-BDs), are limited by the very low number statistics available.}
    {We aim to identify and characterize a set of pre- and proto-BDs as well as Class II BDs in the Lupus 1 and 3 molecular clouds to test their formation mechanism.}
    {We performed ALMA band 6 (1.3 mm) continuum observations of a selection of 64 cores previously identified from AzTEC/ASTE data (1.1 mm), along with previously known Class II BDs in the Lupus 1 and 3 molecular clouds. Surveyed archival data in the optical and infrared were used to complement these observations. We expect these ALMA observations prove efficient in detecting the youngest sources in these regions, since they probe the frequency domain at which these sources emit most of their radiation.}
    {We detected 19 sources from 15 ALMA fields. Considering all the pointings in our observing setup, the ALMA detection rate was $\sim$23\,\% and the derived masses of the detected sources were between $\sim$0.18 and 124 $\mathrm{M_{Jup}}$. We classified these sources according to their spectral energy distribution as 5 Class II sources, 2 new Class I/0 candidates, and 12 new possible pre-BD or deeply embedded protostellar candidates. We detected a promising candidate for a Class 0/I proto-BD source (ALMA J154229.778-334241.86) and inferred the disk dust mass of a bona fide Class II BD. The pre-BD cores might be the byproduct of an ongoing process of large-scale collapse. The Class II BD disks follow the correlation between disk mass and the mass of the central object that is observed at the low-mass stellar regime.}
    {We conclude that it is highly probable that the sources in the sample are formed as a scaled-down version of low-mass star formation, although disk fragmentation may be responsible for a considerable fraction of BDs.}

    \keywords{brown dwarfs -- stars:formation -- techniques:interferometric -- submillimeter:stars}
    \titlerunning{Early stages of substellar formation in the Lupus 1 and 3 molecular clouds}
    \authorrunning{Santamar\'ia-Miranda et al.}
    \maketitle

\section{Introduction}
More than two decades have passed since the first brown dwarf (BD) object was discovered \citep{Reboloetal95-1, Oppenheimer1995, Basri1995}. Since then a large number of BDs have been detected in several star-forming regions (SFRs), including Serpens \citep{Lodieu02}, Taurus-Auriga \citep{White&Basri03-1, Luhman18-1}, $\lambda$ Orionis  \citep{Barradoetal04-1,Bayoetal11-1}, $\sigma$ Orionis \citep{Caballeroetal07-1, Pena-ramirez15-1} Upper Scorpius \citep{Boutetal07-1, Lodieu18-1},   $\rho$ Ophiuchi \citep{Alvesetal12-1}, NGC 1333 \citep{scholzetal12-1}, Lupus \citep{Comeron2009-1, Sanchis20-1}, IC 348 in Perseus \citep{Alves13-1},  Chamaeleon I \citep{Luhman2007,Muzicetal15-1}, NGC 2244 \citep{Muzic2019}, and others in the field. However, their formation mechanism(s) are not fully understood.
\\

\smallskip
Several mechanisms have been proposed to explain the formation of BDs, such as an ejection from filaments, disk fragmentation, photoevaporation, eroding outflows, and turbulent fragmentation. The ejection mechanism is based on the fragmentation of a molecular cloud and a subsequent close dynamical interaction between the objects, provoking the ejection of the less massive objects \citep{Reipurthandclarke01-1}. In this picture, \citet{Bateetal02-1} used hydrodynamical simulations and found that $\sim$25\% of the BDs are ejected  from the filaments from which they were born, while the rest are formed via disk fragmentation.  Hydrodynamical simulations by \citet{Bate14} showed no statistical difference between the properties of simulated BD and observed BD. \\
The theory of disk fragmentation suggests that during the formation of planetary-mass companions and substellar companions in a protoplanetary disk, mainly around Sun-like stars, about two thirds of the BDs  is expected to be ejected from the disk in the process \citep{Stamatellosandwhitworth09-1} and those that remain bound to the central star should be able to retain their own disk and accrete. Evidence for accretion in substellar companions on wide orbits (> 330 au) has been found in several objects such as  FW\,Tau\,B, CT\,Cha\,b, GSC 06214-210\,B, and SR\,12\,C \citep{Bowler14-1,Bowler11-1,Wu15-1,Santamaria-Miranda18-1}. 

Another BD formation mechanism,namely, photoevaporation, was proposed by \citet{Hesteretal96-1} and developed in detail by \citet{WhitworthandZinnecker04-1}. In this scenario, a massive star triggers the formation of a stable pre-stellar core while the ionizing radiation erodes the outside part of the core. Due to the mass loss, the final object is expected to be either a very low-mass star, a BD or a planetary-mass object. Several observational works studying SFRs containing hot stars have identified proto-BD candidates that could have been formed by this mechanism \citep[see e.g.,][]{Huelamo2017,Barrado18-1}.

\citet{Machida2009} have proposed that BDs are formed as a scaled-down version of low-mass stars from a core that is less massive and compact with an extreme accretion rate ($\sim$10$^{-5}-$10$^{-6}$M$_\odot$ yr$^{-1}$). This decreases the mass transfer from the envelope to the protostar obtaining a BD as a final object instead of a star. \citet{Machida2008b} postulate that the molecular outflow is solely responsible for reducing the mass transfer, lowering the accretion rate and the star formation efficiency (SFE). 

Ultimately, there are certain alternative perspectives. For instance, turbulent fragmentation theory \citep{Padoan2004, Chabrier14}  explains that a turbulent environment driven by supersonic turbulence can naturally yield the formation of substellar fragments, which should be gravitationally unstable and collapse to form pre- and proto-BDs. On the other hand, the Global Hierarchical Collapse scenario \citep{Vazquez19} proposes that Hoyle-like gravitational fragmentation \citep{Hoyle1953} in the presence of low levels of turbulence can continue down to substellar-mass fragments, which finally end up in BDs. In both cases, it is expected that BDs are formed as a scaled-down version of low-mass stars.

One of the main goals in the field of the formation of substellar objects is to test the different formation mechanisms and to determine which of them is dominant. If BDs are formed as a scaled-down version of low-mass stars, they should have cold envelopes and outflows similar to those observed in Class 0/I protostars. To test this scenario, we need to increase the number of detected sources at the earliest evolutionary stages in the (sub)millimetre (mm) regime when they are still embedded in the parental cloud -- namely, the pre- and proto-BDs. So far, the only confirmed pre-BD is Oph B-11 \citep{Andre2012} and there are pre-BD candidates both in Taurus \citep{Palau2012-1, Tokuda2019} and Barnard 30 \citep{Huelamo2017, Barrado18-1}. Regarding proto-BDs, one excellent Class 0/I candidate has been identified in Perseus \citep{Palau14-1} and other candidates are proposed in Taurus \citep{Apai05, Barrado2009, Palau2012-1, Morata15-1, Dang-duc16}, Chamaeleon II \citep{deGregorio2016}, Ophiuchus \citep{Whelan18, Riaz18, Kawabe18}, Serpens \citep{Riaz16,Riaz18}, and $\sigma$ Orionis \citep{Riaz15, Riaz2017, Riaz19}, although many of these may also be more evolved Class I objects. Finally, the search and study of very low luminosity objects (VeLLOs) have revealed sources that show proto-BD characteristics (\citealt{Bourke2006,Lee2009, Lee13, Lee18,Kauffmann11-1, Kim19}; also see the references in Table 4 of \citealt{Palau14-1}).

Class II BDs are also expected to exhibit phenomena similar to those seen in very low mass (VLM) protostars, including disks and jets in the optical and infrared. Indeed, circumstellar disks around BDs were identified by \citet{Natta01-1} and \citet{Natta02-1} through the study of spectral energy distributions (SEDs) and the confirmed presence of jets, Herbig-Haro objects, outflows, and accretion \citep{Jayawardhana03-1,Natta04-1, Whelan05-1, Phan-Bao08-1, Riaz2017}. In particular, (sub)millimetre observations are important to study the properties of BD disks. The first sub(mm) study was presented by \citet{Klein2003}, who reported the detection of millimeter dust emission from BDs. Then, \citet{Scholz06-1} in a sample of 20 BD disks in Taurus inferred dust settling towards the disk midplane. Following these pioneering works, BD disks were observed using ALMA, constraining their main parameters, such as the dust and gas disk masses or their radii, which provide information about the formation of these objects. The first ALMA study of three BD disks was performed in Taurus \citep{Riccietal14-1}. Subsequent studies \citep{Testietal16-1} showed the possible environmental influence on the formation of BDs in Ophiuchus when compared to Taurus disks. A similar survey in Upper Scorpius \citep{vanderplasetal16-1} revealed a new relation between the stellar luminosity and the temperature in the substellar regime. The TBOSS survey with ALMA \citep{Ward-Duong18} included BD disks in Taurus, and IC 348, and Chamaeleon that have been studied by \citet{Ruiz-rodriguez18} and \citet{Pascucci16-1}. Finally, Lupus BDs have been also observed with ALMA \citep{Ansdell16-1, Ansdell18-1, Sanchis20-1}.

The Soul of Lupus with ALMA Consortium (SOLA, \citealt{Saito15}) is focused on the study of the Lupus clouds. The Lupus molecular complex \citep{Barnard1927} is composed of nine molecular clouds. In this work, we focus on the Lupus 1 and 3 clouds. Despite both clouds belonging to the same complex, they show different properties. The mass of Lupus 1 is an order of magnitude larger than that of Lupus 3. \citet{cambresyetal99-1} derived a mass $2.6 \times 10^{4}$ $\mathrm{M_{\odot}}$ for Lupus 1 and $3 \times 10^{3}$ $\mathrm{M_{\odot}}$ for Lupus 3 from extinction values; a similar mass ratio was obtained using $^{13}$CO \citep{tachiharaetal96-1} with masses of $1.3 \times 10^{3} \mathrm{M_{\odot}}$ and $3.4 \times 10^{2}$ $\mathrm{M_{\odot}}$, respectively. Lupus 1 is more isolated in the complex than is Lupus 3, the latter of which is surrounded by Lupus clouds 2, 4, 5, and 6. \citep[and references therein]{comeron08-1}. The stellar populations are also different: Lupus 1 is dominated by early M dwarfs \citep{comeronetal09-2} and Lupus 3 is abundant in T Tauri stars. Furthermore, Lupus 3 has the highest star formation rate and column density of all the Lupus clouds \citep[and references therein]{comeron08-1}.

\label{sec:p2_intro}
In this work, we present the main results from the SOLA project. The objective is to shed light on the main formation mechanism of BDs. We have complemented the ALMA data with optical and infrared archival observations to build the SED of all the objects. In Section \ref{observations}, we present the AzTEC and ALMA observations performed for this SOLA project. In Section \ref{p2_analysis}, we present the data analysis, including ALMA continuum detections, mass estimates, the optical/infrared counterpart association to the ALMA detections, SEDs, bolometric temperatures and luminosities, and a brief description of each source. In Section \ref{p2_discussion}, we present a comparison of our results with previous AzTEC results and assess them against the main BD formation theories. We also discuss the detection rate and the evolution of the cores and compare the  disk masses with those in other SFRs. Finally, our conclusions and summary are presented in Section \ref{p2_conclusion}.

\section{Observations}
\label{observations}

\subsection{The sample}
\label{sec:sample}
\begin{figure*}
\centering
\subfloat{\includegraphics[width=0.5\textwidth]{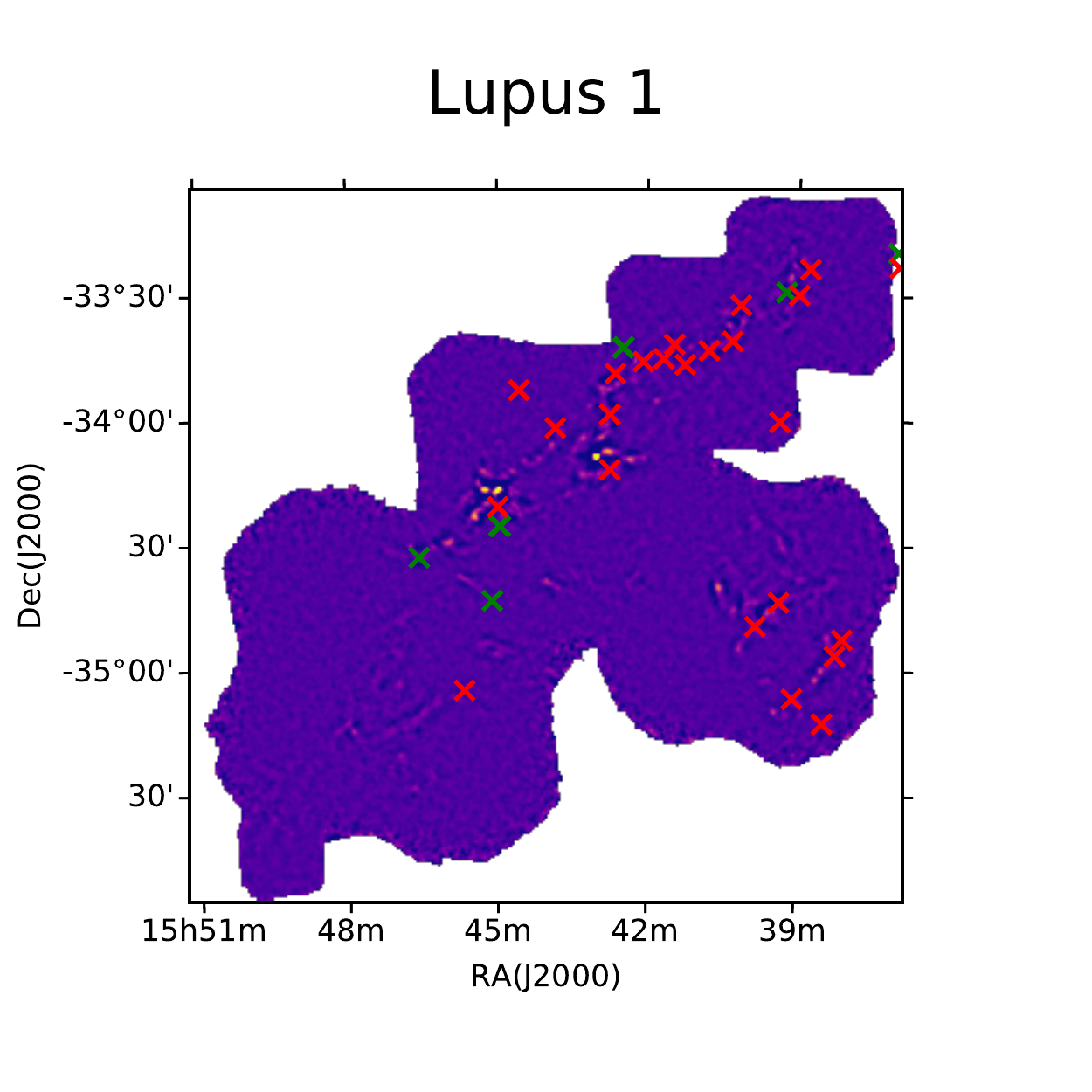}}
\subfloat{\includegraphics[width=0.5\textwidth]{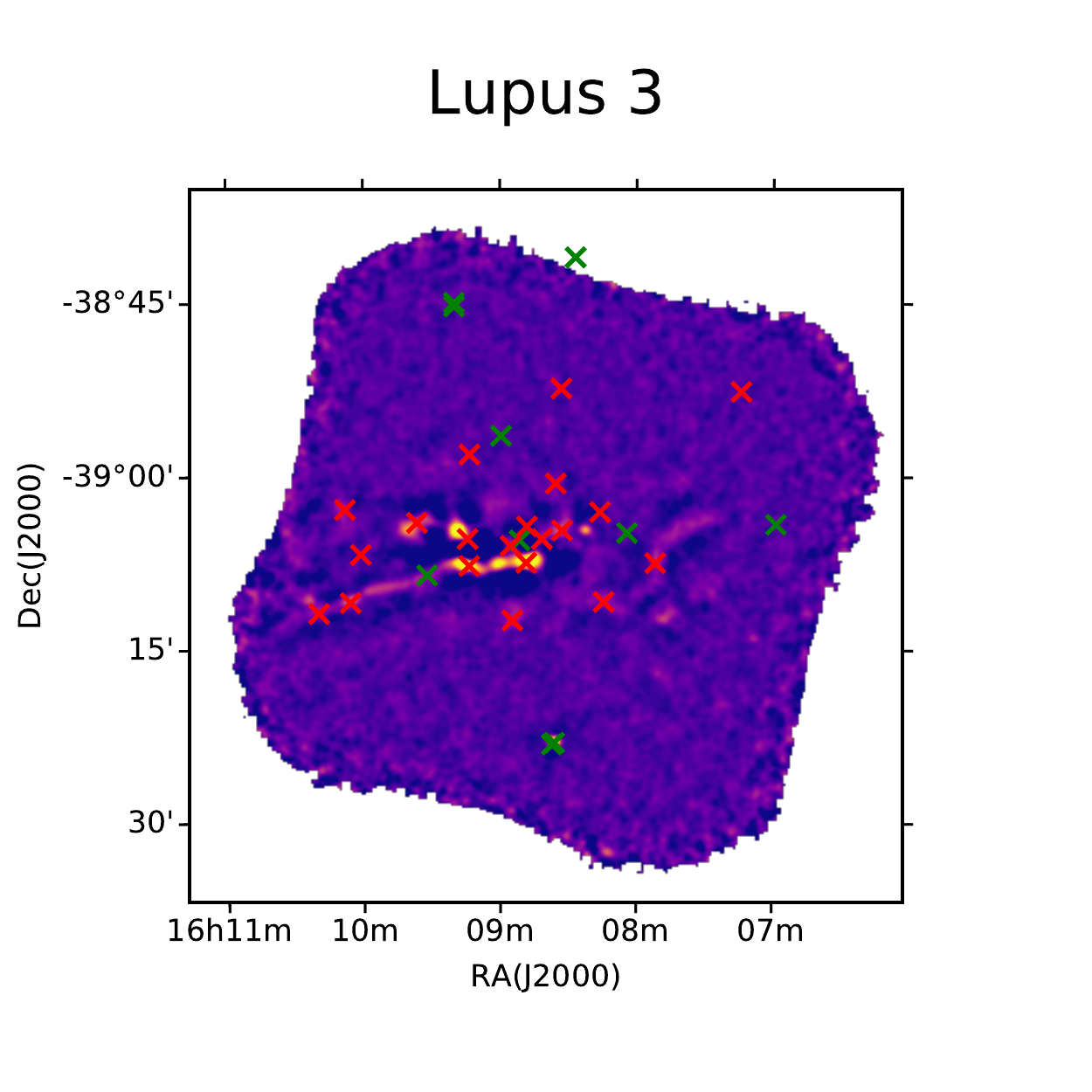}}
\caption{AzTEC 1.1 mm maps of Lupus 1 (left panel) and Lupus 3 molecular clouds (right panel) overlaid with the ALMA pointings. Green crosses represent ALMA detections and red ones the AzTEC-identified sources that were not detected with ALMA.}
\label{aztec_detections}
\end{figure*}

The sample is based on two maps of Lupus 1 and 3 clouds  (Figure \ref{aztec_detections}), taken with the AzTEC 1.1 mm array camera at the Atacama Submillimeter Telescope Experiment (ASTE) telescope located in the Atacama Desert in northern Chile. A subset of these data were presented by \citet{tsukagoshietal11-1} and \citet{tamuraetal15-1}, who studied V1094 Sco (discussed later in this work) and bright submillimetre galaxies, respectively. 

The SOLA Consortium was given access to these two maps of millimetre continuum observations and in using them, we have revealed hundreds of continuum sources embedded within filamentary structures. The angular resolution of the AzTEC observations was $28\arcsec$ with a FoV of $7.8\arcmin$ and a rms noise level of 5 mJy beam$^{-1}$. More information about these observations is found in \citet{tsukagoshietal11-1} and \citet{tamuraetal15-1}. 

To select dust cores in AzTEC maps,  we first used the \textit{clumpfind} algorithm \citep{williams94}. As part of the selection criteria for our source sample, we selected AzTEC clumps with peak intensities in the range  30--100 mJy, which show a mass below 0.225 M$_{\odot}$ assuming a temperature of 10 K and an emissivity index of $\beta$=2. Considering a SFE 30 $\%$ \citep{motteetal98-1} these sources should form substellar objects. Using this methodology, we obtained 39 AzTEC cores that were not previously reported in the literature.

Then, we cross-correlated the AzTEC substellar cores with optical and infrared archival observations (see Section \ref{ancillary data}) to build the SEDs. The SEDs were consistent with young (sub)stellar objects (YSO). These cores were classified as 33 pre-BD objects along with 6 Class 0 and I objects. A detailed explanation of classification methodology used in this work using the SED  is in Section \ref{clasificacion}.  

We complemented the AzTEC sample with previously known substellar objects in Lupus 1 and 3 in order to fully characterize the substellar population at different evolutionary stages. We selected 22 already known Class II BD and 3 Class I/II BD from  \citet{merinetal08-1, comeronetal09-2,kora14-1}.

In summary, the final sample of 64 objects was divided in a preliminary classification of 33 pre-BD objects, 6 Class 0 and I objects, 3 Class I/II, and 22 Class II objects.  The initial classification, phase centre coordinates, and noise rms are given in Appendix \ref{tabla_phase_rms}.

\subsection{Ancillary data}
\label{ancillary data}
\begin{figure*}
\centering
\includegraphics[width=\textwidth]{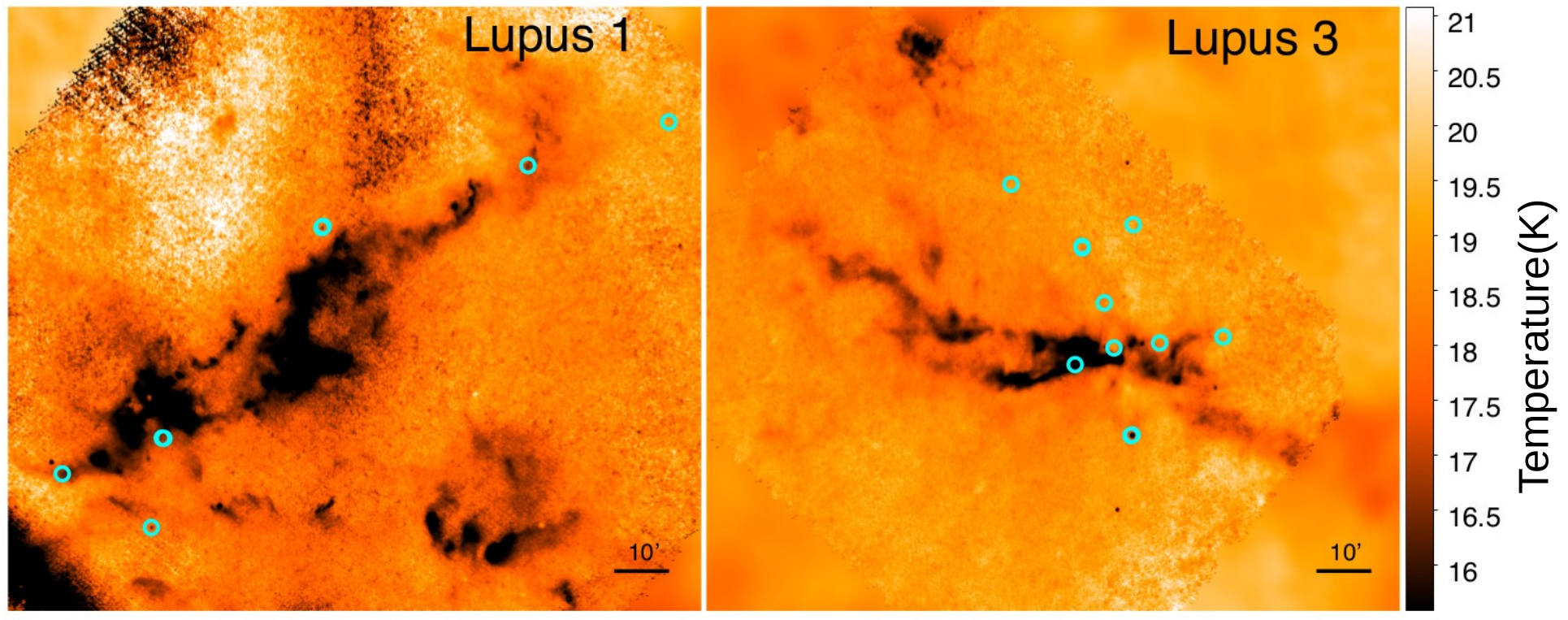}
\caption[Herschel dust temperature map of Lupus 1 and 3]{\label{Herschel}Herschel temperature maps for Lupus 1 and 3 clouds (Teixeira et al., in prep.). North is up and East to the left. The ALMA detections are displayed as cyan circles and some circles overlap.}
\end{figure*}

The combination of the AzTEC maps and a subsequent search of a catalog derived from literature and archival data (L\'opez et al. in prep) enabled us to obtain a set of good-quality pre- and proto-BDs candidates. This catalog includes data from optical to the (sub)millimetre wavelengths. In order to classify the objects detected with ALMA (see Figure \ref{Herschel} to see the position of the detected sources in a Herschel map), we searched in the archives of several observatories. The filters used and the respective astrometric accuracy for each telescope/instrument are given in Table \ref{Tab:instrumentos}.

\begin{table*}
\caption{Telescopes and filters for archival data.}\label{Tab:instrumentos}
\small
\begin{tabular}{lcccc c}
\hline \hline
Telescope/ & Filter$_{1}$ ($\lambda _{1}$)& Filter$_{2}$ ($\lambda _{2}$) & Filter$_{3}$ ($\lambda _{3}$) & Filter$_{4}$ ($\lambda _{4}$) & Astrometric accuracy \\
instrument or survey & & & & & [arcsec] \\
\hline
\hline  \\  
ESO 2.2m/WFI & R$_c$ (657.1 nm) &  I$_c$(826.9 nm) & Z (964.8 nm) & & $\sim$2 \\
ESO 1m/DENIS & I (0.82 $\mu$m) & J (1.25 $\mu$m) & K (2.15 $\mu$m) & & $\sim$1 \\ 
CTIO/2MASS & J (1.235 $\mu$m) & H (1.662 $\mu$m) & K$_s$ (2.159 $\mu$m) & & 0.1 \\ 
WISE & W1 (3.4 $\mu$m) & W2 (4.6 $\mu$m) & W3 (12$\mu$m) & W4 (22 $\mu$m) & 0.50 \\ 
Spitzer/IRAC & 1 (3.6 $\mu$m) & 2(4.5 $\mu$m) & 3 (5.8 $\mu$m) &  4 (8.0 $\mu$m)  & 0.25 -- 0.50 \\
Spitzer/MIPS & 1 (23.675 $\mu$m) & 2 (71.42 $\mu$m) & 3 (155.9 $\mu$m)& & $\sim$0.5 \\
Akari & f9 (9 $\mu$m) & f18 (18 $\mu$m) & & & 0.15 \\ 
Herschel/PACS & 70 $\mu$m & 100 $\mu$m & 160 $\mu$m & & 1.6  \\ 
Herschel/SPIRE & 250 $\mu$m & 350 $\mu$m & 500 $\mu$m & & 1.6 \\ 
APEX/LABOCA & 868.9 $\mu$m & & & &4.5 \\ 
AzTEC/ASTE & 1.1\,mm & & & & 2  \\ 
\end{tabular}
\end{table*} 

Optical data were obtained from \citet{comeronetal09-2} using the Wide Field Imager (WFI) at the 2.2m MPG/ESO Telescope \citep{Baadeetal99-1}. 

The near-infrared wavelength range data was covered by two surveys: the Deep Near Infrared Survey of the Southern Sky (DENIS) at the 1m ESO telescope  \citep{Epchteinetal94-1} and the Cerro Tololo Inter-American Observatory 2MASS telescope (CTIO/2MASS) \citep{Skrutskieetal06-1}. 

Mid-infrared wavelength data were obtained from the Wide-Field Infrared Survey Explorer (WISE, \citealt{Wrightetal10-1} and the InfraRed Array Camera on the Spitzer Space Telescope (IRAC) \citep{Faddaetal04-1}, both NASA space missions. Spitzer had another instrument, the Multiband Imaging Photometer for SIRTF(MIPS), that provided far-infrared wavelength coverage. Additionally, two more space telescopes provided far-infrared data: ASTRO-F \citep{Murakamietal07-1} from JAXA and Herschel \citep{Pilbrattetal10-1} from the European Space Agency (ESA). From Herschel we obtained data from two instruments: the Photodetector Array Camera and Spectrometer (PACS) \citep{Poglitschetal10-1} and the Spectral and Photometric Imaging Receiver (SPIRE) \citep{Griffinetal10-1}.

Millimetre and submillimetre observations of Lupus 3 were also available from the Large Apex BOlometer CAmera  \citep[LABOCA,][]{Siringoetal09-1}  at the APEX Observatory.

Finally, data from other telescopes including the Australia Telescope Compact Array (ATCA), the Swedish-ESO Submillimetre Telescope (SEST), and the Submillimeter Array (SMA) were examined, but no detections were obtained from any of these telescopes for Lupus 1 and 3.  

\subsection{ALMA observations}
ALMA observations were performed between 31 March and 1 April 2016 as part of the ALMA Cycle 3 program 2015.1.00512.S. We observed all 64 targets in our sample using single field interferometry in Band 6, with a field of view (FoV) of $23\arcsec$. The number of antennas for each data set ranged between 42 and 44. The total allocated time for this program was 6.7 hours including overheads, with an average time on each science source of 4.7 minutes. Data were taken under good and stable weather conditions (precipitable water vapour 1.1 {\textendash} 1.7 mm).  

The correlator was set up to observe five different spectral windows in dual polarization mode with the primary aim of detecting continuum emission, while also detecting possible molecular emission lines. One of the basebands was configured to observe two spectral windows of 0.469\,GHz bandwidth and 0.488\,MHz channel width ($\sim$0.3 km/s velocity resolution) each, one of them centred at 230.538\,GHz, CO(2--1) rest frequency, and the other one at 231.150\,GHz. The other three basebands were configured to observe three different spectral windows of 1.875\,GHz each, and at a spectral resolution of 1.938\,MHz ($\sim$2.5 km/s velocity resolution) centred at 233.5\,GHz, 217.0\,GHz and 219.25\,GHz for possible detections of C$^{18}$O(2--1), SiO(5--4), and DCN(3--2) and to have the best atmospheric transmission. 

QSO J1517-2422 was used as bandpass calibrator for the whole sample. This QSO was used as a flux calibrator for one of the datasets while Titan was used for the remaining execution blocks. Two phase calibrators were used: for the three datasets of Lupus 1 QSO J1610-3958 was used, while QSO J1534-3526 was used for the observations towards Lupus 3. 

The Common Astronomy Software Applications package (CASA) \citep{McMullin2007} was used to process the data. Most of the datasets used the pipeline version 4.5.3 for calibration and image processing and only one of the executions was reduced using the ALMA standard calibration mode (not pipelined) in CASA version 4.6. The task CLEAN was used to produce continuum and spectral line images. We selected a Briggs weighting with Robust parameter value of 2 to produce all the images. Primary beam correction was applied before inferring physical parameters from the images. The achieved noise rms at the phase centre was on average $\sim$70 $\mu$Jy/beam for the continuum and $\sim$6 mJy/beam in a 0.3 km/s channel width for the spectral line data. The angular resolution of the images was $\sim$0.9$\arcsec$, corresponding to $\sim$138 au at 154 pc (see Section \ref{distancia derivada}). The absolute positional accuracy of the ALMA images in our sample spans between 4 and 270 mas with an average value of 148 mas\footnote{ALMA Technical Handbook, Chapter 10.5.2 Astrometric Observations, https://help.almascience.org/index.php?/Knowledgebase/Article/
View/319/6/what-is-the-astrometric-accuracy-of-alma}.

We detected 19 sources from the 15 ALMA fields containing 64 science targets. Detailed information for each source detection is given in Table \ref{table:tabla_ppal}. The main properties of the detected sources are described in Section \ref{p2_analysis}.

We adopted the naming convention recommended by ALMA for newly discovered sources, but for convenience in the discussion we shorten the source names to J\#\#\#\#\#\#, for example ALMA J154228.675-334230.18 is referred to as J154228, with a minor modification in the case of ALMA J160920.089-384515.92 (J1609200) and ALMA J160920.171-384456.40 (J1609201) in order to distinguish them.

\subsection{Distances}
\label{distancia derivada}
We used the parallaxes from Gaia Data Release 2 \citep{Gaiadr2} to obtain an accurate average distance estimate of the Lupus\,1 and 3 clouds. We used a sample of spectroscopically confirmed members \citep{comeronetal09-2} of Lupus 1 and 3 and applied the \textit{kalkayotl}  \footnote{\url{https://github.com/olivares-j/Kalkayotl}} python code \citep{Olivares20-1} that uses a Bayesian approach to obtain a precise distance, as recommended in \citet{Bailer-Jones15-1}. We also used this methodology to obtain the distance for single sources in both Lupus 1 and 3 when their parallaxes were available (see Table \ref{distances_gaia_dr2}). Otherwise we assumed the mean distance for each cloud that we obtained,namely, 153.4 $\pm$ 4.6 and 154.8 $\pm$ 9.6 pc for Lupus 1 and 3, respectively. There are two Class II BDs (153701.1-332 and 160545.8-385)  in our survey whose distances are not the expected ones for Lupus 1 and 3 using 3 $\sigma$. Therefore, we do not consider these sources as part of the complex. 
In order to make comparisons among source characteristics in Lupus and other SFRs in the literature, we recalculated distances to the regions that we will reference in Section \ref{sec:disk masses}. Chamaeleon I distances for individual sources are included in \citet{Manara19-lupuscam}. For the sources without Gaia parallaxes, we assumed the distance to Chamaeleon I cloud, 179$^{+11}_{-10}$  \citep{Voirin18}. Upper Scorpius distances for individual sources and the mean distance of the cloud are included in \citet{Galli18-1}; we assumed the mean distance to the cloud for the sources without Gaia parallaxes. Taurus distances are included in \citet{Galli19-1}. In Ophiuchus, we derived the distances from Gaia DR2 parallaxes \citep{Gaiadr2} using the mentioned \textit{kalkayotl} code. Using a selection of confirmed members \citep{natta+randich06-1} in the Ophiuchus SFR, we obtained an average distance to Ophiuchus SFR of 139.26 $\pm$ 14.57 pc. The distance for individual sources in Ophiuchus using the same methodology are in Table \ref{distances_gaia_dr2}. The distance to Lupus sources in \citet{Sanchis20-1} that were not part of our sample are included in \citet{Manara19-lupuscam}.

\newgeometry{margin=0.1cm} 
\begin{landscape}
\begin{table*} 
{\fontsize{7.5}{9.5} \selectfont 
\caption{\label{table:tabla_ppal}Properties of the ALMA continuum detections at 1.3 mm wavelength} 
\begin{threeparttable}
$$ 
\begin{array}{*{16}{c}}
\hline
\noalign{\smallskip}	
\mathrm{Name} & \mathrm{RA (J2000)\footnote[1]{}} & \mathrm{Dec (J2000)\footnote[1]{}} & \mathrm{Astrometric\footnote[2]{}}  & \mathrm{Separation\,from}  & \mathrm{rms} &  \mathrm{Flux\footnote[3]{}} & \mathrm{Peak} & \mathrm{Integrated\footnote[4]{}} & \multicolumn{3}{c}{\mathrm{Image\,component\,size\footnote[4]{}}} & \mathrm{Peak\,flux\footnote[4]{}} & \multicolumn{3}{c}{\mathrm{Beam \,size\footnote[4]{}}}  \\
\, & \, &\, & \mathrm{accuracy} &\mathrm{phase\,center} &\, & \mathrm{density}& \mathrm{intensity} & \mathrm{flux} & \multicolumn{3}{c}{\mathrm{deconvolved\,from\,beam}}  & \, & \, &\, &\, \\
\cmidrule(lr){10-12}
\cmidrule(lr){14-16}
\noalign{\smallskip}
       		\, & \mathrm{[h\,m\,s]} & \mathrm{[^{\circ}\,\,'\,\,"]} & \mathrm{[mas]} & \mathrm{[arcsec]} & \mathrm{[mJy/beam]} & \mathrm{[mJy]} & \mathrm{[mJy} & \mathrm{[mJy]}  & \mathrm{Major}\,\mathrm{axis} & \mathrm{Minor\,  axis} &  \mathrm{Position \, angle} & \mathrm{[ mJy/beam ]} & \mathrm{Major} & \mathrm{Minor} & \mathrm{PA}  \\
          &  \, & \, & \, & \, & &  & \mathrm{/beam]} & \, & \mathrm{[mas]} & \mathrm{[mas]} & \mathrm{[deg]} & \, & \mathrm{[arcsec]} &\mathrm{[arcsec]} & \mathrm{[deg]} \\
            \noalign{\smallskip}
            \hline
            \noalign{\smallskip}
      		\mathrm{ALMA\, J153702.653-331924.92} & 15:37:02.653 & -33:19:24.92 & 251 & 4.82 &  0.096  & 0.45 & 0.45 & 0.43 \pm 0.09 & - & - & -  & 0.48 \pm 0.06  & 0.94 &	0.93 &	32 \\     		
     		\mathrm{ALMA\, J153914.996-332907.62} & 15:39:14.996 & -33:29:07.62 & 151 & 9.62 &  0.094 & 0.73 & 0.73 & 0.82 \pm 0.08 &  -  & - & - & 0.74 \pm 0.04 & 0.91 &	0.83 &	-84 \\
     		\mathrm{ALMA\, J154228.675-334230.18} & 15:42:28.675 & -33:42:30.18 & 75  & 9.71 &  0.14 & 2.84 & 2.25 & 3.33 \pm 0.16  & - &	-  & - & 2.23 \pm 0.07 & 0.94 &	0.93 &	9\\
      		\mathrm{ALMA\, J154229.778-334241.86} & 15:42:29.778 & -33:42:41.86 & 44  & 10.12 & 0.11 & 4.06 & 2.81 & 4.50 \pm 0.12 & 770 \pm 40 &	640 \pm 40 &	57 \pm 13 & 2.85 \pm 0.05  & 0.94	& 0.93 &	9  \\ 
     		\mathrm{ALMA\, J154456.522-342532.99} & 15:44:56.522 & -34:25:32.99 & 42  & 6.24 & 0.068 & 2.50 & 1.92 & 2.79 \pm 0.11  & 840 \pm 60 & 280 \pm 80 &79 \pm 4 & 1.94 \pm 0.05  & 0.91 &	0.83 &	-83 \\ 
     		\mathrm{ALMA\, J154458.061-342528.51} & 15:44:58.061 & -34:25:28.51 & 80  & 13.36 &0.097 & 1.42 & 1.42 & 1.49 \pm 0.06  & - & - & - & 1.44 \pm 0.03  & 0.91 &	0.83 &	-83 \\     		
     		\mathrm{ALMA\, J154506.515-344326.15} & 15:45:06.515 & -34:43:26.15 & 180 & 11.97 & 0.089	 & 0.58 & 0.58 & 0.49 \pm 0.02  & - & - & -  &  0.59 \pm 0.01  & 0.95 & 	0.82 &	-81\\
     		\mathrm{ALMA\, J154634.169-343301.90} & 15:46:34.169 & -34:33:01.90 & 130 & 14.77 & 0.083 & 0.75 & 0.75 & 0.83 \pm 0.02 &  - & -  & - & 0.76 \pm 0.01  & 0.96 &	0.82 &	-82\\     		
     		\mathrm{ALMA\, J160658.604-390407.885} & 16:06:58.604 & -39:04:07.88  & 152 & 7.60 & 0.062 & 0.48 & 0.48 & 0.66 \pm 0.04  & - & - & - & 0.49 \pm 0.02  & 0.93 & 0.85 & 87  \\
     		\mathrm{ALMA\, J160804.168-390452.84} & 16:08:04.168 & -39:04:52.84 & 126 &4.31 & 0.093 & 0.87 & 0.87 & 1.01 \pm 0.05  & - & - & - & 0.87 \pm 0.03  & 0.93 & 0.85 & 86 \\     		
     		\mathrm{160826.8-384101} & 16:08:26.773 & -38:41:01.48 & 44 & 9.06 & 0.059 & 1.57 & 1.57 & 1.58 \pm 0.05 &- & - & - & 1.63 \pm 0.03  & 0.93 & 0.85 & 87 \\     		
     		\mathrm{V1094 Sco} & 16:08:36.157 &  -39:23:02.74  & 4 &15.82  & 0.23 & 244.05  & 73.64 & 218 \pm 6 & 1760 \pm 60 & 1090 \pm 40 & 108\pm 4 & 61.8 \pm 1.4 &0.92 & 0.84 & 84\\  
     		\mathrm{Lup 706}  & 16:08:37.316 & -39:23:11.38 & 226 &0.50  & 0.090 & 0.47  & 0.47 & 0.53 \pm 0.04  & - & - & - & 0.46 \pm 0.02 & 0.93 & 0.85 &  85\\ 
     		\mathrm{Par-Lup3-4} & 16:08:51.426 & -39:05:30.82 & 196 & 0.05  & 0.031 & 0.31  & 0.31 & 0.41 \pm 0.07 & - & - & - & 0.28 \pm 0.04 & 0.93  & 0.84 &  84\\ 
     		\mathrm{SONYC-Lup3-7} & 16:08:59.530 & -38:56:27.96 & 270 & 6.26&  0.055 & 0.32 & 0.32  & 0.45 \pm 0.04  & - & - & - & 0.32 \pm 0.02  & 0.93	& 0.84& 83  \\     		
     		\mathrm{ALMA\, J160920.089-384515.92} & 16:09:20.089 & -38:45:15.92 & 113 & 6.01 & 0.078 & 0.90 & 0.90 & 1.18 \pm 0.11  & -	&	- & - & 0.82 \pm 0.05  & 0.93	& 0.85 &	86 \\     		
     		\mathrm{ALMA\, J160920.171-384456.40} & 16:09:20.171 & -38:44:56.40 & 226 & 13.75 & 0.18 & 0.91 & 0.91 & 1.2 \pm 0.2  & - & - & -  & 0.90 \pm 0.10  &  0.93	& 0.85 &	86\\     		
     		\mathrm{ALMA\, J160932.167-390832.27} & 16:09:32.167 & -39:08:32.27 & 145 & 11.89 & 0.13 & 0.62  & 0.62 & 0.47 \pm 0.04 & - & - & - &  0.65 \pm 0.03  & 1.12 &	0.82 &	85.61 \\     		
           \mathrm{ALMA\, J161030.273-383154.52} & 16:10:30.273 & -38:31:54.52 & 108 &4.76 & 0.064 & 0.70 & 0.70 & 0.72 \pm 0.03  & -  & -  & -& 0.71 \pm 0.02 & 0.93 &  0.85 & 86.24\\
			\hline \hline
     	\end{array}   
$$
\begin{tablenotes}
\item[1] J2000.0 positions of the peak intensity in right ascension and declination
\item[2]{\url{https://help.almascience.org/index.php?/Knowledgebase/Article/View/319/6/what-is-the-astrometric-accuracy-of-alma}, ALMA Technical Handbook, Chapter 10.5.2 Astrometric Observations)}
\item[3]Flux density measured inside the contour at the 3 $\sigma$ level 
\item[4]Integrated flux, deconvolved size, peak flux, and beam size obtained from a Gaussian fitting using the CASA task \textit{imfit} 
\end{tablenotes}
\end{threeparttable}
}
\end{table*}
\end{landscape}
\restoregeometry

\begin{table} 
\caption{Distances using Gaia Data Release 2}             
\label{distances_gaia_dr2}      
\centering                          
\begin{tabular}{|c c|}        
\hline\hline                 
Name & Distance [pc] \\    
\hline
\hline  
\multicolumn{2}{|c|}{\textbf{Lupus BDs detected with ALMA}} \\                      
\hline 
160826.8-384101	&   165 $\pm$ 4  \\
V1094 Sco	&   154.7 $\pm$ 1.1  \\
Lup 706	&     191 $\pm$ 29\\
Par-Lup 3-4	&  155 $\pm$ 14   \\
SONYC-Lup 3-7	&  151 $\pm$ 6   \\
\hline
\multicolumn{2}{|c|}{\textbf{Lupus BDs not detected with ALMA}} \\
\hline
153701.1-332255	&  94 $\pm$ 4   \\
153709.9-330129	&  175 $\pm$ 7   \\
153921.8-340020	&  152 $\pm$ 9   \\
154140.8-334519	&  151 $\pm$ 8   \\
160545.8-385454	&  44 $\pm$ 1 \\
160714.0-385238	&  125 $\pm$ 4  \\
160816.0-390304		&  164 $\pm$ 5  \\
160833.0-385222	&  158 $\pm$ 4   \\
160848.2-390920		&  184 $\pm$ 5  \\
161144.9-383245	&  169 $\pm$ 5  \\
161225.6-381742	&  159 $\pm$ 3  \\
161210.4-390904	&  233 $\pm$ 30   \\
\hline
\multicolumn{2}{|c|}{\textbf{Ophiuchus}} \\
\hline
ISO-Oph030	&  137 $\pm$ 7   \\
ISO-Oph032	&  150 $\pm$ 9   \\
ISO-Oph042	&  154 $\pm$ 19    \\
ISO-Oph102	&  142 $\pm$ 6   \\
GY92-264	&  139 $\pm$ 5 \\
ISO-Oph160	&  142 $\pm$ 22  \\
ISO-Oph164	&  142 $\pm$ 16  \\
GY92-320    &  141 $\pm$ 18   \\
ISO-Oph176	&  140 $\pm$ 19  \\
ISO-Oph193	&  147 $\pm$ 17  \\
\hline
\end{tabular}
\end{table} 

\section{Results}
\label{p2_analysis}
\subsection{ALMA: continuum detections}
\begin{figure*}
\centering
\subfloat{\includegraphics[width=0.46\textwidth]{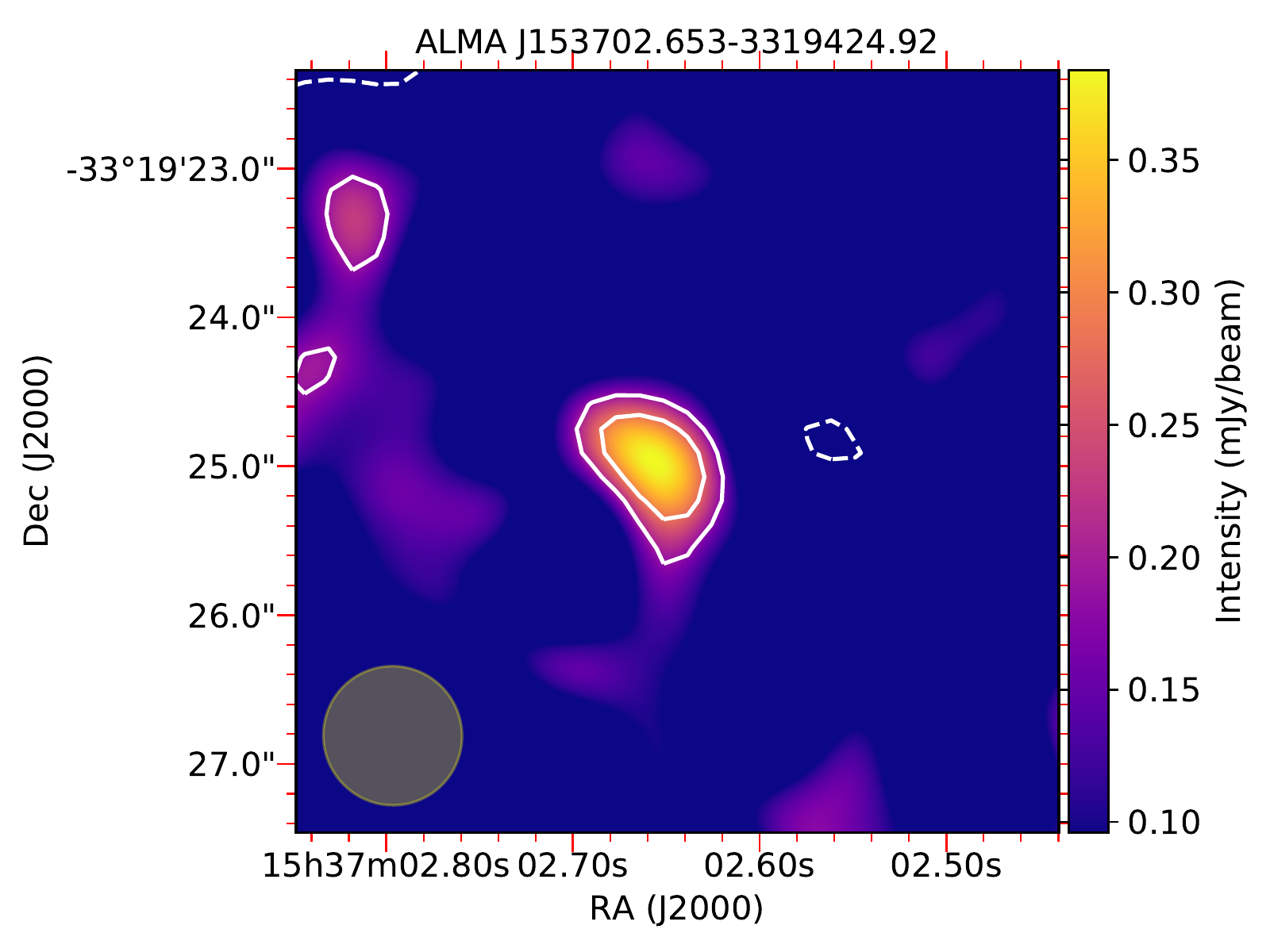}}
\subfloat{\includegraphics[width=0.46\textwidth]{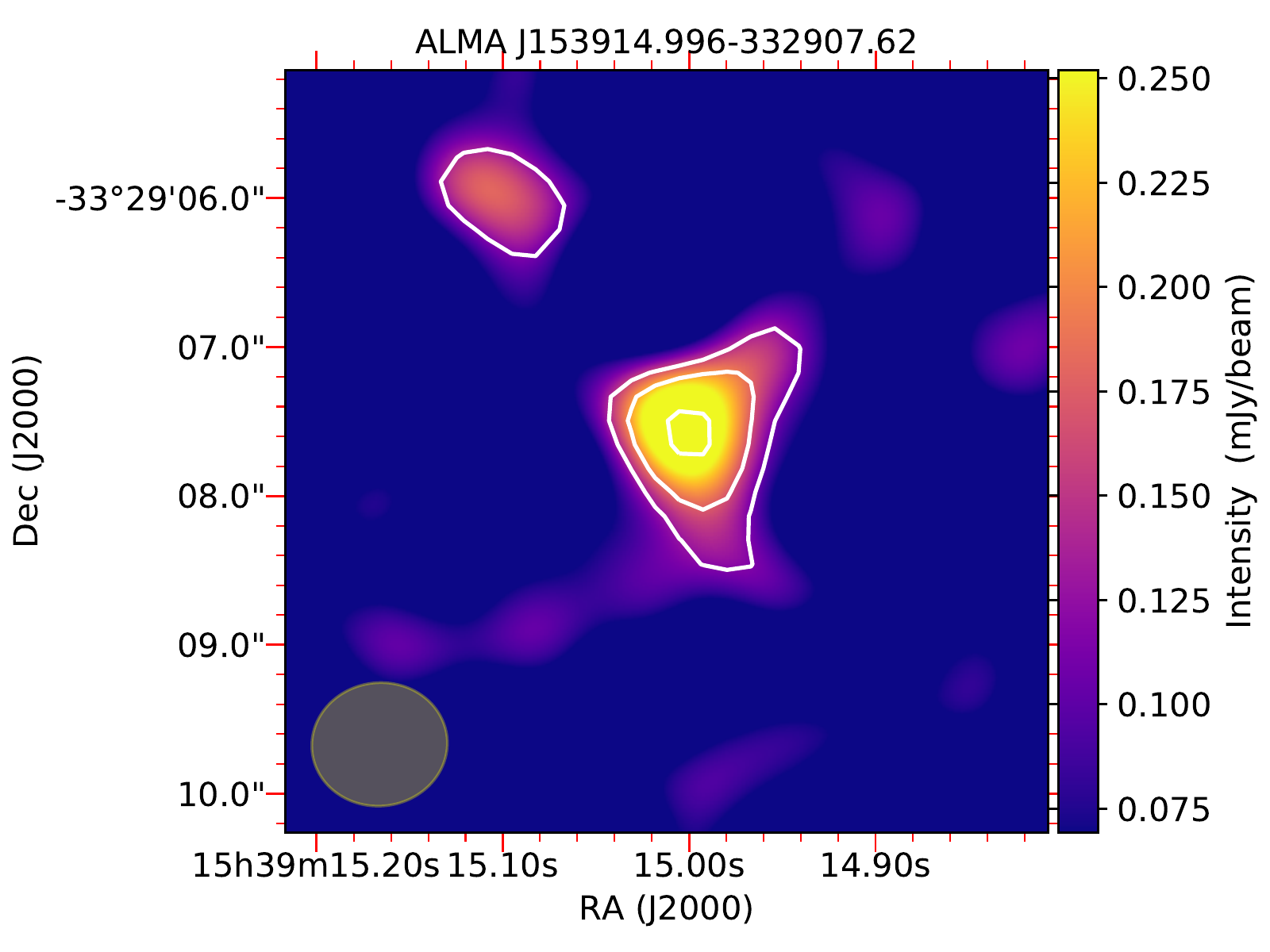}}
\qquad
\subfloat{\includegraphics[width=0.46\textwidth]{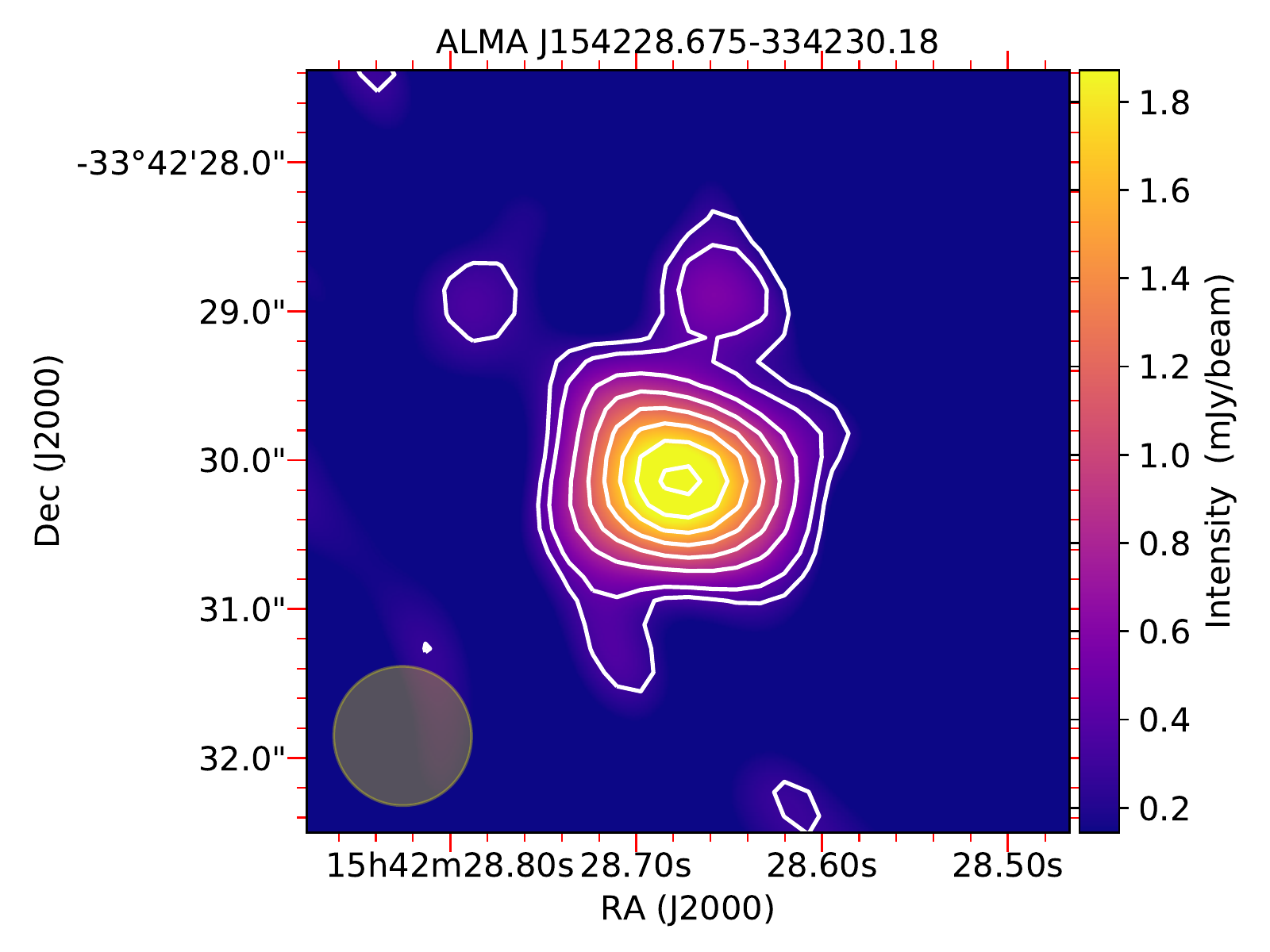}}
\subfloat{\includegraphics[width=0.46\textwidth]{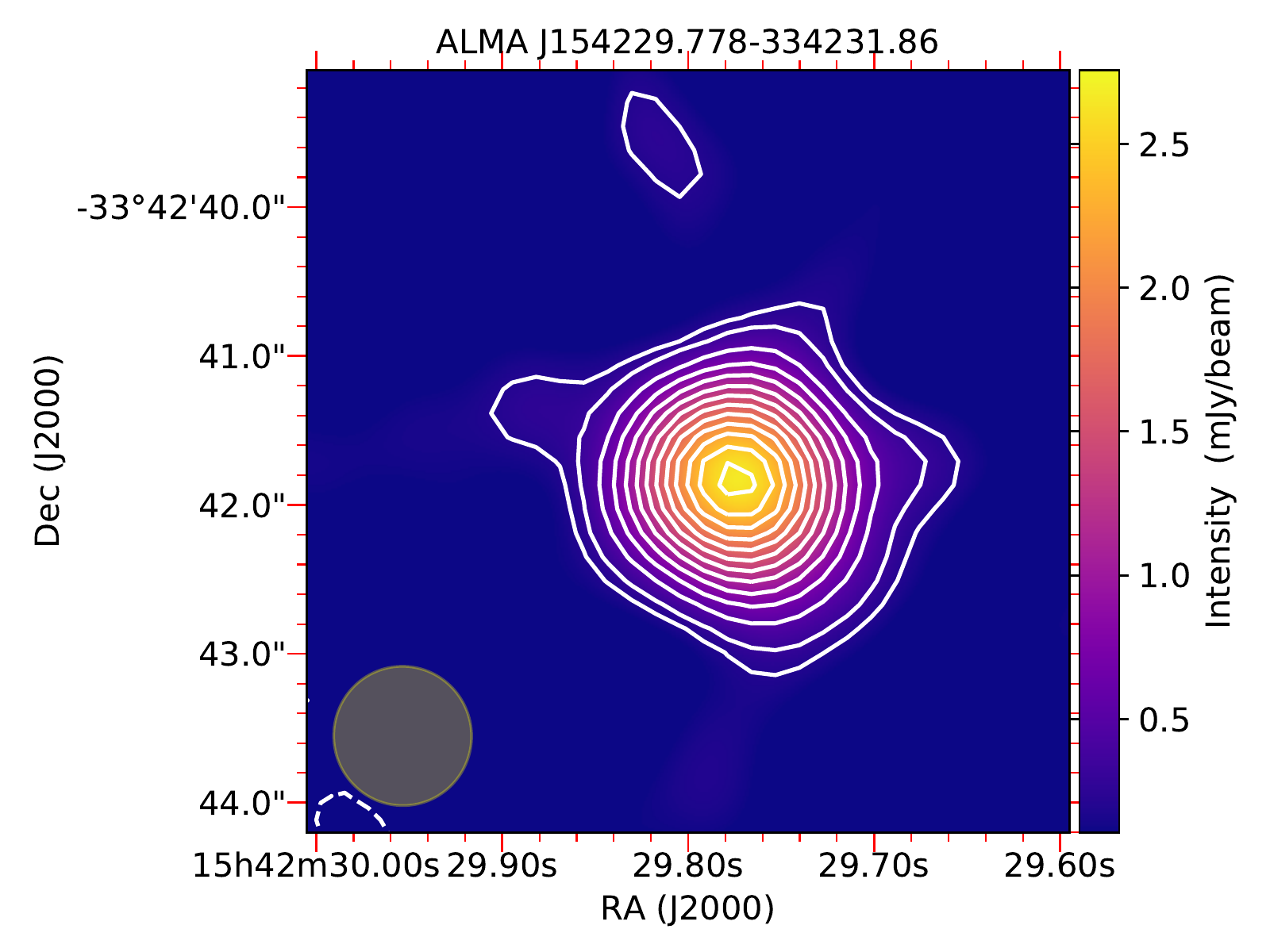}}
\qquad
\subfloat{\includegraphics[width=0.46\textwidth]{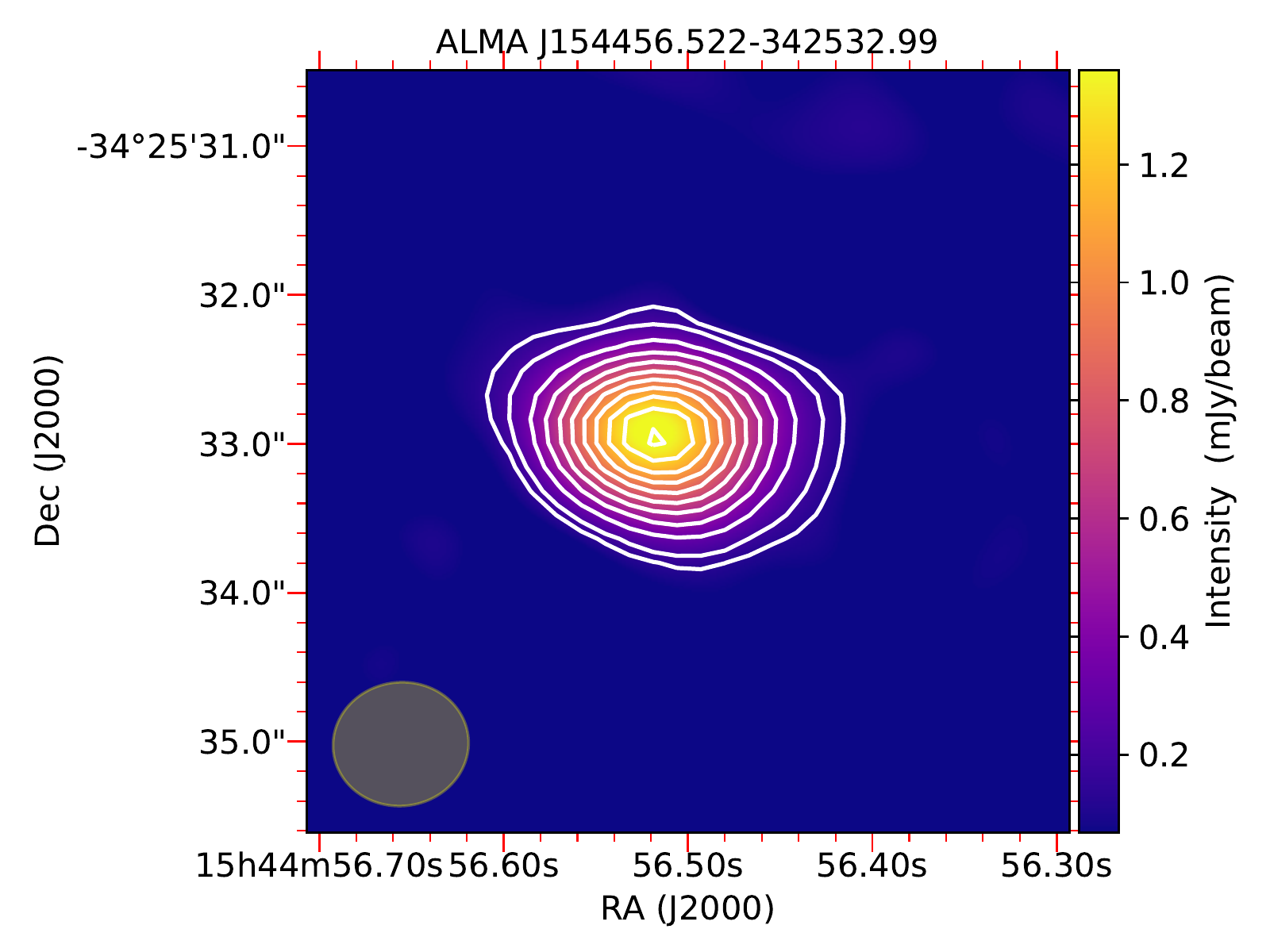}}
\subfloat{\includegraphics[width=0.46\textwidth]{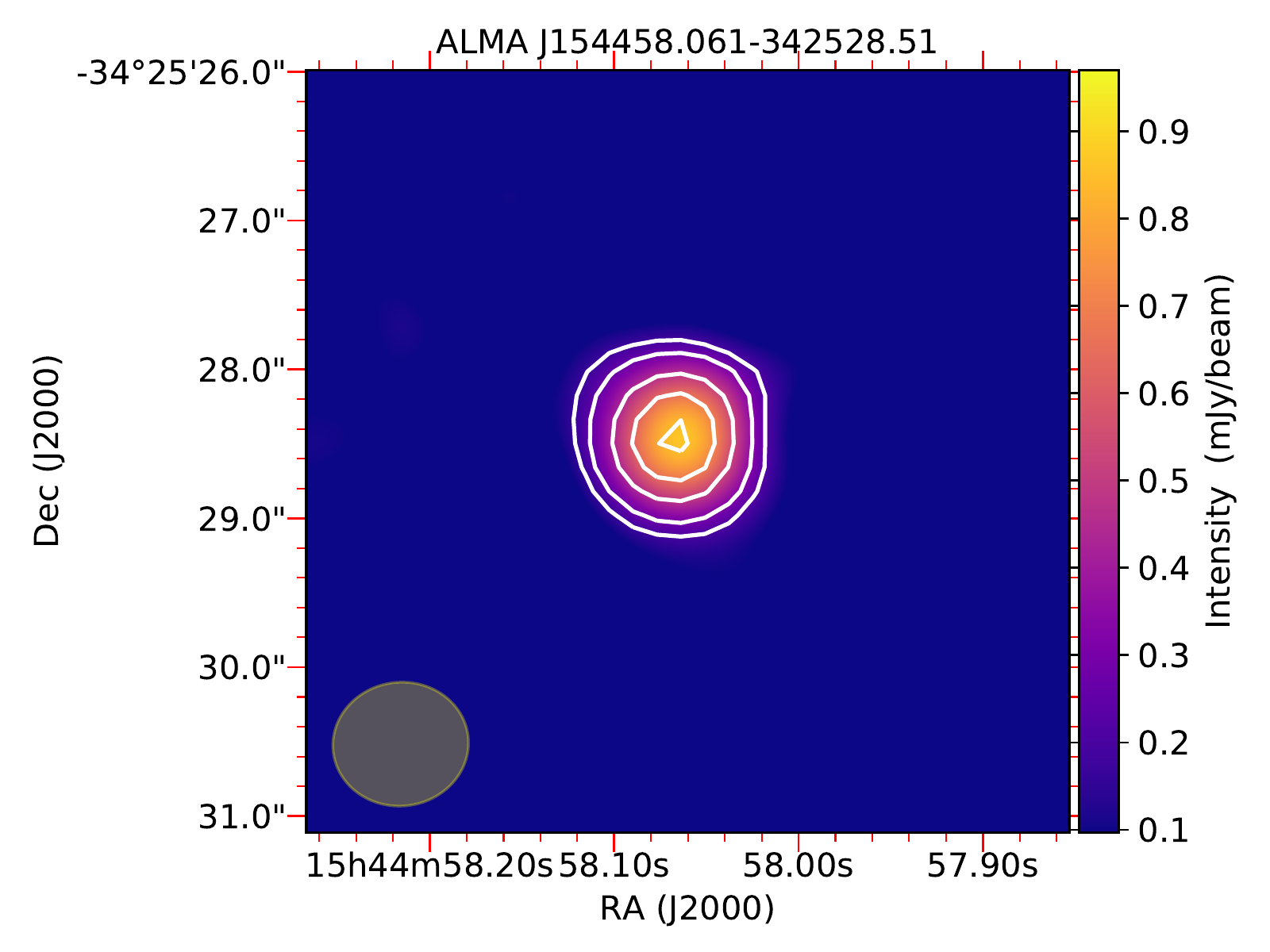}}
\caption{\label{panel_alma} 1.3 mm ALMA continuum images of the 19 detected sources. White contours are 2, 3, 5, 7, 9, 11, 13, 15, 17, 19, 21, 23, 25, 27, 29, 50, 100, 150 $\sigma$. Here, $\sigma$ is the rms noise level of each respective map, given in Table \ref{table:tabla_ppal}. Dashed contours are negative emission at -2 $\sigma$. Beam size is represented by the grey ellipse in the bottom left corner. Continued on Appendix \ref{apendix_panel_alma}}
\end{figure*}

The ALMA continuum images for each detection are shown in Figure \ref{panel_alma}. In total, 19 sources were detected in continuum emission in 15 pointings among the 64 science targets. The flux density for detections ranged between 0.22 and 244.05 mJy (see Table\,\ref{table:tabla_ppal}).

There are four ALMA pointings where we detect two sources inside the primary beam (J154228 and J154229, J154456 and J154458, J1609200 and J160920171, Lup 706 and V1094 Sco) as can be seen in Appendix \ref{dobles}. Finally, we note that most of the detected sources (16 out of 19) are spatially unresolved in our observations. 

\begin{figure*}
\centering
\includegraphics[width=\textwidth]{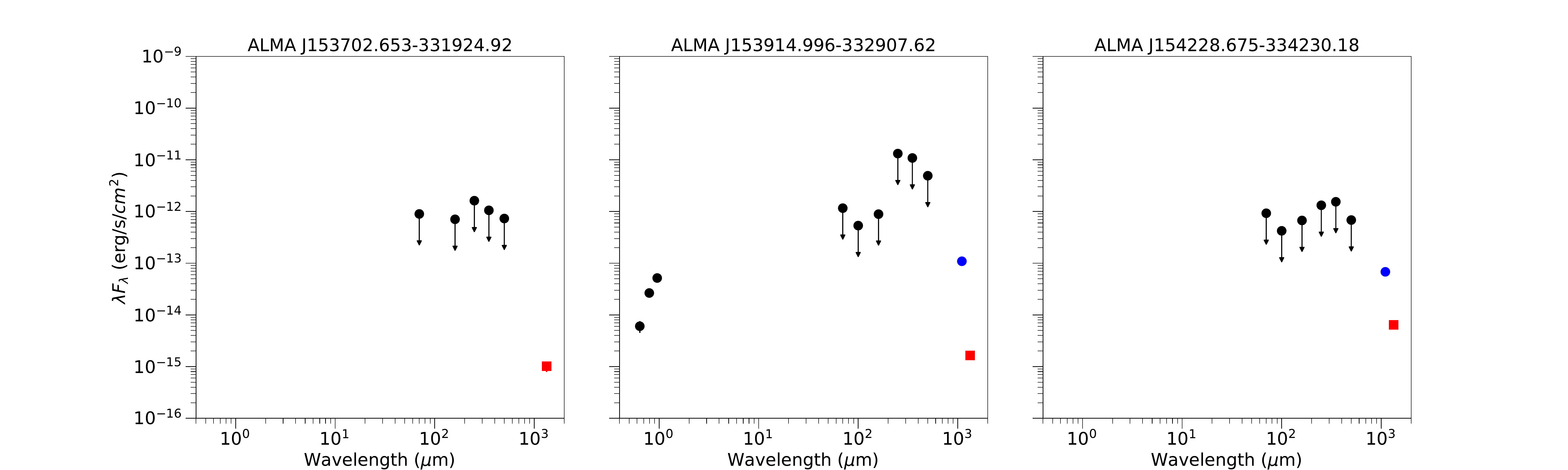}
\caption{SED for sources detected with ALMA at 1.3 mm. Red squares show the ALMA fluxes. Blue circles show the AzTEC fluxes. Black circles show the fluxes from other telescopes: WFI, DENIS, 2MASS, WISE, Spitzer, Akari, Herschel, and APEX/LABOCA. Upper limits are included as arrows. Continued on Appendix \ref{apendix_panel_sed}}
\label{panel_sed}
\end{figure*} 

All the sources detected in the continuum with ALMA are included in Table \ref{table:tabla_ppal}. We provide the position, the separation from the phase centre, the rms, the flux density, and the peak intensity. Other quantities such as the integrated flux, size, peak flux, and beam size  obtained from a Gaussian fitting using the task \textit{imfit} in CASA are also included in the table. 
\subsection{ALMA: Gas emission detections}
Although obtaining continuum emission images was the main observational goal of the ALMA observations, the CO(2--1) emission line was also included in the spectral setup. Detecting gas emission close to the cloud velocity for our sources would be a confirmation of their cloud association. Unfortunately, we only have a clear gas detection in two sources. The first one is V1094 Sco and it was studied by \citet{Ansdell16-1,Ansdell18-1}. The second one is Par-Lup 3-4 where we discovered the base of a bipolar molecular outflow traced by CO gas \citep{Santamaria-Miranda20-1}. Extended cloud emission is seen in $^{12}$CO(2--1) surrounding J154634 at velocities between +3.4 to +5.3 km/s, but no spatially compact gas emission was detected at the position of this source. These velocities are consistent with those of the Lupus 3 cloud, +4.1 km/s with $\Delta$V = 1.7 km/s \citep{tachiharaetal96-1}. We note that several disks in Lupus 3 show systemic velocities in that range (e.g. EX Lup at +4.4 km/s \citealt{Hales18-1} or the sources in \citealt{Ansdell18-1}). The rest of the ALMA sources were not detected in CO and have an upper limit to their CO(2--1) fluxes of $\sim$6 mJy/beam (1 $\sigma$) in a 0.3 km/s channel.
Additionally, there is no detection of the other emission lines (C$^{18}$O(2--1), SiO(5--4)) in any of the sources with a rms of $\sim$4 mJy/beam, $\sim$1 mJy/beam, respectively, with  both of them in a 2.7 km/s channel.

\subsection{Spectral energy distribution}

\subsubsection{Optical/IR/radio counterpart association}
To associate the ALMA detections with optical/IR/radio sources, we first inspected the archival images. A total of seven counterpart sources were identified by searching within a separation smaller than the total accuracy error. 
The total accuracy error is calculated as the square root of the quadratic sum of ALMA astrometric accuracy and the astrometric accuracies of the counterpart's instrument. In the case of the ESO 2.2m/WFI, for which there is no astrometric accuracy, we use a conservative value of $\sim$2$\arcsec$. The astrometric accuracies were between 1 and 2 \,arcsec, depending on the instrument, and are listed in Table\,\ref{Tab:instrumentos}. The SEDs of all the ALMA detections are consistent with YSOs and are displayed in Figure \ref{panel_sed}. 

\subsubsection{Classification}
We classified the sources in the following categories: ALMA detections without optical/IR counterpart (pre-BD phase), Class 0, Class I, and Class II sources. For this classification, we use the $\alpha _{IR}$ slope \citep{Adams87}, as well as the T$\mathrm{_{bol}}$ and L$\mathrm{_{bol}}$ values.
We used the $\alpha _{IR}$ slope values in \citet{Greene94} where Class 0 protostars have $\alpha _{IR}$ slope values exceding 0.3, Class I/II sources have $\alpha _{IR}$ slope values between -0.3 and 0.3, Class II sources  have $\alpha _{IR}$ slope values between -1.6 and -0.3, and Class III sources  have $\alpha _{IR}$ slope values below -1.6. We used different combinations of photometric bands (2.16 $\micro$m, 12 $\micro$m), (2.16 $\micro$m, 23.67 $\micro$m) and (3.4 $\micro$m, 23.67 $\micro$m) because not all the photometry was available for all sources.  

\label{tbol_lbol}
We calculated the bolometric luminosity (L$_\mathrm{bol}$) and the bolometric temperature (T$_\mathrm{bol}$) for each source that has at least a counterpart at three different wavelengths (see Table \ref{Tab:tbol_lbol}). This group of sources includes: J154229, 160826, V1094 Sco, Lup706, Par-Lup 3-4, and SONYC-Lup 3-7. We also included J153914 to obtain a lower limit, although this source does not have an infrared counterpart. To calculate L$_\mathrm{bol}$, we used formula (1) from \citet{Enochetal09-1}, and for T$_\mathrm{bol}$, we used formula (2) from the same paper and the mean frequency from \citet{Myersetal93-1}. In addition, to classify Class 0, I or II/III consistently, we used the definition in \citet{Chen1995} where Class 0 protostars have T$_{\mathrm{bol}}$ below 70 K, Class I sources have temperatures between 70 and 650 K, and Class II-III sources temperatures exceeding 650\,K. \\ 

\begin{table}
\caption{Bolometric temperature and bolometric luminosity for those ALMA detections with more than three points in their SED.}     
\label{Tab:tbol_lbol}      
\centering                          
\begin{tabular}{l c c}        
\hline\hline                 
Name & Temperature & Luminosity \\    
& [K] &  [L$_{\odot}$] \\
\hline
\hline  \\                
J154229 & 64.6$^{+2.4}_{-1.7}$ & 0.0044 $\pm$ 0.0008\\
J153914 & > 46.92 & > 0.04012 \\ 
160826 & 1411$^{+9}_{-5}$ & 0.0268$^{+0.0009}_{-0.0005}$ \\
V1094 Sco & 2019$^{+25}_{-139}$ & 0.5600$^{+0.0009}_{-0.0005}$ \\
Lup706 & 415$^{+10}_{-14}$ & 0.0290 $\pm$ 0.003 \\
Par-Lup 3-4 & 406$^{+21}_{-24}$ & 0.0190$\pm$0.0006 \\
SONYC-Lup 3-7 & 1959$^{+16}_{-14}$ & 0.0095$\pm$ 0.0004 \\
\hline                                   
\end{tabular}
\end{table} 

\begin{figure*} 

\centering
\includegraphics[width=\textwidth]{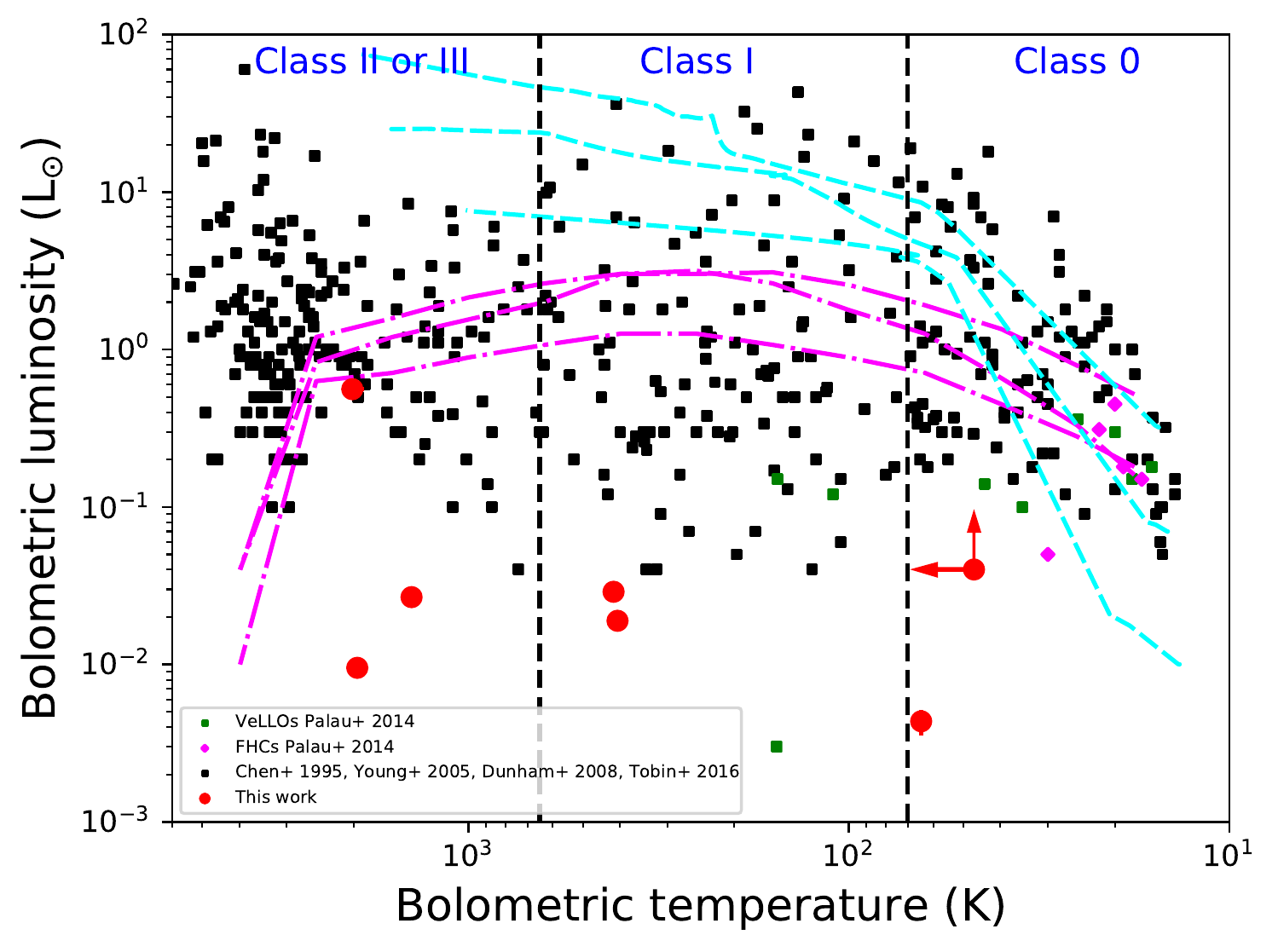}
      \caption[Bolometric luminosity versus bolometric temperature of the Lupus 1 and 3 sample]{Bolometric luminosity versus bolometric temperature. Black squares represent sources from \citet{Young&Evans05-1} and \citet{Dunham2008} which show evidence of embedded low luminosity sources. Sources in Perseus from \citet{Tobinetal16-1} are represented as open squares. Young Taurus \citep{Chen1995} sources are displayed as black crosses. Green squares are known VeLLOs and the magenta diamonds are the First Hydrostatic Cores from  \citet{Palau14-1}. The vertical dashed lines mark the Class 0–I and Class I–II bolometric temperature boundaries from \citet{Chen1995}. The cyan short-dashed lines represent the evolutionary tracks for the three models with different masses considered by \citet{Young&Evans05-1}. The magenta, dotted, and dot-dashed lines show the evolutionary tracks for three models considered by \citet{Myersetal98-1}. Red points including uncertainty represent values obtained in this papers. Lower limits are included as red arrows shows for J153914.}  
         \label{tbol_lbol_plot}          
\end{figure*}

One source (J154229) is at the boundary between Class 0 and Class I objects, according to its T$_\mathrm{bol}$ value. Using the T$_\mathrm{bol}$ classification method Par-Lup 3-4 and Lup 706 appear to be Class I objects, although both sources have been previously classified as Class II in the literature. These two sources are mentioned in  \citet{alcalaetal14-1} as having the lowest luminosities among all the Lupus YSOs. The subluminous nature of Par-Lup 3-4 has been investigated by \citet{Huelamo10-1} concluding that it could be explained by the presence of a close to edge-on disk (inclination of $\sim$81 degrees). A detailed study of this source has been presented by \citet{Santamaria-Miranda20-1}, confirming the high inclination of the system. In the case of Lup706, the $\alpha _{IR}$ slope shows values of Class I/II or Class II using different combination of photometric bands, it may be in transition to Class I to Class II. The remaining sources (160826, V1094 Sco, and SONYC-Lup 3-7) can be classified as Class II or III objects and we maintain their previous classifications. Figure \ref{tbol_lbol_plot} shows the position in the T$_\mathrm{bol}$-L$_\mathrm{bol}$ diagram of the mentioned sources along with objects from previous studies in the regions of Perseus and Taurus. 

The sources we study in this work have lower luminosities than other BD candidates from previous works, as shown in Figure \ref{tbol_lbol_plot}, with the exception of the proto-BD candidate J041757 of \citet{Barrado2009} and \citet{Palau2012-1}. Thanks to the sensitivity of ALMA and the close proximity of the Lupus complex, we have detected the least luminous pre-BD candidate so far (J160658). We also detected a new Class 0/I substellar candidate (J154229), a type of object for which confirmed detections are very few in number (e.g. ICM348-SSM2E, \citealt{Palau14-1}). Class I sources are also difficult to detect, and here we add one new candidate to the list of Class I substellar candidate sources such as L1148-IRS \citep{Kauffmann11-1}, and J042118 and J041757 \citep{Palau2012-1}. The ALMA detection of the Class II sources provides a new point in the SED that helps us to better constrain their position in the T$\mathrm{_{bol}}$- L$\mathrm{_{bol}}$ diagram. 

Additionally, we identified the whole sample as VeLLOs, with the exception of V1094 Sco. These VeLLOs are sources that have an internal luminosity (L$\mathrm{_{int}}$) below 0.1 L$_{\odot}$. To estimate L$\mathrm{_{int}}$, we used the formula from \citet{Dunham2008}:
\begin{equation}
L\mathrm{_{int}} (\mathrm{L_{\odot}}) = 3.3 \times 10^{8}\bigg(\mathrm{F}_{70}\bigg(\mathrm{\frac{d}{140}}\bigg)^{2}\bigg)^{0.94}   
\end{equation}
where $\mathrm{F}_{70}$ is the flux at 70\,$\mu$m. For almost all the objects in our sample we only have upper limits (see Appendix \ref{wavelength_fluxes_SED}) for that specific wavelength. Using these upper limits we find that all our sources are below the VeLLO luminosity threshold. 

Finally, we double-checked previous classifications in the literature, if available, for all the sources. Table \ref{tabla_clasificatoria} shows the classification for each source using the different methods. 
\label{clasificacion}

\begin{table*}[h]
\caption{Source classification using different methods}    
    \centering
    \begin{tabular}{c c c c c c}
\hline    
Source & Previous classification & Counterpart   & $\alpha _{IR}$ slope & T$_\mathrm{bol}$   & Final classification   \\
\hline
ALMA J153702.653-331924.92 & No & No  & - & -  & Pre-BD phase candidate \\
ALMA J153914.996-332907.62 & No & Yes  & - & > Class 0  & Class I candidate \\
ALMA J154228.675-334230.18 & No & No  & - & -  & Pre-BD phase candidate \\
ALMA J154229.778-334241.86 & No & Yes  & Class I  & Class 0/I  & Class 0/I candidate \\
ALMA J154456.522-342532.99 & No &  No & - & -  & pre-BD phase candidate\\
ALMA J154458.061-342528.51 & No & No  & - & -  & Pre-BD phase candidate \\
ALMA J154506.515-344326.15 & No &  No & - & -  & Pre-BD phase candidate\\
ALMA J154634.169-343301.90 & No & No  & - & -  & Pre-BD phase candidate \\
ALMA J160658.604-390407.88 & No & No  & - & -  & Pre-BD phase candidate \\
ALMA J160804.168-390452.84 & No & No  & - & -  & Pre-BD phase candidate\\
160826.8-384101 & Class II & Yes  & Class II & Class II  & Class II \\
V*V1094 Sco & Class II & Yes  & Class II & Class II  & Class II\\
Lup706 & Class II & Yes  & Class I/II & Class I  & Class I/II \\
Par-lup3-4 & Class I/II & Yes  & Class I & Class I  & Class I/II\\
SONYC-Lup3-7 & Class II & Yes  & Class I/II & Class II  & Class II\\
ALMA J160920.089-384515.92  & No & No  & - & -  & Pre-BD phase candidate \\
ALMA J160920.171-384456.40 & No  & No  & - &  - &Pre-BD phase candidate \\
ALMA J160932.167-390832.27& No & No  & - & -  & Pre-BD phase candidate \\
ALMA J161030.273-383154.52 & No & No  & - & -  & Pre-BD phase candidate \\
    \end{tabular}
    \label{tabla_clasificatoria}
\end{table*}

\subsection{ALMA: Mass estimates}
\label{alma continuum masses}
The total mass of the ALMA detections (see Table \ref{Tab:masas_medidas}) is calculated assuming that the observed emission is optically thin and the gas-to-dust mass ratio is 100 \citep{Bohlin78}. Given these assumptions, we use the following formula from \citet{Hildebrand1983} to estimate the dust and gas mass: 
	\begin{equation}
      M =  \frac{S_{\nu} D^{2}}{B_{\nu}(T_{d}) \kappa_{\nu}}.  
   \end{equation}

Here $S_{\nu}$ is the flux density in the region inside a 3 $\sigma$ contour level, D is the distance to the source,  or the average cloud distance if not in Table \ref{distances_gaia_dr2}, $B_{\nu}(T_{d})$ is the Planck function at the observed frequency (225 GHz) at temperature $T_{d}$, and $\kappa_{\nu}$ is the absorption coefficient obtained from Table 1 in \citeauthor{Ose94}(1994; column for thin ice mantles and density of $10^{6}$ cm$^{-3}$), interpolated for a frequency of 225 Ghz, which provides a value of $\kappa_{\nu}$=8.5 $\times 10^ {-3}$ cm${^2}$g${^{-1}}$ taking into account the aforementioned gas-to-dust ratio.

\begin{table*}
\footnotesize
\caption{Derived properties from ALMA and AzTEC detections}
\label{Tab:masas_medidas}
\centering
\begin{tabular}{cccccc }
\hline \hline
Name & Mass (ALMA)& Mass (AzTEC) &  Size (AzTEC) & Lupus  & Missing  \\
 &  [M$\mathrm{_{Jup}}$] &  [M$\mathrm{_{Jup}}$]& [AU] & cloud & flux [\%]\\
\hline
\hline  \\  
ALMA J153702.653-331924.92 & 0.9 $\pm$ 0.3 &  - & - &  1 & - \\
ALMA J153914.996-332907.62 & 0.63 $\pm$ 0.11 & 18 $\pm$ 9 & 3270 & 1 & 97\\
ALMA J154228.675-334230.18 & 5.4 $\pm$ 1.1 &  41 $\pm$ 10 & < 2680 &1 & 82\\
ALMA J154229.778-334241.86 & 3.5 $\pm$ 1.7 &  11 $\pm$ 6 & 3580 &1 & 83\\
ALMA J154456.522-342532.99 & 4.8 $\pm$ 1.0  & 31 $\pm$ 8 & 3800 & 1 & 87\\
ALMA J154458.061-342528.51 & 2.7 $\pm$ 0.6  & 18 $\pm$ 4 & 3800 &1 & 86\\
ALMA J154506.515-344326.15 & 1.1 $\pm$ 0.3  &  51 $\pm$ 12 & 3380 &1 & 99\\
ALMA J154634.169-343301.90 & 1.4 $\pm$ 0.3 & 180 $\pm$ 40 & 7580 &1 & 99\\
ALMA J160658.604-390407.88 & 1.0 $\pm$ 0.2 & - &- & 3 & - \\
ALMA J160804.168-390452.84 & 1.7 $\pm$ 0.4 & - &- & 3 & -\\
160826.8-384101 & 1.1 $\pm$ 0.4 & - & -& 3 & -\\
V1094 Sco & 124 $\pm$ 35 & 82 $\pm$ 10 & 5030 & 3 & -\\
Lup 706 & 0.20 $\pm$ 0.12 & - & - & 3 & - \\
Par-lup 3-4 & 0.19 $\pm$ 0.08 & - & -& 3 & -\\
SONYC-Lup 3-7 & 0.18 $\pm$ 0.07 & - & -& 3 & - \\
ALMA J160920.089-384515.92  & 1.8 $\pm$ 0.4 & - & -& 3 & -\\
ALMA J160920.171-384456.40 & 1.8 $\pm$ 0.5 & - &- &3 & -\\
ALMA J160932.167-390832.27& 1.2 $\pm$ 0.4  & 92 $\pm$ 25 & 5000 & 3 & 99\\
ALMA J161030.273-383154.52 & 1.4 $\pm$ 0.3 & - &- &3 & -\\

\hline
\end{tabular}
\end{table*}  

We built the SED (Figure \ref{panel_sed}) of all the ALMA detected sources to classify their evolutionary state, and then we adopted a dust temperature depending on class. For Class 0/I candidates, we used a temperature of 15 $\pm$ 5 K based on the average temperature maps over the Class 0/I candidates in Lupus~1 and 3 (Figure~\ref{Herschel}). These maps have been obtained from Herschel data (Teixeira et al., in prep.) following the methodology described in \citet{Lombardi14-1}. Most of the ALMA detections in Lupus~1 lie in regions with temperatures close to 16.5\,K, with only one source (J153702) showing a value close to 19\,K. In the case of the Lupus~3 sources, the reported temperatures vary between $\sim$13 and 19\,K, depending on their location within the cloud. Other examples of similar temperatures for Class 0/I in the literature can be found in \citeauthor{Stutz10-1}(2010; 17 $\pm$ 1 K)  for low-mass stars, or for BDs \citep{Barrado18-1}, using a temperature convention of 15 K. For less-evolved (starless) sources (ALMA detections without optical/IR counterpart), we use a temperature of 9 $\pm$ 1 K as an intermediate value between $\sim$8.5-10 \citep[and references therein]{Andre2012} or 7-13\,K from earlier work \citep{Evans01-1}. For spectroscopically confirmed Class II BDs, we used a temperature of 20 K \citep{Pascucci16-1}. The temperature for V1094 Sco came from the formula T=25(L$_{*}$/L$_{\odot})^{0.25}$K \citep{Andrews2013} as it is in the solar-mass regime (see Section \ref{vscop}). 

Mass uncertainties are calculated using the computed distance error, the flux error, and the temperature error. Opacity uncertainty is not included although we are aware that it is a major source of uncertainty for mass estimates. Using the opacity formula from \citet{Ward-Thompson10-1}, the masses would be larger by 60$\%$. 

The range of masses of the ALMA detections is between 0.18 and 124 M$\mathrm{_{Jup}}$. All the masses are in the planetary mass regime, except that of V1094 Sco, which is a well-known protostar with a protoplanetary disk surrounding it \citep{Ansdell16-1}. 
\label{sec:continuum}

\subsection{AzTEC: Mass estimates}
\label{aztec_masses_sec}
Masses derived from the AzTEC data are estimated in the same fashion as described in Section\,\ref{alma continuum masses}. In Table\,\ref{Tab:masas_medidas}, we show the mass estimates based on the ALMA continuum detections together with the mass estimates and the sizes of the AzTEC clumps, as well as an estimate of the missing flux in our ALMA observations. We also show the cloud membership for each object (11 in Lupus 3 and 8 in Lupus 1). 

In order to calculate the missing flux in the ALMA observations we first calculated the expected flux of the AzTEC clumps at the frequency of the ALMA observations, assuming a dust emissivity index of $\beta$=1.8.
The missing flux estimate for V1094 Sco is not included in the table given the complexity of the  dust emissivity index for this source \citep{vanTerwisga18-1}. There is a large mass difference between the AzTEC clumps and the compact sources detected with ALMA, the latter providing much smaller mass values. The percentage of missing flux with ALMA ranges between $\sim$82\,\% to more than $\sim$99\,\%. In Section \ref{detection_rate} we discuss the flux discrepancy.

There are five ALMA sources (J160658, J160804, Par-Lup 3-4, J1609200, and J1609201) not associated with any AzTEC clump. With the exception of Par-Lup 3-4, all these sources were serendipitously discovered in the ALMA field of view (see Section \ref{no_counterparts}),  where we expected to find another BD. In the case of Lup 706, we assumed that most of the emission originates from V1094 Sco. The rest of the ALMA detections are associated with AzTEC spatially resolved clumps except for J154228, which is associated with a point-like source in the AzTEC maps (see Appendix \ref{aztec_maps}). SONYC-Lup 3-7 and J161030, detected with ALMA, were not covered by the AzTEC map. J153702 and 160826.8-384101 were noisy maps and discarded.

The masses of the AzTEC cores with no ALMA continuum detections are also given in Table \ref{table:estabilidad_AzTEC}.  

\begin{table*}
\caption{AzTEC: Densities (critical and observed) and radii (critical and observed) at a temperature of 9K for the ALMA non-detections}     
\label{table:estabilidad_AzTEC}      
\centering                          
\begin{tabular}{c c c r c c c}        
\hline\hline                 
Name & Mass & $\mathrm{n_{crit}}$ & $\mathrm{n_{obs}}$ & $\mathrm{R_{max}}$ & $\mathrm{R_{obs}}$ & Dynamical state \\    
& [$\mathrm{M_{\odot}}$] & [$\mathrm{cm^{-3}}$] &  [$\mathrm{cm^{-3}}$] & [AU] & [AU] \\
\hline
\hline  \\                
AzTEC-lup1-99 & 0.055 & 3.6$\times$10$^{6}$ & 2.4$\times$10$^{4}$ & 760 & 4000 & stable\\
AzTEC-lup1-103 & 0.07 & 2.2$\times$10$^{6}$ & 1.7$\times$10$^{4}$ & 960 & 4930 & stable\\
AzTEC-lup1-109 & 0.085 & 1.5$\times$10$^{6}$ & 1.5$\times$10$^{4}$ & 1170 & 5460 & stable \\
AzTEC-lup1-111 & 0.068 & 2.4$\times$10$^{6}$ & 1.5$\times$10$^{4}$ & 930 & 4980 & stable \\
AzTEC-lup1-57 & 0.085 & 1.5$\times$10$^{6}$ & 1.9$\times$10$^{4}$ & 1170 & 5060 & stable \\
AzTEC-lup1-67 & 0.11 & 9.2$\times$10$^{5}$ & 1.4$\times$10$^{4}$ & 1490 & 6100 & stable \\
AzTEC-lup1-114 & 0.10 & 1.1$\times$10$^{6}$ & 1.1$\times$10$^{4}$ & 1390 & 6450 & stable \\
AzTEC-lup1-84 & 0.048 & 4.7$\times$10$^{6}$ & 3.2$\times$10$^{4}$ & 660 & 3510 & stable\\
AzTEC-lup1-104 & 0.10 & 1.0$\times$10$^{6}$ & 1.3$\times$10$^{4}$ & 1430 & 6090 & stable \\
AzTEC-lup1-101 & 0.11 & 9.0$\times$10$^{5}$ & 1.2$\times$10$^{4}$ & 1510 & 6390 & stable\\
AzTEC-lup1-119 & 0.062 & 2.9$\times$10$^{6}$ & 1.9$\times$10$^{4}$ & 850 & 4510 & stable \\
AzTEC-lup1-124 & 0.11 & 9.7$\times$10$^{5}$ & 1.1$\times$10$^{4}$ & 1460 & 6580 & stable \\
AzTEC-lup1-52 & 0.15 & 5.2$\times$10$^{5}$ & 1.4$\times$10$^{4}$ & 1990 & 6580 & stable \\
AzTEC-lup1-54 & 0.28 & 1.4$\times$10$^{5}$ & 9.6$\times$10$^{3}$ & 3780 & 9320 & stable \\
AzTEC-lup1-94 & 0.12 & 7.6$\times$10$^{5}$ & 1.1$\times$10$^{4}$ & 1640 & 6680 & stable \\
AzTEC-lup1-123 & 0.048 & 4.7$\times$10$^{6}$ & 2.3$\times$10$^{4}$ & 660 & 3890 & stable \\
AzTEC-lup3-15 & 0.12 & 7.6$\times$10$^{5}$ & 1.1$\times$10$^{4}$ & 1640 & 6720 & stable \\
AzTEC-lup3-20 & 0.077 & 1.8$\times$10$^{6}$ & 1.3$\times$10$^{4}$ & 1060 & 5540 & stable \\
AzTEC-lup3-12  & 0.28 & 1.0$\times$10$^{5}$ & 1.3$\times$10$^{4}$ & 4260 & 8390 & stable\\
AzTEC-lup3-10  & 0.034 & 6.7$\times$10$^{6}$ & 4.9$\times$10$^{4}$ & 520 & 2710 & stable \\
AzTEC-lup3-5   & 1.3 & 4.6$\times$10$^{3}$ & 3.2$\times$10$^{4}$ & 20160 & 10480 & unstable\\
AzTEC-lup3-14 & 0.13 & 6.1$\times$10$^{5}$ & 1.1$\times$10$^{4}$ & 1830 & 6950 & stable \\
AzTEC-lup3-19  & 0.037 & 5.7$\times$10$^{6}$ & 5.3$\times$10$^{4}$ & 570 & 2710 & stable\\
AzTEC-lup3-8  & 1.0 & 7.7$\times$10$^{3}$ & 4.3$\times$10$^{4}$ & 15510 & 8710 & unstable \\
AzTEC-lup3-4   & 1.0 & 7.7$\times$10$^{3}$ & 3.0$\times$10$^{4}$ & 15460 & 9790 & unstable\\
AzTEC-lup3-13  & 0.54 & 2.8$\times$10$^{4}$ & 1.7$\times$10$^{4}$ & 8200 & 9680 & stable\\
AzTEC-lup3-9  & 0.22 & 1.7$\times$10$^{5}$ & 2.0$\times$10$^{4}$ & 3300 & 6790 & stable\\
AzTEC-lup3-16 & 0.079 & 1.8$\times$10$^{6}$ & 1.5$\times$10$^{4}$ & 1080 & 5350 & stable \\

\hline                                   
\end{tabular}
\end{table*}
 
\label{sec:aztec continuum}

\subsection{ALMA sources without optical/IR counterpart}
\label{no_counterparts}
The following 12 sources detected in our ALMA survey have no optical or infrared counterpart: J153702, J154228, J154456, J154458, J154506, J154634, J160658, J160804, J1609200, J1609201, J160932, and J161030. Six of these sources (J153702, J160658, J160804, J1609200, J1609201, and J161030) were serendipitously detected inside the ALMA primary beam at a distance greater than 3$\arcsec$ from the nominal position of the spectroscopically confirmed Class II sources originally targeted. The remaining six sources (J154228, J154456, J154458, J154506, J154634, and J160932) are detected inside the ALMA primary beam of the observed pre-BD candidates from our AzTEC catalog.

J153702 is located at a distance of 6.9$\arcsec$ from the Class II object 153703.1-331927 (ALMA phase centre). J160658 is located at a distance of 3.1$\arcsec$ from the Class II object 160658.7-390405, that is located at 157 $\pm$ 3 pc according to the Gaia detection. J160804 is at a distance of 8.0$\arcsec$ from the Class II source 160804.8-390449, the source that was originally targeted in our survey. J161030 is at a distance of 5.0$\arcsec$ from 161030.6-383151, the source originally targeted by ALMA. J1609200 and J1609201 are two sources detected in the same ALMA field of view (see Appendix \ref{dobles}), located at 10.5$\arcsec$ and 15.7$\arcsec$ distance, respectively from the phase centre, which was the position of the Class II source 160920.8-384510. 
There is an optical/infrared counterpart close to J1609201 at 1.7$\arcsec$, including Gaia measurements, but the high precision astrometry of Gaia as well as the accurate absolute astrometry for this source reveals that the counterpart are not associated with J1609201. The 2MASS counterpart in Gaia is 16092031-3844568 \citep{Cutri2003}.

The remaining sources are seen inside the ALMA primary beam centred at the AzTEC pre-BD candidates of our sample. J154228 is located 9.7$\arcsec$ from the ALMA phase centre. The NASA/IPAC Extragalactic Database (NED) catalogue lists the closest object to the source as extragalactic and located at a distance of 3.7$\arcsec$, which makes a physical association unlikely given the absolute position accuracy of ALMA. J154456 and J154458 fall inside the same primary beam. The separation between these two sources is 19.6$\arcsec$, and they are located at 6.2$\arcsec$ and 13.4$\arcsec$, respectively, from the phase centre. The closest optical object found is SSTSL2 J154456.77-342532.3 at 3.1$\arcsec$ and 16.4$\arcsec$ distance from J154456 and J154458, respectively. The ALMA detection of J154456 is spatially resolved with a Gaussian deconvolved size of 0.84$\arcsec$ $\pm$ 0.06$\arcsec$ $\times$ 0.28 $\pm$ 0.08$\arcsec$. J154506, J160932, and J154634 are located at distances of 12.0$\arcsec$, 11.9$\arcsec$, and 14.8$\arcsec$, respectively, from their phase centres.  

At the distance of Lupus, we found that the ALMA compact continuum emission mentioned above implies substellar masses below 10 M$\mathrm{_{Jup}}$, which are well inside the substellar regime independent of the values of temperature or opacity coefficient that we adopted. We checked for possible extragalactic contaminants in the NED and we found no clear detections associated with any of the detected objects. All these sources except J153702 and J161030, are located well within or quite close to the Lupus dust filaments as seen in the Herschel and AzTEC maps (see Figure \ref{aztec_detections} and \ref{Herschel}). Therefore, they are probably associated with the Lupus molecular clouds and we classify them as pre-BD candidates or deeply embedded proto-BD candidates. Future gas observations with better sensitivity at the positions of these candidates should help to confirm or reject their Galactic nature and the membership of each source to the complex.


\subsection{ALMA sources with optical/infrared counterpart}
\label{ALMA sources with counterparts}
The sources described in this subsection have optical or infrared counterparts, or a combination of both. Using the classification tools described above (counterpart presence, $\alpha \mathrm{_{IR}}$ slope, and T$\mathrm{_{bol}}$), we make an attempt to classify their state of evolution, considering the uncertainty in their physical association to the Lupus clouds, and the incomplete SEDs for most of them. We identify one new Class 0/I proto-BD candidate, one Class I proto-BD candidate (we note that these sources may instead be pre-BDs if their optical counterpart are not in fact associated with them), and confirm five Class II substellar objects previously known. 

\subsubsection{Class 0/I}
J154229 (Figure \ref{panel_alma}) is detected and spatially resolved in this work for the first time. We report a 26 $\sigma$ detection and a deconvolved size of $0.77\arcsec \times 0.64\arcsec$, with a mass of 3.5 $\pm$ 1.7 $\mathrm{M_{Jup}}$. We found a counterpart at several wavelengths at a distance between 0.1$\arcsec$ and 0.4$\arcsec$ (WFI, WISE, Spitzer/IRAC, Spitzer/MIPS). The counterpart in other wavebands are positionally consistent with a single source. 
The J154229 $\alpha \mathrm{_{IR}}$ slope is compatible with a Class I source, but the bolometric temperature indicates that it is most likely a Class 0 object that is very close to the boundary with Class I. Therefore we classify it as a proto-BD candidate in a transition stage between Class 0 to Class I (see Table \ref{Tab:tbol_lbol}). The SED of J154229 is similar to that of the Taurus proto-BD Class 0/I candidate J041757-B found by \citet{Barrado2009}. The pre-BD candidate J154228 is located inside the same primary beam at a distance of 17.9$\arcsec$. 

\subsubsection{Class I}
J153914 has not been previously reported in the literature. We detect an ALMA continuum point source corresponding to a mass of 0.63 M$\mathrm{_{Jup}}$. This source has an optical counterpart 0.7$\arcsec$ distant from the ALMA detection.The source is detected at three different optical bands showing a steep positive slope. There are no infrared counterpart associated with this source, which complicates the confirmation of the physical association to the WFI source with the ALMA source. If confirmed, this source would be a Class I proto-BD candidate or a background object. On the contrary, if there is not such an association then it would be classified as a pre-BD or deeply embedded proto-BD candidate. Future infrared observations are needed to clarify the exact nature of this source.

\subsubsection{Class I/II and Class II} 
160826.8-384101 is a spectroscopically confirmed Class II source in Lupus 3 \citep{comeronetal09-2}. The central object has $\mathrm{T_{eff}}$=2900 K and a mass of 0.06 $\mathrm{M_{\odot}}$  \citep{comeronetal09-2}. 
ALMA continuum emission at 1.3 mm is detected at 26 $\sigma$ (Appendix \ref{apendix_panel_alma}), and it is not spatially resolved. The disk mass is 1.1 $\pm$ 0.4 $\mathrm{M_{Jup}}$ at a distance of 165 $\pm$ 4 pc. The bolometric temperature is $\sim$1411 K, in the Class II range. 

\label{vscop}
V1094 Sco is a Class II YSO in the stellar mass regime \citep{frascaetal17-1} discovered by \citet{krautteretal97-1}. Our high signal-to-noise ratio image spatially resolves the dust disk. We estimated a disk mass of 124 $\pm$ 35 $\mathrm{M_{Jup}}$ and the disk temperature is 22.5 K, at a distance of 154.7 $\pm$ 1.1 pc. The object is extensively discussed in several papers \citep{Baraffe15-1, Alcala17-1, frascaetal17-1}. The central source has $\mathrm{T_{eff}}$= 4205 $\pm$ 193 K with a mass of 1.10  $\mathrm{M_{\odot}}$ and it is classified as K6 \citep{frascaetal17-1}. 
The SED is very complete including detections from the optical to the infrared at different separations (WFI, DENIS, 2MASS, WISE, Spitzer-MIPS, Spitzer-IRAC, Spitzer-MIPS, SPIRE, PACS, LABOCA, and Akari). 

Lup 706 is a spectroscopically confirmed Class II substellar object, previously discovered and classified as a BD \citep{lopezetal05-1}. We report a non-spatially resolved detection (Appendix \ref{apendix_panel_alma}). The mass of the disk is 0.20 $\pm$ 0.12 $\mathrm{M_{Jup}}$ at a distance of 191 $\pm$ 29 pc. The nominal distance is on the high side but the large error bar of the distance does not rule out cloud membership. The projected separation between Lup 706 and V1094 Sco is 16$\arcsec$.

The central object has $\mathrm{T_{eff}}$ = 2750 K and a mass of 0.06$^{+0.03}_{-0.02}M_{\odot}$  and is classified as an M7.5 BD \citep{alcalaetal14-1, kora14-1}. Lup 706 has been classified as a Class II object \citep{alcalaetal14-1,Alcala17-1}, the same classification that we obtained from the SED. Although the bolometric temperature is $\sim$415\,K, below the Class II source limit, it is possible that, as we see in Par-Lup 3-4, the inclination of the surrounding disk is responsible for this low temperature estimate. Both sources, Par-Lup 3-4 and Lup 706 were classified as subluminous by \citet{alcalaetal14-1}. Future SED modeling could give us more information about the inclination angle. A recent study \citep{Sanchis20-1}
reports a lower dust disk mass of Lup 706 from ALMA Band 7 data, but their results are consistent with ours within 1 $\sigma$ using the absorption coefficient in \citet{Ose94}. 

Par-Lup 3-4 (Appendix \ref{apendix_panel_alma}) was discovered by \citet{nakajimaetal00-1} and confirmed spectroscopically as a Lupus 3 member by \citet{Comeron03-1}. It is a VLM star with a mass of 0.13 $\mathrm{M_{\odot}}$, M4.5 spectral type and $\mathrm{T_{eff}}$= 3089 $\pm$ 246 K \citep{Alcala17-1,manaraetal13-1,frascaetal17-1}. We report a spatially unresolved detection. The mass of the disk is 0.19 $\pm$ 0.08 $\mathrm{M_{Jup}}$ at a distance of 155 $\pm$ 14 pc. We also detect CO(2--1) gas emission associated with this source at velocities close to the V$\mathrm{_{LSR}}$ of Lupus, which confirms its association to the molecular cloud. Par-Lup 3-4 is classified as a Class II source in the literature, with an almost edge-on disk \citep{Huelamo10-1}. Interestingly, additional ALMA data on this source in Band 7 have revealed the presence of a bipolar molecular outflow (Santarmar\'ia-Miranda et al. 2020). The bolometric temperature is $\sim$405\,K, in the Class I regime, but the source is under-luminous \citep{Comeron03-1}. 

SONYC-Lup 3-7 is a spectroscopically confirmed Class II substellar source in Lupus 3 discussed in \citet{alcalaetal14-1} and \citet{kora14-1}. Our ALMA data (Appendix \ref{apendix_panel_alma}) show an unresolved object with a disk mass of 0.18 $\pm$ 0.07 $\mathrm{M_{Jup}}$ at a distance of 151 $\pm$ 6 pc. The bolometric temperature obtained from the SED is $\sim$1959\,K.
\citet{alcalaetal14-1} and \citet{kora14-1} found that the central object has a temperature of 2600--2850 K and a mass of 0.03 $\pm$ 0.01 $\mathrm{M_{\odot}}$. We confirm the source as a Class II, based on its SED. A recent study \citep{Sanchis20-1} reported a dust disk mass of SONYC-Lup 3-7 in Band 7 and that is compatible with our results within 2 $\sigma$. 

\section{Discussion}
\label{p2_discussion}

\subsection{Spatial distribution: Detections versus non-detections}
The spatial distribution of the Lupus 1 objects detected with AzTEC is shown in Figure \ref{aztec_detections}. The AzTEC data reveal two filaments, the primary extending more than 1.5 degrees in the Northern part of the map (left panel Figure \ref{aztec_detections}), and a second smaller filament located southwest from the primary. Almost all the sources are in the primary filament, although there are several non-detections in the secondary filament. There are three AzTEC cores outside the two filaments. Of these three cases outside of the filaments, only one (J153702) is  detected with ALMA, and this source was serendipitously detected when we pointed to a spectroscopically confirmed Class II BD. 
For Lupus 3, the distribution is slightly different (right panel in Figure \ref{aztec_detections}). There is only one filament that has several sources along it, including ALMA detections without an optical/IR counterpart, such as J160804 or 160932. However, there are also some ALMA detections located at the outskirts of the filament (J1609200, J1609201 and J161030). Finally, there are nine AzTEC detections are located at a considerable distance from the centre of the filament, that are distributed across the whole map. For comparison, we checked that some proto-BD candidates in other regions are also detected outside the main filaments. For example, J041757 is at the outskirts of the B213 main filament of Taurus \citep{Palau2012-1}. We conclude that there is no specific preferred cloud environment where AzTEC sources detected with ALMA tend to cluster. 

\subsection{Detection rate of pre- and proto-BD candidates and large-scale core properties}
\label{detection_rate}	
The ALMA detection rate of the whole sample was $\sim$23\,\% (15 of the 64 pointings), or 18\,\% (5 of 40) if we reduce the sample to the initial classification of 33 pre-stellar and 7 Class 0 and I objects we surveyed. A similar work in Barnard 30 \citep{Huelamo2017} using LABOCA and ALMA obtained a detection rate of 17 $\%$. Both ALMA experiments were designed to detect faint sources even in the worst-case scenario where all the emission was extended in an area equivalent to the ALMA largest angular scale, which in our case was $\sim$11$\arcsec$.
The selection strategy (see Section \ref{sec:sample}) was built using an AzTEC beam size of $\sim$30$\arcsec$ and cores with peak intensities between 30--100 mJy. We adopted a very conservative approach whereby the emission from the dusty envelope or disk surrounding a very young brown dwarf could be as large as the ALMA largest angular size (LAS) of 11$\arcsec$ for our observations, that is, $\sim$1650 au at $\sim$150 pc, and a uniform distribution over that envelope.
Considering the ALMA synthesized beam of our observations ($\sim$0.85$\arcsec$ or $\sim$130 AU) and the fact that our data are sensitive enough to detect sources at the 5$\sigma$ level based on the average AzTEC rms, this means that non-detections are probably related to the source size and not to the sensitivity, further suggesting the AzTEC clump material is distributed mainly on large scales. This implies that we are filtering most of the extended emission in our ALMA observations.

\subsection{Evolution of the ALMA starless cores: Final mass of the pre-BD candidates}
\begin{figure}
\centering
\includegraphics[width=0.5\textwidth]{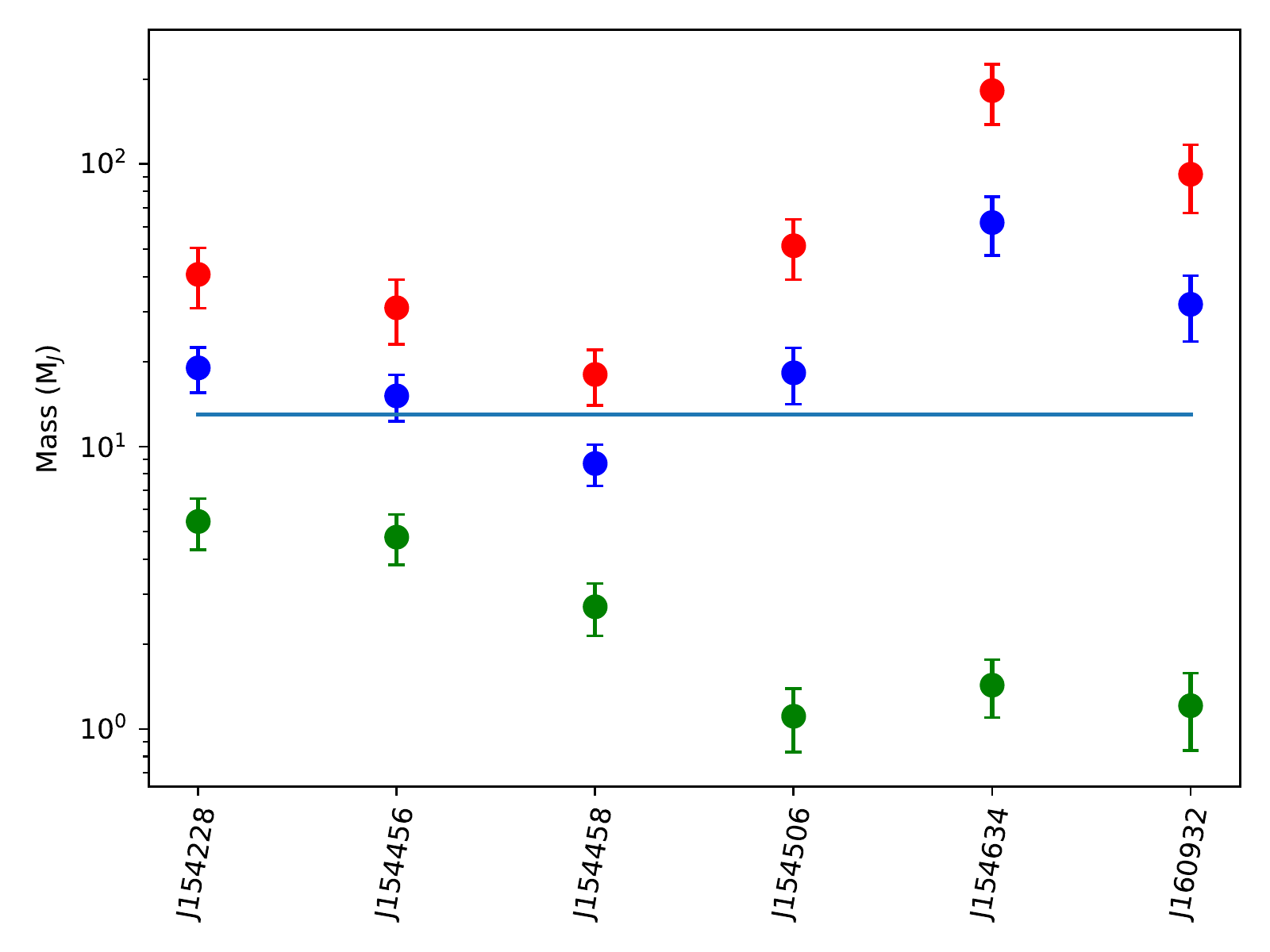}
      \caption[Mass of the ALMA sample pre-BDs candidates in the Lupus sample]{Mass of the ALMA pre-BDs candidates  without an optical/infrared counterpart at a temperature of 9K. Red points are the AzTEC masses. Green points are the ALMA masses. Blue points are the assumed final masses as a combination of the ALMA mass and the AzTEC mass assuming a core SFE of 30\,\%. Horizontal line marks the deuterium-burning limit.}    
\label{final_mass}
\end{figure}

For the ALMA pre-BD candidates, we estimated the final masses of the central compact object by assuming a core formation efficiency in the substellar regime of 30\,\% (and a temperature of 9 K), which is similar to that inferred for low-mass cores \citep{motteetal98-1}. Therefore, for the pre-BD candidates detected with ALMA, we added 30\,\% of the AzTEC clump mass to the ALMA mass. As a result we find that all of these pre-BD candidates have the capacity to evolve to form substellar objects with final masses between 9 to 60 M$\mathrm{_{Jup}}$ (see Figure \ref{final_mass}). We conclude that the six ALMA compact sources detected inside the six AzTEC clumps, and without counterpart at shorter wavelengths, may end up as substellar objects.  

\label{Fragmentation}

\subsection{Nature of ALMA detections without optical/infrared counterparts}
In this section, we discuss the nature of the ALMA detections without counterparts based on the assumption that they are starless cores (pre-BD). We note, however, that we cannot exclude the possibility that they are embedded proto-BDs without any counterparts detected in the infrared images.

\subsubsection{Infall according to turbulent fragmentation}

The theory of turbulent fragmentation \citep{Padoan2004} is based on collapse due to externally driven supersonic turbulence. The critical mass for the collapse of a Bonnor-Ebert sphere \citep{Bonnor1956} is defined as: \\
\begin{equation}
\mathrm{M_{BE}[M_{\odot}] =3.3 \bigg(\frac{T}{10[K]}\bigg)^{3/2}\bigg(\frac{n_{crit}}{10^{3}[cm^{-3}]}\bigg)^{-1/2}},
\end{equation}
where T is the temperature in K and n$\mathrm{_{crit}}$ is the critical density in cm$^{-3}$. For the following calculations, we assume the Bonnor-Ebert mass is the one from AzTEC cores or ALMA compact detections. The uncertainty in the Bonnor-Ebert sphere mass is estimated using the mass uncertainty in Table \ref{Tab:masas_medidas} and the temperature error with a fixed value of 1 K, as done in Section \ref{alma continuum masses}. The critical radius (R$_{\mathrm{crit}}$) can be simplified to
\begin{equation}
\mathrm{R_{crit}[AU]= \bigg(\frac{2.83\times10^{16}M[M_{\odot}]}{{n_{crit}}[cm^{-3}]}\bigg)^{1/3}},
\end{equation}
where M is the mass of the source.  

The observed density ($\mathrm{n_{obs}}$) is defined as the ratio of the hydrogen column density to the linear size of the source, and it is calculated using the mass of the source estimated from the observations and the size of the emitting region. For those ALMA sources that are not spatially resolved we use the synthesized beam as an upper limit to their size. 

According to this theory, cores with an observed density that is lower than the critical value ($\mathrm{n_{obs}}$ < $\mathrm{n_{crit}}$) are expected to be transient cores. On the contrary, if the density of the core is greater than the critical density ($\mathrm{n_{obs}}$ > $\mathrm{n_{crit}}$), the core should be gravitationally unstable and is expected to be in the pre-stellar phase. 
A comparison of the average radius of a source ($\mathrm{R_{obs}}$) to the critical radius ($\mathrm{R_{crit}}$) corresponding to the critical density is also used to infer the energetic state of sources. Sources that are spatially unresolved, with $\mathrm{R_{obs}}$ smaller than $\mathrm{R_{crit}}$ indicate that the core is unstable. For resolved sources, $\mathrm{R_{crit}}$ sets a boundary between stable cores ($\mathrm{R_{obs}}$ > $\mathrm{R_{crit}}$) and unstable ($\mathrm{R_{obs}}$\,<\,$\mathrm{R_{crit}}$) cores.

We calculate the critical density and the observed density for the ALMA pre-BD candidates using two values of temperature: T=9 K, which is more characteristic of pre-BD cores (Table \ref{table:estabilidad_ALMA_10k}); and T=15 K for deeply embedded protostars (Table \ref{table:estabilidad_ALMA}), as we did in Section \ref{alma continuum masses}. We also computed these parameters for the AzTEC cores with ALMA detections (see Tables \ref{table:estabilidad_AzTEC_10K} and \ref{table:estabilidad_AzTEC_15K}), and for the AzTEC clumps without ALMA detections Table \ref{table:estabilidad_AzTEC}.   

\begin{table*}
\caption{ALMA: Densities (critical and observed) and radii (critical and observed) at a temperature of 9 K}             
\label{table:estabilidad_ALMA_10k}      
\centering                          
\begin{tabular}{c c r c r c}        
\hline\hline                 
Name & $\mathrm{n_{crit}}$ & $\mathrm{n_{obs}}$ & $\mathrm{R_{max}}$ & $\mathrm{R_{obs}}$ & Dynamical state \footnote[1]{} \\    
& [$\mathrm{cm^{-3}}$]\,$\pm$\,\% &  [$\mathrm{cm^{-3}}$]\,$\pm$\,\% & [AU] & [AU] \\
\hline
\hline  \\                      
J153702 & 1.2$\times$10$^{10}$ $\pm$    66 & > 6.3$\times$10$^{7}$ &     13 $\pm$      3 & <    72 & stable?\\
J154228 & 3.0$\times$10$^{8}$ $\pm$    53 & > 1.2$\times$10$^{9}$ &     79 $\pm$     15 & <    49 & unstable\\
J154456 & 3.8$\times$10$^{8}$ $\pm$    52 & 1.7$\times$10$^{9}$ $\pm$  78 &     70 $\pm$     13 &     43 & unstable \\
J154458 & 1.2$\times$10$^{9}$ $\pm$    54 & > 2.5$\times$10$^{8}$ &     40 $\pm$      8 & <    67 & stable?\\
J154506 & 7.1$\times$10$^{9}$ $\pm$    61 & > 9.6$\times$10$^{7}$ &     16 $\pm$      4 & <    68 & stable?\\
J154634 & 1.0$\times$10$^{10}$ $\pm$    62 & > 7.8$\times$10$^{7}$ &     13 $\pm$      3 & <    68 & stable?\\
J160658 & 4.1$\times$10$^{9}$ $\pm$    59 & > 1.2$\times$10$^{8}$ &     21 $\pm$      4 & <    69 & stable?\\
J160804 & 3.0$\times$10$^{9}$ $\pm$    60 & > 1.4$\times$10$^{8}$ &     25 $\pm$      5 & <    69 & stable?\\
J1609200 & 2.8$\times$10$^{9}$ $\pm$    59 & > 1.1$\times$10$^{8}$ &     26 $\pm$      5 & <    75 & stable? \\
J1609201 & 2.8$\times$10$^{9}$ $\pm$    68 & > 1.5$\times$10$^{8}$ &     26 $\pm$      6 & <    69 & stable? \\
J160932 & 1.1$\times$10$^{10}$ $\pm$    80 & > 7.4$\times$10$^{7}$ &     13 $\pm$      4 & <    69 & stable? \\
J161030 & 4.7$\times$10$^{9}$ $\pm$    59 & > 1.1$\times$10$^{8}$ &     20 $\pm$      4 & <    69 & stable?\\
\hline                                   
\end{tabular}

\begin{tablenotes}
\item[1]$^{1}$ Stable vs. infalling described in Section \ref{Fragmentation}, the question mark indicates upper limits.
\end{tablenotes}

\end{table*} 
\begin{table*}
\caption{ALMA: Densities (critical and observed) and radii (critical and observed) at a temperature of 15 K}             
\label{table:estabilidad_ALMA}      
\centering                          
\begin{tabular}{c c r c r c}        
\hline\hline                 
Name & $\mathrm{n_{crit}}$ & $\mathrm{n_{obs}}$ & $\mathrm{R_{max}}$ & $\mathrm{R_{obs}}$ & Dynamical state \\    
& [$\mathrm{cm^{-3}}$]\,$\pm$\,\%&  [$\mathrm{cm^{-3}}$]\,$\pm$\,\% & [AU] & [AU] \\
\hline
\hline  \\                      
J153702 & 2.6$\times$10$^{11}$ $\pm$    50 & > 2.9$\times$10$^{7}$ &      3 $\pm$      1 & <    72 & stable?\\
J154228 & 6.6$\times$10$^{9}$ $\pm$    32 & > 5.5$\times$10$^{8}$  &     22 $\pm$      2 &    < 49 & stable?\\
J154456 & 8.6$\times$10$^{9}$ $\pm$    30 & 7.6$\times$10$^{8}$ $\pm$  69 &     19 $\pm$      2 &     43 & stable? \\
J154458 & 2.7$\times$10$^{10}$ $\pm$    33 & > 1.1$\times$10$^{8}$ &     11 $\pm$      1 & <    67 & stable?\\
J154506 & 1.6$\times$10$^{11}$ $\pm$    43 & > 4.3$\times$10$^{7}$ &      4 $\pm$      1 & <    68 & stable?\\
J154634 & 2.3$\times$10$^{11}$ $\pm$    45 & > 3.5$\times$10$^{7}$ &      4 $\pm$      1 & <    68 & stable?\\
J160658 & 9.2$\times$10$^{10}$ $\pm$    40 & > 5.5$\times$10$^{7}$ &      6 $\pm$      1 & <    69 &stable? \\
J160804 & 6.8$\times$10$^{10}$ $\pm$    42 & > 6.4$\times$10$^{7}$ &      7 $\pm$      1 & <    69 & stable?\\
J1609200 & 6.4$\times$10$^{10}$ $\pm$    41 & > 5.1$\times$10$^{7}$ &      7 $\pm$      1 & <    75 & stable? \\
J1609201 & 6.2$\times$10$^{10}$ $\pm$    53 & > 6.6$\times$10$^{7}$ &      7 $\pm$      1 & <    69 & stable?\\
J160932 & 2.4$\times$10$^{11}$ $\pm$    68 & > 3.4$\times$10$^{7}$ &      4 $\pm$      1 & <    69 & stable? \\
J161030 & 1.1$\times$10$^{11}$ $\pm$    41 & > 5.1$\times$10$^{7}$ &      5 $\pm$      1 & <    69 & stable? \\
 
\hline                                   
\end{tabular}
\end{table*}  
\begin{table*}
\caption{AzTEC: Densities (critical and observed) and radii (critical and observed) at a temperature of 9 K}     
\label{table:estabilidad_AzTEC_10K}      
\centering                          
\begin{tabular}{c c r c r c}        
\hline\hline                 
Name & $\mathrm{n_{crit}}$ & $\mathrm{n_{obs}}$ & $\mathrm{R_{max}}$ & $\mathrm{R_{obs}}$ & Dynamical state  \\    
& [$\mathrm{cm^{-3}}$]\,$\pm$\,\% &  [$\mathrm{cm^{-3}}$]\,$\pm$\,\% & [AU] & [AU] \\
\hline
\hline  \\                
J154228 & 5.3$\times$10$^{6}$ $\pm$    59 & > 5.7$\times$10$^{4}$ &    590 $\pm$    130 & <  2680 &  stable\\
J154456 & 9.0$\times$10$^{6}$ $\pm$    59 & 1.5$\times$10$^{4}$ $\pm$  30 &    450 $\pm$     100 &   3800 & stable \\
J154558 & 2.6$\times$10$^{7}$ $\pm$    59 & 8.9$\times$10$^{3}$ $\pm$  30 &    270 $\pm$     60 &   3800 & stable \\
J154506 & 3.3$\times$10$^{6}$ $\pm$    59 & 3.6$\times$10$^{4}$ $\pm$  30 &    750 $\pm$    160 &   3380 & stable\\
J154634 & 2.6$\times$10$^{5}$ $\pm$    59 & 1.1$\times$10$^{4}$ $\pm$  30 &   2650 $\pm$    560 &   7580 & stable\\
J160932 & 9.3$\times$10$^{5}$ $\pm$    62 & 2.1$\times$10$^{4}$ $\pm$  39 &   1410 $\pm$    320 &   5000 & stable\\
\hline                                   
\end{tabular}
\end{table*}  

\begin{table*}
\caption{AzTEC: Densities (critical and observed) and radii (critical and observed) at a temperature of 15 K}     
\label{table:estabilidad_AzTEC_15K}      
\centering                          
\begin{tabular}{c c r c r c}        
\hline\hline                 
Name & $\mathrm{n_{crit}}$ & $\mathrm{n_{obs}}$ & $\mathrm{R_{max}}$ & $\mathrm{R_{obs}}$ & Dynamical state  \\    
& [$\mathrm{cm^{-3}}$]\,$\pm$\,\% &  [$\mathrm{cm^{-3}}$]\,$\pm$\,\% & [AU] & [AU] \\
\hline
\hline  \\ 
J154228 & 1.4$\times$10$^{8}$ $\pm$    37 & > 2.4$\times$10$^{4}$ &    150 $\pm$     20 & <  2680 & stable?\\
J154456 & 2.3$\times$10$^{8}$ $\pm$    37 & 6.5$\times$10$^{3}$ $\pm$  22 &    120 $\pm$     15 &   3800 & stable\\
J154558 & 6.8$\times$10$^{8}$ $\pm$    37 & 3.8$\times$10$^{3}$ $\pm$  22 &     70 $\pm$      10 &   3800 & stable\\
J154506 & 8.5$\times$10$^{7}$ $\pm$    37 & 1.5$\times$10$^{4}$ $\pm$  22 &    190 $\pm$     30 &   3380 & stable\\
J154634 & 6.8$\times$10$^{6}$ $\pm$    37 & 4.8$\times$10$^{3}$ $\pm$  22 &    680 $\pm$     90 &   7580 & stable\\
J160932 & 2.4$\times$10$^{7}$ $\pm$    43 & 8.9$\times$10$^{3}$ $\pm$  32 &    360 $\pm$     60 &   5000 & stable\\
\hline                                   
\end{tabular}
\end{table*}  

An unresolved ALMA sources is presumed to be collapsing if its radius is smaller than the value given in Table \ref{table:estabilidad_ALMA_10k}. For the resolved source J154456, we obtained density and size values that suggests it is collapsing. 

For the spatially resolved larger AzTEC cores, we find observed densities that are one to two orders of magnitude smaller than the critical densities, regardless of the adopted temperature (9 or 15 K). This remains true even if we include the 1$\sigma$ uncertainties. Their radii are also larger than the $\mathrm{R_{crit}}$. All this suggests they are transient cores. The unresolved AzTEC core hosting J154228 could be unstable if its radius is smaller than $\sim$530 au.

We carried out the calculation of the energetic states of the rest of the AzTEC cores where no ALMA detections were obtained and found that all the cores in the substellar regime seem to be stable.  

\begin{figure}
\includegraphics[width=0.5\textwidth]{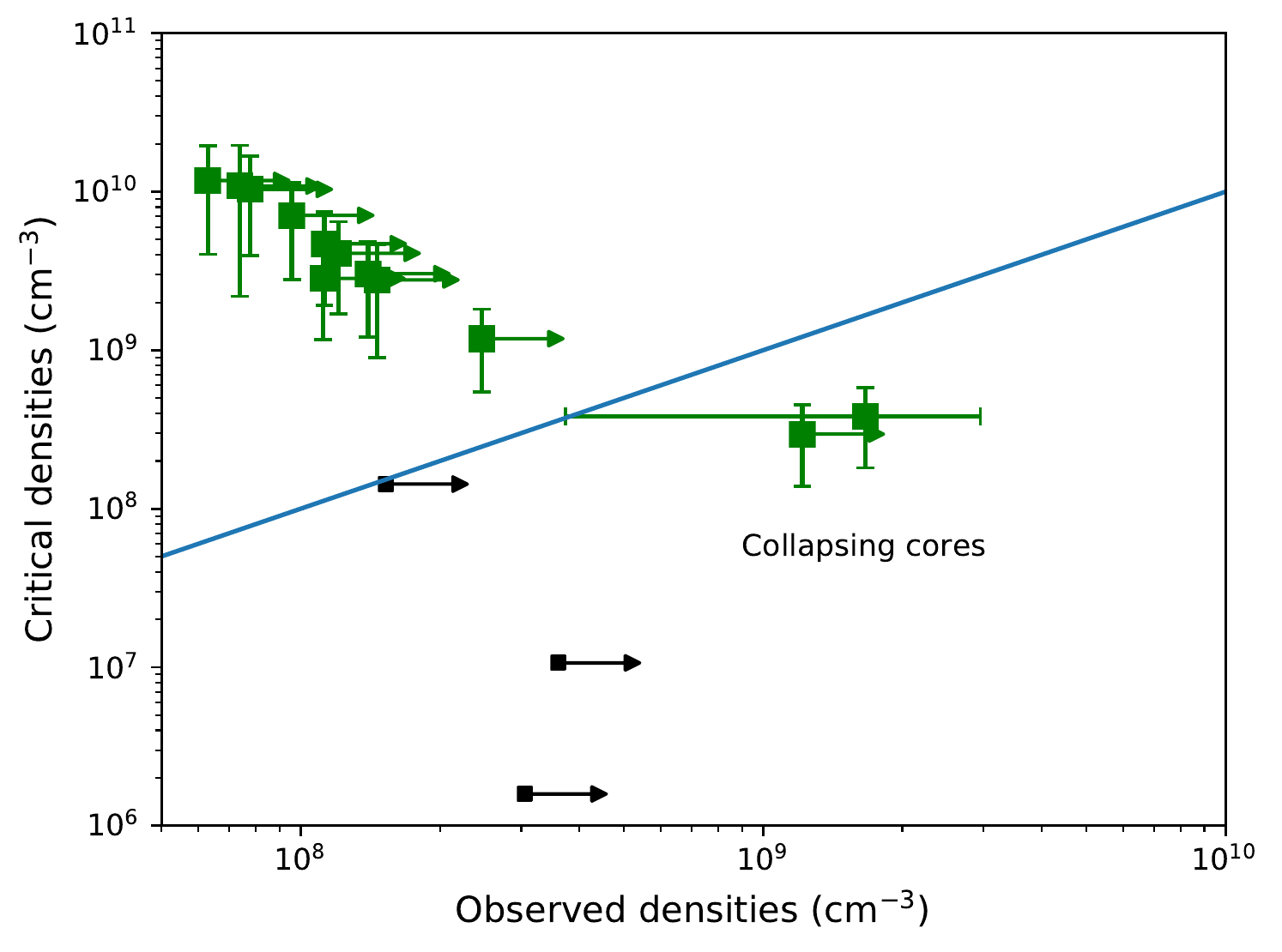}
      \caption{Critical densities  vs observed densities  at an adopted temperature T = 9 K using ALMA. Green symbols shows the sources studied in this work and the square is J154456. Black symbols shows the sources in \citet{Huelamo2017}. Lower limits are included as arrows. Below the blue line the cores should be unstable, and therefore collapsing.}
\label{nurvsal}
\end{figure}

In Figure \ref{nurvsal}, we compare the derived $\mathrm{n_{crit}}$ and $\mathrm{n_{obs}}$ for the ALMA sources, assuming a temperature of 9 K. As seen, two of our pre-BD candidates are in the infalling regime. The rest of the cores are not resolved with ALMA and therefore we cannot exclude that they also may be unstable. We compared our results with those from  \citet{Huelamo2017} by computing the source masses in their work using a temperature of 9 K instead of the 15 K assumed in the original paper. With a temperature of 15 K and the uncertainties related to the mass (of a factor of 4) it is not entirely clear whether the cores in Barnard 30 are in collapse. However, using a temperature of 9 K, the pre-BD core candidates in Barnard 30 clearly lie in the unstable regime. 
On the other hand, our Cycle 3 observations were more sensitive than the Cycle 1 observations in \citet{Huelamo2017}, which might explain our ability to make detections in the apparently non-collapsing regime.

\subsubsection{r$^{-2}$ density profile}
\label{profile}
The analysis in the previous section shows that most of the AzTEC pre-BD cores with ALMA detections are stable. However, this result is in conflict with the fact that we already detect very compact sources at these core centers, which indicates ongoing collapse. \citet{Huelamo2017} proposed that this kind of configuration in pre-BD objects (stable large-scale cores with a compact source inside) could be the “tip of the iceberg”  of a larger-scale  collapse. Thus, ALMA compact sources would have been formed as the product of gravitational contraction, where an r$^{-2}$ density profile is naturally developed with a finite infall velocity in a gravitationally unstable background \citep{Naranjo-Romeroetal15-1,gomezetal07-1, mohammadpourtandstahler13-1,larson69-1}. This is naturally expected in the scenario of global hierarchical collapse \citep{Vazquez19}.

Using the r$^{-2}$ density profile and the flux detected with AzTEC we estimate the expected mass inside the average ALMA beam having a FWHM of 0.91$\arcsec$ ($\sim$139\,au). Results can be seen in Table \ref{table:r2densityprofile}, where we give the ratio between the estimated and the observed masses. There are three sources that lie on the 1:1 relation. Hence, it seems that this set of pre-BD candidates could be the product of gravitational collapse. Three of the sources are not close to a ratio of 1 and two of them are the same sources that seem to be unstable when applying the turbulent fragmentation theory. The difference between the sources with a ratio close to 1 and the other three might be related to their evolutionary stage: ratios between 0.14 and 0.07 would suggest density profiles steeper than r$^{-2}$ and this would indicate that they have already accreted more mass onto the central object, which is consistent with their being more evolved. The other three sources show a Bonnor-Ebert isothermal profile, therefore indicating that they are less evolved.

\begin{table*}
\caption{Estimated masses of the ALMA pre-BD core candidates within an ALMA beam, assuming a r$^{-2}$ density profile from AzTEC}     
\small
\label{table:r2densityprofile}      
\centering                          
\begin{tabular}{c c r c c c}        
\hline\hline                 
	& \multicolumn{2}{c}{AzTEC (9 K)} & \multicolumn{2}{c}{ALMA\,Mass(9 K)} \\
\cmidrule(lr){2-3}
\cmidrule(lr){4-6}	
	
Name & Mass & Radii & Estimated & Observed & Ratio \footnote[1]{} \\    
& [$\mathrm{M_{Jup}}$] &  [AU] & [$\mathrm{M_{Jup}}$] & [$\mathrm{M_{Jup}}$] \\
\hline
\hline  \\                
ALMA J154228.675-334230.18 & 41 $\pm$ 10 & <2680 & 0.75 $\pm$ 0.18 & 5.4 $\pm$ 1.1 & 0.14 \\
ALMA J154456.522-342532.99 & 31 $\pm$ 8 & 3800 & 0.35 $\pm$ 0.09  & 4.8 $\pm$ 1.0 & 0.07\\
ALMA J154458.061-342528.51 & 18 $\pm$ 4 & 3800 & 0.20 $\pm$ 0.05 & 2.7 $\pm$ 0.6 & 0.07 \\
ALMA J154506.515-344326.15 & 51 $\pm$ 12 & 3380 & 1.0 $\pm$ 0.3 & 1.1 $\pm$ 0.3 & 0.94\\
ALMA J154634.169-343301.90 & 180 $\pm$ 40 & 7580 & 1.6 $\pm$ 0.4 & 1.4 $\pm$ 0.3 & 1.15\\
ALMA J160932.167-390832.27 & 100 $\pm$ 30 & 5000 & 1.5 $\pm$ 0.40 & 1.2 $\pm$ 0.4 & 1.21\\
\hline                                   
\end{tabular}

\begin{tablenotes}
\item[1]$^{1}$ Estimated ALMA mass/ Observed ALMA mass
\end{tablenotes}
\end{table*}  

\begin{table}
\caption{Predicted pre-BD cores radii containing the mass measured with ALMA and assuming a r$^{-2}$ density profile, versus the observed radii with ALMA}     
\small
\label{table:r2radiusprofile}      
\centering                          
\begin{tabular}{c c c}        
\hline\hline                 
Name & Estimated & Observed \\    
& [AU] & [AU] \\
\hline
\hline  \\                
ALMA J154228.675-334230.18  & 48  & < 49  \\
ALMA J154456.522-342532.99  &  42   & 43    \\
ALMA J154458.061-342528.51  &  24   &  < 67  \\
ALMA J154506.515-344326.15  &  10  &  < 68  \\
ALMA J154634.169-343301.90  &  12  & < 69  \\
ALMA J160932.167-390832.27  &  11    & < 69  \\
\hline                                   
\end{tabular}
\end{table}  

The assumed r$^{-2}$ density profile predicts sizes smaller than the ALMA synthesized beam, with the exception of J154456 (the only pre-BD spatially resolved in the sample). This is in agreement with the unavailability of resolving spatially most of these compact structures with this interferometer configuration (see Table \ref{table:r2radiusprofile})

\begin{figure}
\centering
\includegraphics[width=0.50\textwidth]{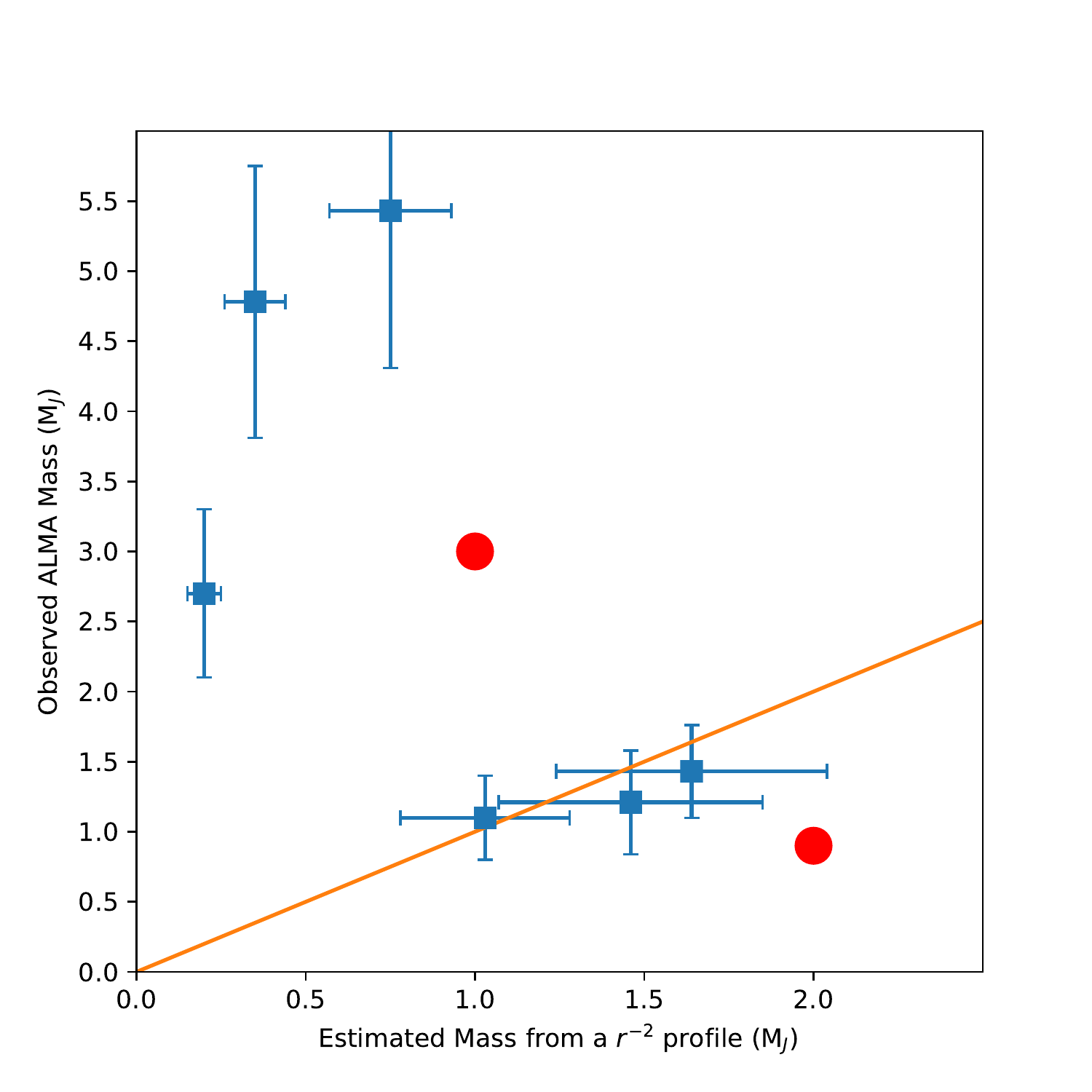}
      \caption{Observed mass for the pre-BD vs. the estimated mass using a  r$^{-2}$ density profile. Using that profile and the flux detected with AzTEC we estimate the expected mass inside the ALMA beam. Blue squares are sources in this work detected with ALMA. Red points are sources from \citet{Huelamo2017}. The ratio between the observed and the estimated masses is close to $\sim$1 (orange line) for half of the sources (see Section \ref{profile}). The other three sources with steeper density profiles might be more evolved as they have accreted more mass onto the central object.}        
\label{density_profile}
\end{figure}

In Figure \ref{density_profile}, we compare the mass derived from the ALMA observations (at T = 9\,K) to the mass estimated from an r$^{-2}$ density profile. The estimated mass error is based on the ALMA and AzTEC mass errors. Barnard 30 pre-BD candidates are also included. We note that \citet{Huelamo2017} considered typical uncertainties of a factor of 4 in the mass estimate. The results in Barnard 30 are very similar to those we found in our Lupus sample, and they also seem to be in a state of gravitational contraction. Thus, both studies ultimately reach a very similar conclusion: most of the objects detected with ALMA  seem to be at the beginning of the large-scale contraction expected in a scaled-down version of a low-mass star formation scenario.

\label{nature_counterparts}

\subsection{The nature of J154229.778-334241.86} 
\begin{figure*}
\includegraphics[width=0.5\textwidth]{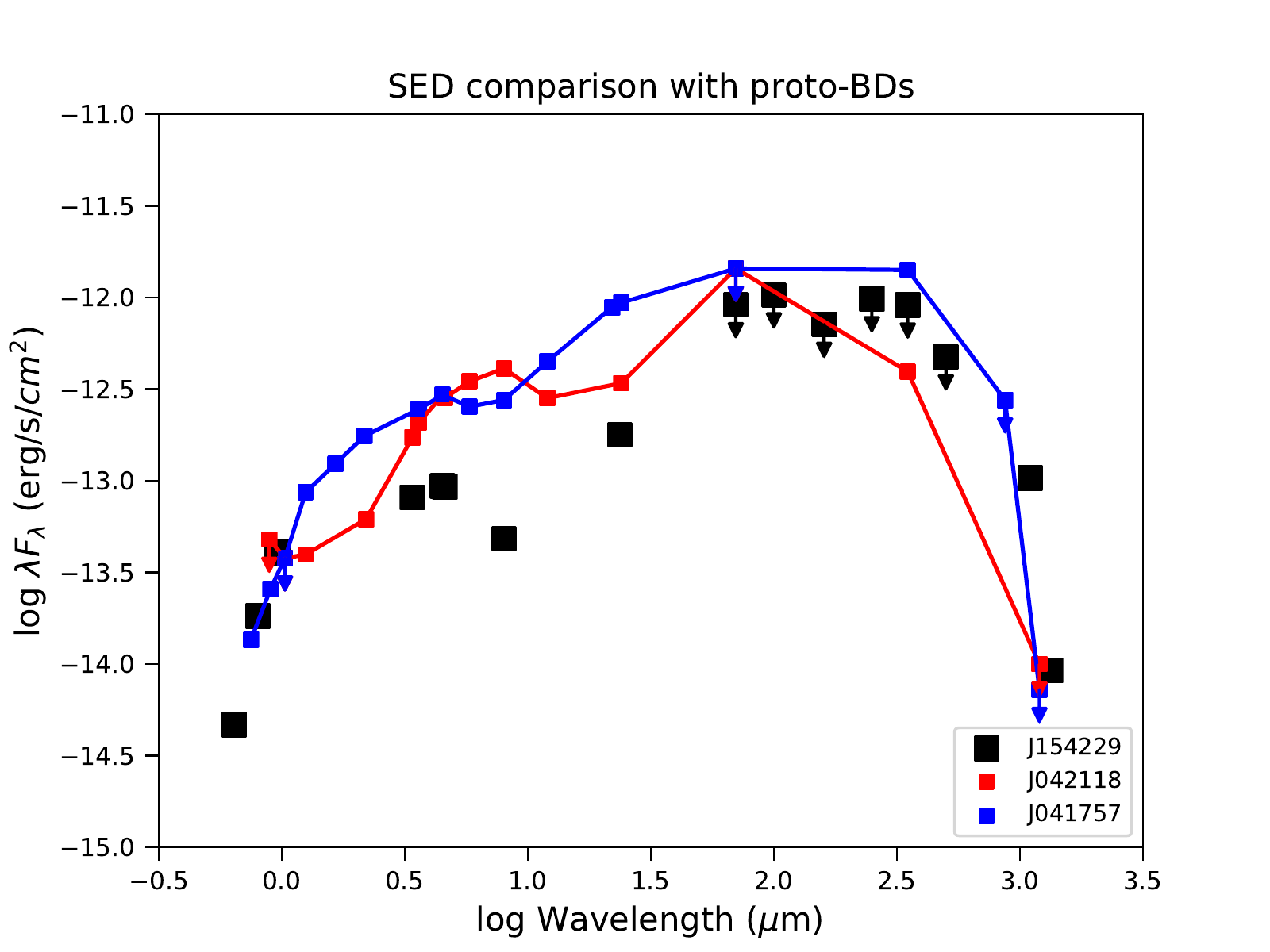}
\includegraphics[width=0.5\textwidth]{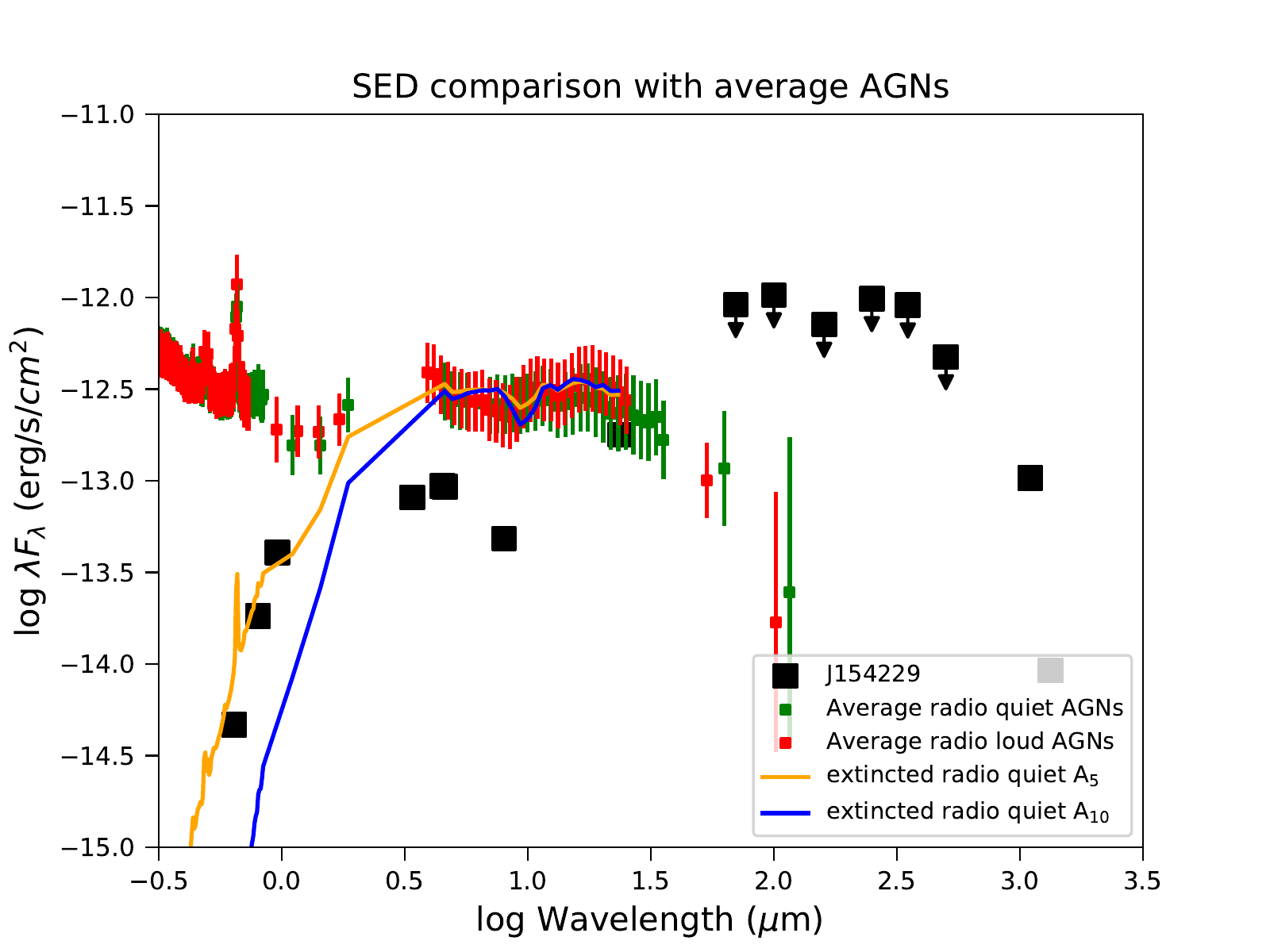}
\caption[SED J154229]{Left panel: SED of J154229.778-334241.86 (black) compared with proto-BD candidates J041757 (blue) and J042118 (red)  including upper limits (marked with arrows). This figure is based on Figure 8 from \citet{Palau2012-1}. Right panel: SED of J154229.778-334241.86 (black) compared with average SEDs for radio-quiet AGN (green) with a V-band extinction of 5 (orange) and  V-band extinction of 10 (blue), and radio-loud AGN (red) from \citet{Shang2011}. This figure is based on Figure 9 from \citet{Palau2012-1}.}    
          
\label{sed_protos_aina}
\end{figure*}

J154229 is the only proto-BD candidate in our study whose SED is well populated. In the NED catalogue, there is a nearby infrared source located 0.24$\arcsec$ distant, whose extragalactic nature has not been confirmed. To rule out the possibility that J154229 is extragalactic, meaning that it is, in fact, an active galactic nucleus (AGN), we compared its SED (see right panel in Figure \ref{sed_protos_aina}) with an average SED of radio-loud and radio-quiet AGN with different extinction values, following \citet{Palau2012-1}. The differences, especially in the optical and in the millimetre, suggest it is not very likely that J154229 is an AGN. We compared the SED of J154229 with two other proto-BD candidates, J041757 and J042118 in Taurus \citep{Barrado2009, Palau2012-1} (see left panel in Figure \ref{sed_protos_aina}). The similar shape of all three SEDs suggests J154229 is very likely a bona fide proto-BD candidate.  One of these sources, J041757, has a cold dust envelope with a size of 1000 au and a mass of 5M$_{\mathrm{Jup}}$, while the envelope of J154229 is three times larger ($\sim$3500 au) and more massive (11 M$_{\mathrm{Jup}}$). This difference may suggest that J15299 is slightly younger than J042118. We conclude that due to the similarities in the SED, J154229 is a promising proto-BD candidate. However, gas detection is necessary to confirm its association with the Lupus molecular cloud. 

\subsection{Class II brown dwarfs: disk masses}
\label{sec:disk masses}
The properties of the disks of more evolved BDs can also shed light on the formation of substellar objects and on the possibility of planet formation in these sources. Theoretical predictions \citep{stamatellos2015} indicate that for central sources with equal masses, the disk masses are higher in sources formed by disk fragmentation compared to sources formed as a scaled-down version of low-mass stars. In order to compare different formation scenarios, we estimated the dust disk masses of the  Class II sources in our sample along with previous Class II BDs observed with ALMA in several SFRs.  

In the case of the Lupus sources, the mass of the central object was obtained from the literature \citep{comeronetal09-1,alcalaetal14-1} and adjusted to the new Gaia DR2 parallaxes. To do this, we rescaled the luminosity values from the literature using the new Gaia distances, then we interpolated the stellar mass from the position in the Hertzsprung-Russel diagram using the evolutionary models of \citet{Baraffe15-1}. We calculated the dust disk mass as described in Section \ref{sec:continuum} using the distances given in Table \ref{distances_gaia_dr2} and a value of $\kappa_{\nu}$ = 8.5$\times$10$^{-1}$ cm$^{2}$g$^{-1}$ \citep{Ose94}. The range of masses we obtained is between 0.58 to 3.4\,M$_{\oplus}$ (see Table \ref{polvo_y_temperatura}).
We calculated upper limits for the Class II sources in Lupus 1 and 3 that were not detected with ALMA using the rms value in Appendix \ref{tabla_phase_rms}. We included objects whose distances are compatible with the average distance derived for Lupus 1 and 3 (see Section \ref{distancia derivada}) as well as BDs in our sample that are probably not Lupus members. 

\begin{table}
\caption{Disk dust masses for the spectroscopically confirmed Class II BDs our sample. Temperature is constant at 20K.}             
\label{polvo_y_temperatura}      
\centering                          
\begin{tabular}{c c }        
\hline\hline                 
Name  & Disk dust mass \\    
  & [M$_{\oplus}$]\\ 
\hline
\hline  \\                      
160826 & 3.4 $\pm$ 0.2\\
Lup 706 & 0.6 $\pm$ 0.2\\
Par-Lup 3-4 & 0.6 $\pm$ 0.2\\
SONYC-Lup 3-7 & 0.58 $\pm$ 0.09\\
\hline                           
\end{tabular}
\end{table} 

Having inferred the disk masses of our Class II substellar sources associated with Lupus, we compared them with BD and VLM star disks previously observed with ALMA. We searched for available Class II substellar sources with published ALMA detections and then we recalculated the masses using the fluxes provided in the literature and the above values of temperature and opacity \citep{Ose94}. Using stellar parameters from the literature, the theoretical evolutionary models from \citet{Baraffe15-1}, and theThe  new Gaia DR2 parallaxes, we calculated the masses for each BD and VLM star. 
\smallskip
We included sources in several SFRs such as Ophiuchus \citep{Testietal16-1}, Upper Scorpius \citep{vanderplasetal16-1}, Chamaeleon I \citep{Pascucci16-1}, Taurus \citet{Riccietal14-1,Ward-Duong18}, and Lupus \citep{Ansdell16-1,Sanchis20-1}. We used the stellar parameters in \citet{Alvesetal12-1, Muzic12-1,Manara15-1} to obtain the stellar masses for the Ophiuchus sources. From Upper Scorpius \citep{vanderplasetal16-1} we chose sources with spectral type later than M4. The stellar parameters from \citet{Scholz07-1,vanderplasetal16-1}.
\smallskip

\begin{figure*}
\includegraphics[width=\textwidth]{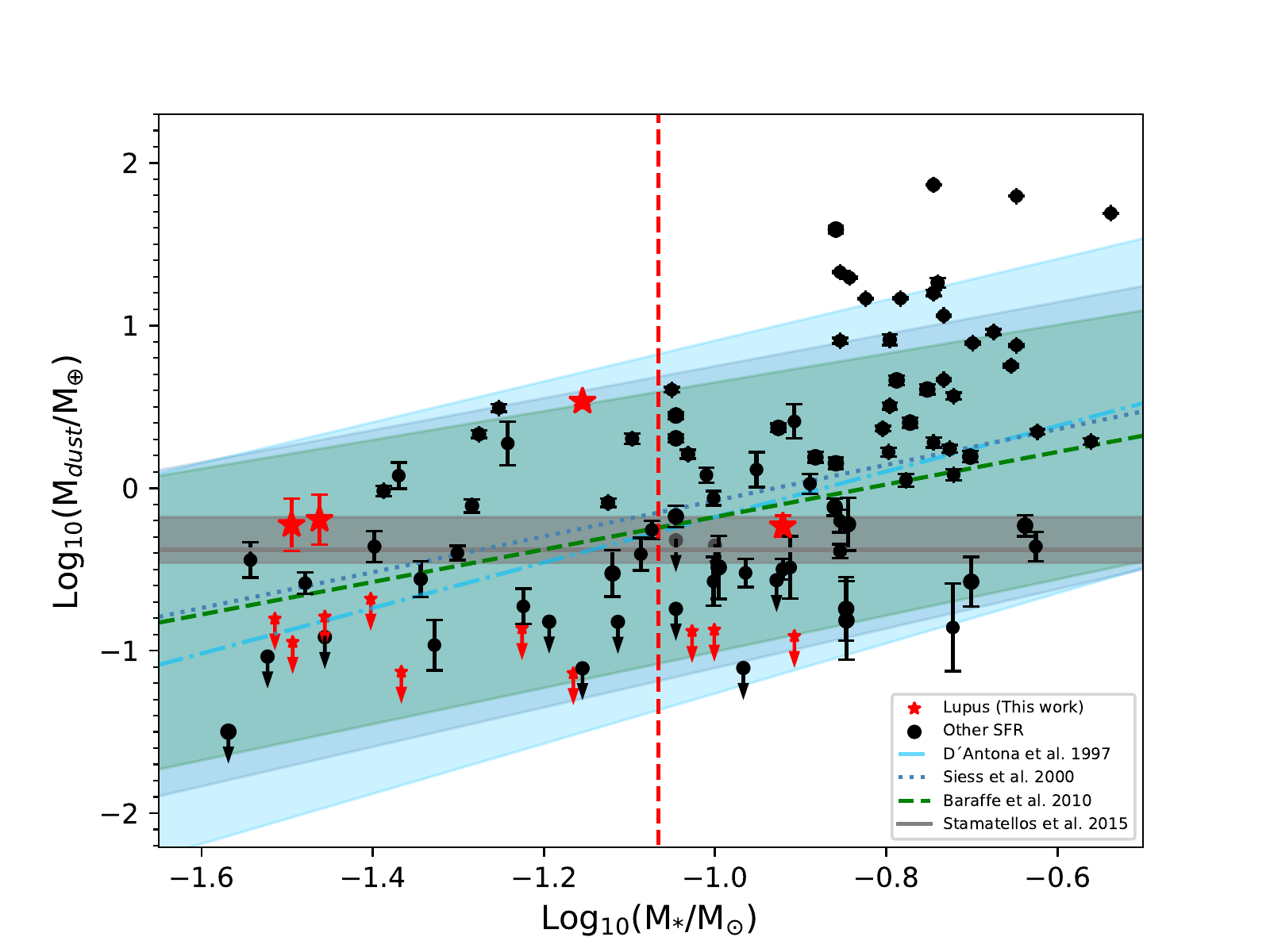}
      \caption{Dust disk masses as a function of the central object mass (M$_{*}$) for Class II BDs. Red stars mark sources in this paper including upper limits (marked with arrows). Black circles correspond to BD and VLM stars in other SFRs such as Ophiuchus \citep{Testietal16-1}, Upper Scorpius \citep{vanderplasetal16-1}, Chamaeleon I \citep{ Pascucci16-1}, Taurus \citep{Riccietal14-1,Ward-Duong18}, and Lupus sources that are not part of our survey \citep{Ansdell16-1,Sanchis20-1}. Upper limits are included as arrows. All the objects were observed with ALMA and masses were recalculated using T=20 K for consistency. The grey area represents the predictions for disk fragmentation in \citet{stamatellos2015}, while the blue and green areas are the scaling relations between disk mass and stellar mass for protoplanetary disks derived by \citet{Andrews2013} using three different sets of model calculations \citep{Antona97-1,Baraffe98-1,siessetal00-1}. The red vertical line represents the limit between BDs and stars.} \label{_testi}
\end{figure*}

For Taurus we collected sources from two different works: \citet{Ward-Duong18} and \citet{Riccietal14-1}. From the latter we included three BDs; however the error in the flux is not provided so we assumed the rms error. We used the stellar parameters from \citet{Ricci2013} and \citet{Andrews2013}. From \citet{Ward-Duong18}, we selected sources with a spectral type later than M4 choosing the flux obtained with the natural weighting algorithm. Stellar properties are indicated in the same work. From Chamaeleon I \citep{Pascucci16-1} we selected sources with spectral types later than M4, and stellar parameters from \citet{Manara14-1,Manara16-1}. Finally, we incorporated other sources in Lupus with spectral type later than M4 that are not part of our survey \citep{Ansdell16-1,Sanchis20-1}; stellar masses for these sources are in \citet{alcalaetal14-1,Alcala17-1,kora14-1}. The stellar and disk masses for all these regions, together with those of Lupus 1 and 3, are shown in Figure \ref{_testi}.

Regarding the formation mechanism, we included the predictions from \citet{stamatellos2015} about disk fragmentation and ejection in Figure \ref{_testi}. About 40\% of all the detected substellar sources in the plot are in the expected range, although there are several sources whose values are even higher than expected for disk fragmentation. The blue and green areas represent the expected scaling relations between disk mass and stellar mass extrapolated from the stellar regime using three different models \citep{Antona97-1,Baraffe98-1,siessetal00-1} and all the BDs, including the upper limits are in agreement with these relations. The two formation methods overlap in the BD regime and part of the detected objects with ALMA are consistent with both formation scenarios. However, the upper limits already indicate that these sources seem inconsistent with the disk fragmentation scenario. On the other hand, upper limits may be formed by the turbulent fragmentation and the whole dataset of detected BD with ALMA seem to favor the star-like scenario as a dominant mechanism. 


The largest dust disk mass detected among the objects in our Lupus sample is 3.4 M$_{\oplus}$. The minimum disk mass expected for planet formation is around 10 M$_{\oplus}$ \citep{Payne07-1,Testietal16-1}. Thus, planet formation in the substellar regime in these objects seems not to be possible unless planets, planetesimals or planetary cores) are already formed in these systems. Recent results from studies of protoplanetary disks \citep{Andrews2018} suggest that planets are formed much earlier than previously thought. This can explain the low dust disk masses that we measured if we assume that  planet formation in BDs follows the core accretion or the disk instability mechanism in the stellar regime. 

Finally, a word of caution should be included. The relation between disk temperature and luminosity is still under debate  \citep{vanderplasetal16-1, Pascucci16-1, Ward-Duong18} given that the models make various assumptions that might not be correct. The luminosity, temperature, and disk radii are intrinsically related. As it was seen for some sources in Ophiuchus \citep{Testietal16-1}, the radii can be smaller than expected. This will affect the disk temperature. We chose a fixed temperature of 20 K for Class II BDs without any uncertainty. If we use T=25(L/L$_{\odot}$)$^{0.25}$ instead, the disk masses are $\sim$3 times larger. 

Future observations with better angular resolution, deeper observations to detect fainter sources, and future modeling are required to gain a fuller understanding of the dominant BD formation mechanism and to decide whether planet formation is possible in these disks, based on the detection similar substructures as those seen in the stellar regime. 

\subsection{Exploring the dominant scenario for BD formation}
In the previous sections, we present a review of the characteristics of pre-BD core candidates and Class II brown dwarfs according to various formation scenarios. 

In conclusion, 40 $\%$ of the detected Class II BDs in all the SFRs, are consistent with a disk fragmentation scenario, according to the theoretical predictions \citep{Stamatellosandwhitworth09-1}. However, the entire sample follows the scaling relation between the stellar mass and the disk mass obtained in the stellar regime. 
Our detection rate is similar to the previous work of \citet{Huelamo2017}. Given that our sample was bigger it is natural that we find a larger total number of candidates. We expect that observations with either increased sensitivity or larger LAS will increase the number of detected sources and help to finally confirm their nature (see Section \ref{detection_rate}). 
We think that photoevaporation is not the dominant formation mechanism in BDs for the two Lupus regions and Barnard 30 because they share similar very young populations, but reside in different large scale environments -- in fact, there are no massive stars in Lupus 1.

As a final remark, we note that although we checked each source in the NED to check their classification as possible extragalactic sources, we cannot fully ensure their Galactic nature except for those sources that were detected by Gaia and spectroscopically confirmed as YSOs. The sources are mainly located in the cloud filament and are probably associated with the cloud, but only future gas detections at the velocity of the cloud would confirm the membership of each source to the complex.

\section{Conclusions}
\label{p2_conclusion}
In this work, we present a search of substellar objects at the earliest stages using high sensitivity ALMA observations in the Lupus\,1\,and\,3 star-forming regions. Our target selection is based on sources found in AzTEC (1.1 mm) maps and complemented with Class II BDs from the literature. 
We supplement our observations with optical/IR archival data. The main objective of this work is to test and constrain BD formation theories by identifying pre- and proto-BDs and classifying them according to their evolutionary state. Our main results as follows.

-- We detected sources in 15 of the 64 ALMA pointings. The total number of compact sources detected is 19. We report masses between 0.33 and 124 M$\mathrm{_{Jup}}$. The ALMA detection rate was $\sim$23\,\%. The ALMA non-detections may be related to the source size, with spatial scales larger than the ALMA LAS dominating the emission. 

-- The number of newly classified candidate objects is 14. We classified 12 sources as pre-BDs or deeply embedded protostar candidates because they do not have optical/infrared counterpart. We classified one Class 0/I and one Class I candidate based on their SED slope and the T$_\mathrm{bol}$.

-- We detected continuum emission in five more Class II sources that were previously selected from the literature and we measured their disk masses, which span values between 0.58 and 3.4\,M$_{\oplus}$. 

-- We studied the nature of the 12 sources that we classified as pre-BD or deeply embedded protostar candidates. For these sources, we compared the estimated density with the critical density from a Bonnor-Ebert isothermal sphere, concluding that one of the sources seems to be collapsing. Second, we explored an alternative path comparing the estimated ALMA masses with the masses that should be obtained from an r$^{-2}$ density profile, and we concluded that the pre-BDs core candidates could be the result of a large-scale gravitational collapse. 

-- We estimated the final mass of the pre-BD candidates, assuming a core efficiency of 30\,\% using the mass derived from AzTEC and adding the ALMA mass, and we find that all the sources could end up as substellar objects.
 
-- The Class 0/I candidate J154229 is at the boundary between Class 0 and I, based on bolometric temperature, and thanks to the efficient sampling of its SED, it has proven to be a promising proto-BD candidate. The SED suggests this object has already formed a BD in its core. Its SED is similar to that of other known proto-BDs and different from the typical SEDs of AGN.

-- We compared the dust disk masses for the objects that we classified as Class II with previous studies of YSOs in other SFRs and obtained similar results. A scaled-down version of low-mass star formation may be the dominant scenario, however, it is fair to note that the disk fragmentation may be responsible for a non-negligible number of BDs. 

Future observations are needed to confirm the nature of the ALMA pre- and proto-BD candidates. Optical/infrared spectroscopic studies for J153914 and J154229 are needed to confirm them as proto-BDs. In addition, the detection of gas associated with any of the candidates in the sample would confirm their membership to the Lupus molecular complex. 


%
\begin{acknowledgements}
This paper makes use of the following ALMA data: ADS/JAO.ALMA$\#$2015.1.00512.S. ALMA is a partnership of ESO (representing its member states), NSF (USA) and NINS (Japan), together with NRC (Canada), MOST and ASIAA (Taiwan), and KASI (Republic of Korea), in cooperation with the Republic of Chile. The Joint ALMA Observatory is operated by ESO, AUI/NRAO and NAOJ. \\
A.S-M, I.G-M, M.R.S and A.B acknowledge support from the ``Iniciativa Cient\'ifica Milenio" via N\'ucleo Milenio de Formaci\'on Planetaria. A.B acknowledges support from FONDECYT Regular grant N. 1190748. \\
IdG-M is partially supported by MCIU-AEI (Spain) grant 
AYA2017-84390-C2-R (co-funded by FEDER) \\
NH acknowledges financial support from the Spanish State Research Agency (AEI) Project No. ESP2017-87676-C5-1-R and from project No. MDM-2017-0737 Unidad de Excelencia “María de Maeztu”- Centro de Astrobiología (INTA-CSIC). \\
The National Radio Astronomy Observatory is a facility of the National Science Foundation operated under cooperative agreement by Associated Universities, Inc.
KM acknowledges funding by the Science and Technology Foundation of Portugal (FCT), grants No. IF/00194/2015 and PTDC/FIS-AST/28731/2017. \\
A.P. acknowledges financial support from CONACyT and UNAM-PAPIIT IN113119 grant, M\'exico
\end{acknowledgements}
\bibliographystyle{aa.bst} 
\bibliography{bibliografiav2} 
\begin{appendix} 
\section{ALMA detections and non detections}
List of the ALMA pointings based on the AzTEC detections. The rms is measured at the phase centre except for the sources with ALMA detections. We included the previous classification based on the AzTEC detections and the SED from the SOLA catalog. Class II and Class I/II sources were obtained from the literature 
\longtab[0]{
\begin{longtable}{cccccc}
\caption{List of the ALMA pointings including detections and non-detections}\label{tabla_phase_rms}\\
\hline
\hline
Name\footnote[1]{Name as it appears in the literature. AzTEC-lup is a denomination for the non ALMA detection pointings based on AzTEC detections.} & Ra(J2000) & Dec(J2000)\footnote[2]{Phase center of the observation}  & rms\footnote[3]{rms is measured at the phase center, except for the ALMA detected sources}(mJy/beam) & Classification\footnote[4]{Previous classification before the ALMA observation based on AzTEC data and optical/infrared counterpart.} & References \footnote[5]{Specific references used to obtain spectroscopically confirmed Class II sources in the ALMA pointing: (1) \citet{comeronetal09-2}, (2) \citet{merinetal08-1}, (3) \citet{kora14-1}} \\
\hline
153701.1-332255&15:37:01.10& -33:22:55.00 & 0.10 & Class II & 1 \\
ALMA J153702.653-331924.92\footnote[6]{\label{newy}Sources detected with ALMA. See table \ref{table:tabla_ppal} } &  15:37:03.10 & -33.19.27.00 & 0.096 & Class II & 1, 3\\
153709.9-330129&15:37:09.90& -33:01:29.00 & 0.085 & Class II & 1\\
AzTEC-lup1-99&15:38:04.40& -34:52:28.24 & 0.054 & Starless core \\
AzTEC-lup1-103&15:38:12.90& -34:56:23.70 & 0.056 & Starless core\\
AzTEC-lup1-109&15:38:27.44& -35:12:40.90 & 0.057 & Starless core\\ 
AzTEC-lup1-72&15:38:46.93& -33:23:36.38 & 0.092 & Class I\\
AzTEC-lup1-111&15:38:59.73& -33:29:50.16 & 0.055 & Starless core\\
AzTEC-lup1-57&15:39:04.25& -35:06:43.0 & 0.069 & Starless core \\ 
ALMA J153914.996-332907.62\footref{newy}   &   15:39:15.84 & -33:28:58.50 & 0.094 & Starless core \\
153921.8-340020&15:39:21.80& -34:00:20.00 & 0.085 & Class II & 1\\
AzTEC-lup1-67&15:39:21.17& -34:43:37.52 & 0.071 & Starless core \\
AzTEC-lup1-114&15:39:49.35& -34:49:26.22 & 0.063 & Starless core\\
AzTEC-lup1-84&15:40:09.15& -33:32:20.16 & 0.057 & Starless core \\
AzTEC lup1-90&15:40:18.86& -33:41:00.09 & 0.061 & Class 0\\
AzTEC-lup1-40&15:40:46.65& -33:43:17.88 & 0.10 & Class I\\
AzTEC-lup1-104&15:41:15.39& -33:46:41.34 & 0.057 & Starless core\\
AzTEC-lup1-101&15:41:28.04& -33:41:51.37 & 0.059 & Starless core\\
154140.8-334519&15:41:40.80& -33:45:19.00 & 0.089 & Class II & 1,2\\
AzTEC-lup1-119&15:42:05.21& -33:45:59.71 & 0.058 & Starless core\\
ALMA J154229.778-334241.86\footref{newy}    &  15:42:29.56 & -33:42:39.94 & 0.11 & Class 0\\
AzTEC-lup1-124&15:42:38.60& -33:48:52.24 & 0.055 & Starless core \\ 
AzTEC-lup1-52&15:42:45.21& -33:58:43.41 & 0.065 & Starless core\\
AzTEC-lup1-54&15:42:45.02& -34:12:01.36 & 0.075 & Starless core\\
AzTEC-lup1-94&15:43:50.32& -34:01:59.60 & 0.067 & Starless core\\ 
154433.9-335254&15:44:33.90& -33:52:54.00 & 0.11 & Class II & 1\\
ALMA J154456.522-342532.99\footref{newy}   &   15:44:57.22 & -34:25:31.55 & 0.068 & Starless core\\
AzTEC-lup1-71&15:44:59.34& -34:20:55.49 & 0.091 & Class 0\\ 
ALMA J154506.515-344326.15\footref{newy}   &   15:45:06.45 & -34:43:16.23 & 0.089 & Starless core\\
AzTEC-lup1-123&15:45:40.24& -35:04:56.61 & 0.062 & Starless core \\
ALMA J154634.169-343301.90 \footref{newy}    & 15:46:33.47 & -34.33.05.10 & 0.083 & Starless core\\
160545.8-385454&16:05:45.80& -38:54:54.00 & 0.062 & Class II & 1\\
ALMA J160658.604-390407.88 \footref{newy}  &   16:06:58.70 & -39:04:05.00 & 0.062 & Class II & 1\\
AzTEC-lup-3-15&16:07:51.70& -39:07:29.50& 0.058 & Starless core \\
160714.0-385238&16:07:14.00& -38:52:38.00& 0.060 & Class II & 1,2\\
ALMA J160804.168-390452.84 \footref{newy}   &  16:08:04.80 & -39:04:49.00 & 0.093 & Class II & 1,3 \\
AzTEC-lup-3-20&16:08:14.40& -39:10:50.89& 0.067 & Starless core\\
160816.0-390304&16:08:16.00& -39:03:04.00& 0.057 & Class II & 1,2,3\\
160826.8-384101\footref{newy}  &   16:08:26.80 & -38:41:01.00 & 0.059 & Class II & 1\\
AzTEC-lup3-12&16:08:32.70& -39:04:39.80& 0.061 & Starless core \\
160833.0-385222&16:08:33.05& -38:52:22.40 & 0.066 & Class II & 1,3\\
160835.5-390035&16:08:35.48& -39:00:35.80 & 0.067 & Class II & 1,3\\
Lup706\footref{newy}        &    16:08:37.33  &-39:23:10.90 & 0.090 & Class II & 1,3\\
AzTEC-lup3-10&16:08:41.60& -39:05:23.91& 0.058 & Starless core\\
160848.2-390920&16:08:48.20& -39:09:19.00& 0.051 & Class II & 1,2,3\\
AzTEC-lup3-5&16:08:48.50& -39:07:27.97& 0.066 & Starless core\\
Par-Lup3-4\footref{newy}     &    16:08:51.44 & -39:05:30.50 & 0.031 & Class I/II & 1,2,3\\
AzTEC-lup-3-14&16:08:54.60& -39:12:26.90& 0.052 & Starless core\\
AzTEC-lup-3-19&16:08:55.40& -39:05:59.89& 0.060 & Starless core\\ 
SONYC-Lup3-7\footref{newy}  & 16:08:59.53 &-38:56:27.80 & 0.055 & Class I/II & 1,2,3\\
AzTEC-lup3-4&16:09:13.60& -39:07:43.95& 0.059 & Starless core\\
SONYC-lup3-10&16:09:13.43& -38:58:04.90& 0.062 & Class I/II & 3\\ 
\newpage
\hline
\hline
Name\footnote[1]{Name as it appears in the literature. AzTEC-lup is a denomination for the non ALMA detection pointings based on AzTEC detections.} & Ra(J2000) & Dec(J2000)\footnote[2]{Phase center of the observation}  & rms\footnote[3]{rms is measured at the phase center, except for the ALMA detected sources}(mJy/beam) & Classification\footnote[4]{Previous classification before the ALMA observation based on AzTEC data and optical/infrared counterpart.} & References \footnote[5]{Specific references used to obtain spectroscopically confirmed Class II sources in the ALMA pointing: (1) \citet{comeronetal09-2}, (2) \citet{merinetal08-1}, (3) \citet{kora14-1}} \\
\hline
AzTEC-lup3-8&16:09:14.30& -39:05:23.95& 0.065 & Starless core\\
AzTEC-lup3-13&16:09:36.60& -39:03:59.64& 0.069 & Starless core\\
ALMA J160920.171-384456.40\footref{newy} & 16:09:20.80 & -38.45.10.00 & 0.18 & Class II & 1,3\\
ALMA J160932.167-390832.27\footref{newy}    &   16:09:32.80 &-39:08:44.11 & 0.13 & Starless core\\
AzTEC-lup3-29&16:10:01.33& -39:06:45.10& 0.057 & Class II & 1,2\\
AzTEC-lup3-9&16:10:05.90& -39:10:54.84& 0.061 & Starless core\\
AzTEC-lup-3-21&16:10:08.22& -39:02:51.68& 0.053 & Class I\\
AzTEC-lup-3-16&16:10:19.80& -39:11:51.07& 0.067 & Starless core\\
ALMA J161030.273-383154.52\footref{newy}   &  16:10:30.60 & -38:31:51.00 & 0.064 & Class II & 1 \\
161144.9-383234&16:11:44.88& -38:32:44.90 & 0.059 & Class II & 1,2,3\\
161225.6-381742&16:12:25.60& -38:17:42.00 & 0.056 & Class II & 1,3\\
161210.4-390904&16:12:10.46& -39:09:04.00& 0.051 & Class II & 1,3\\

\end{longtable}
}

\section{Additional figures}

\begin{figure*}
\includegraphics[width=0.45\textwidth]{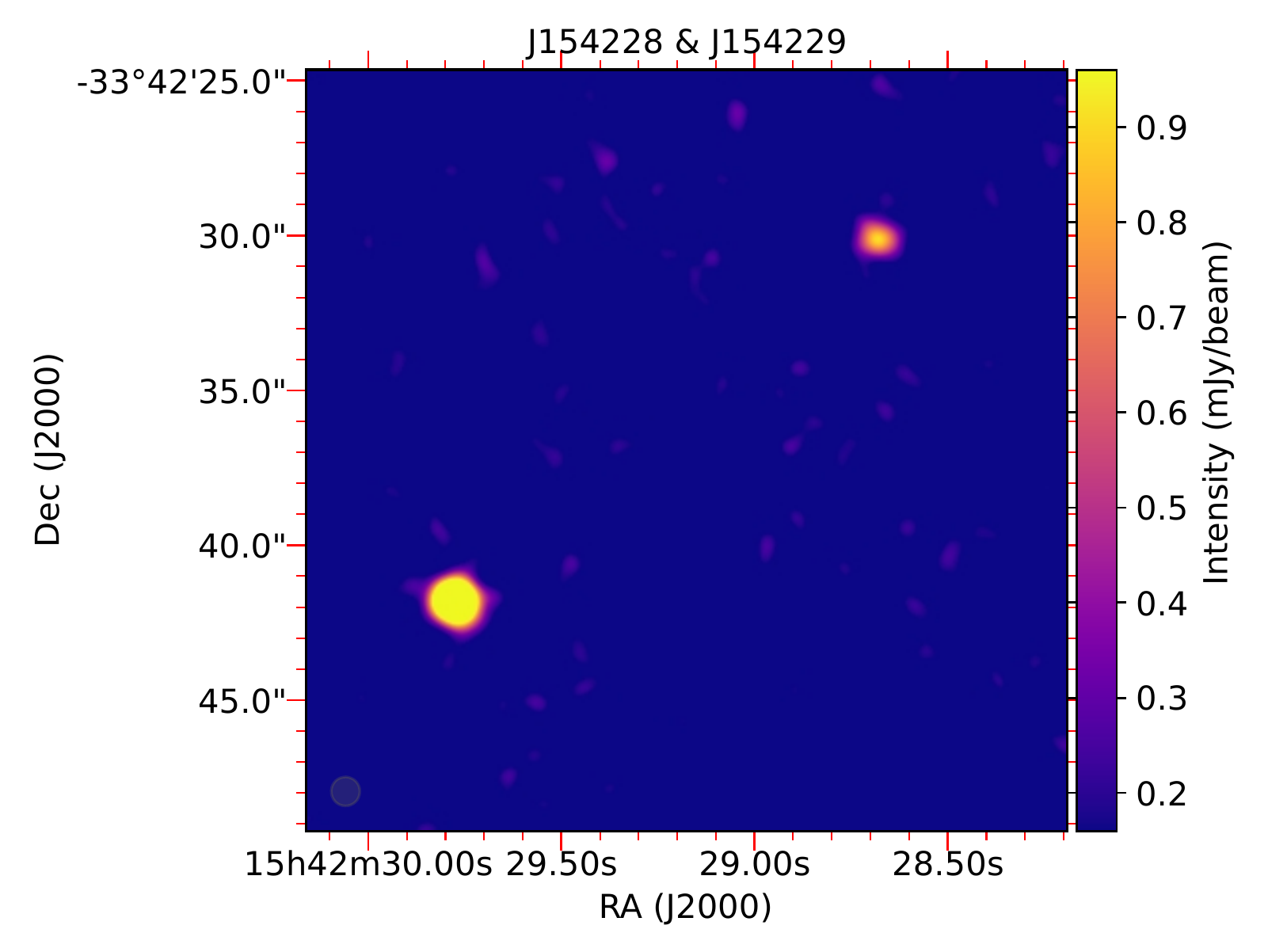}
\includegraphics[width=0.5\textwidth]{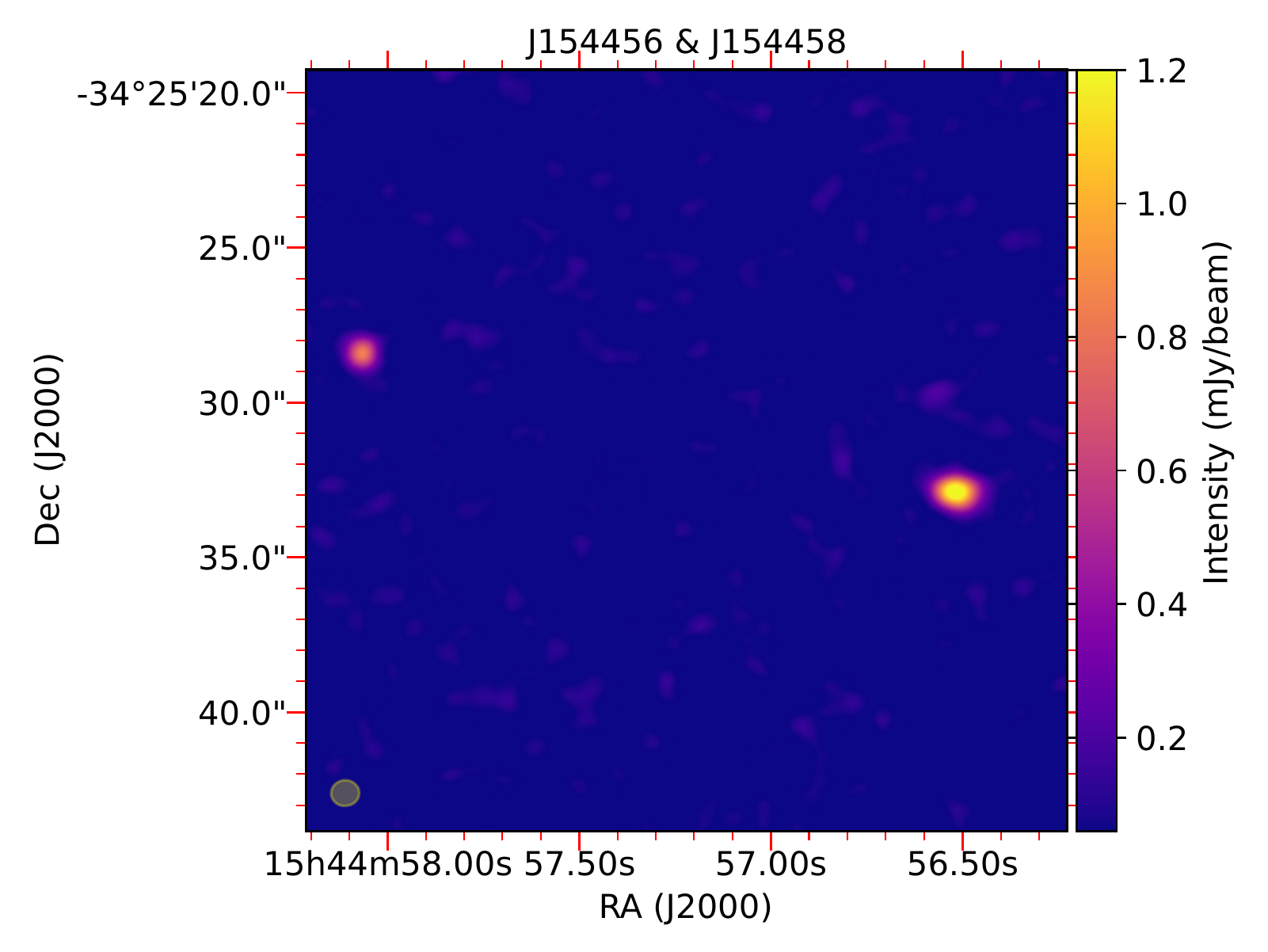}
\qquad
\includegraphics[width=0.45\textwidth]{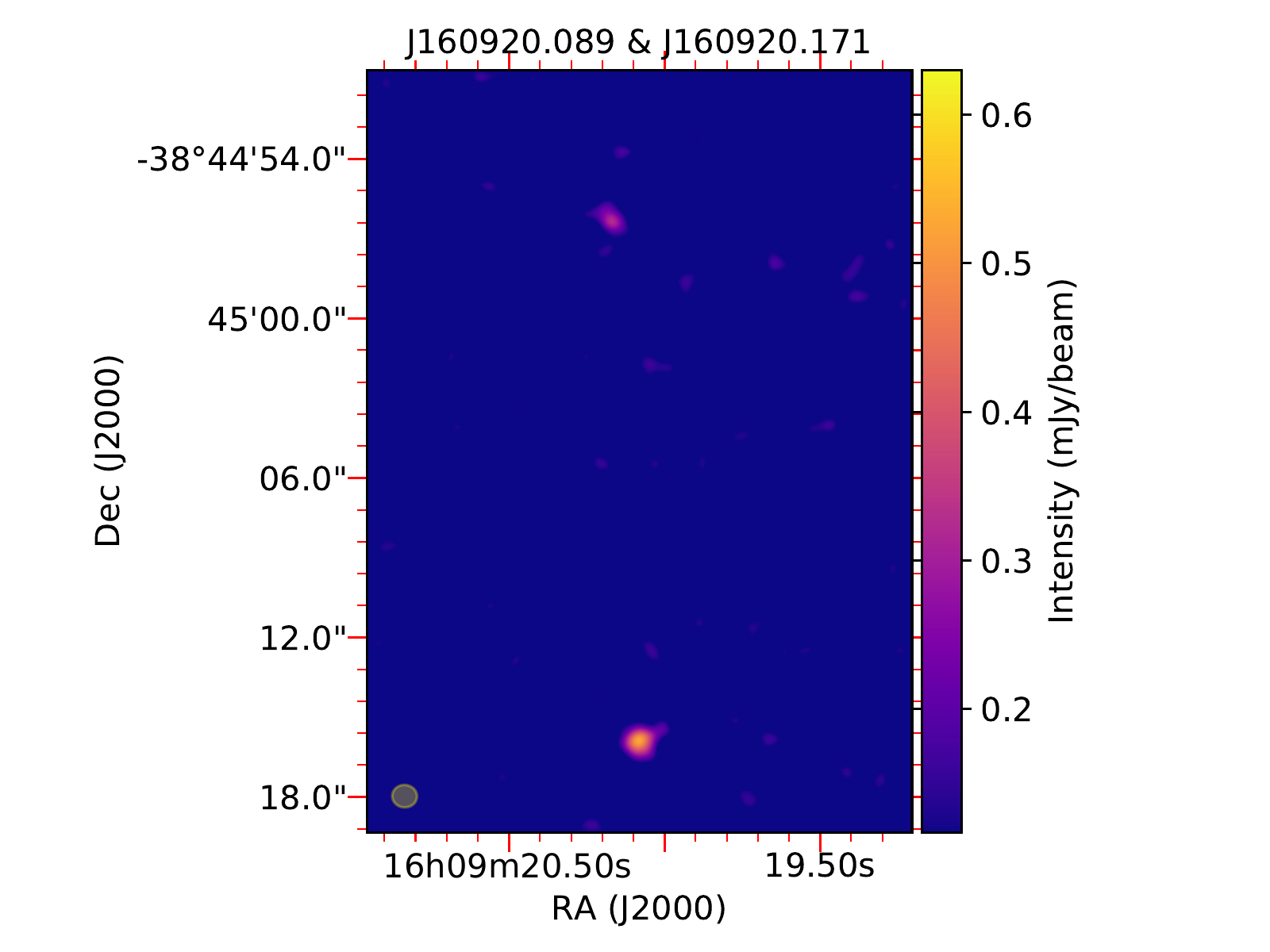}
\includegraphics[width=0.45\textwidth]{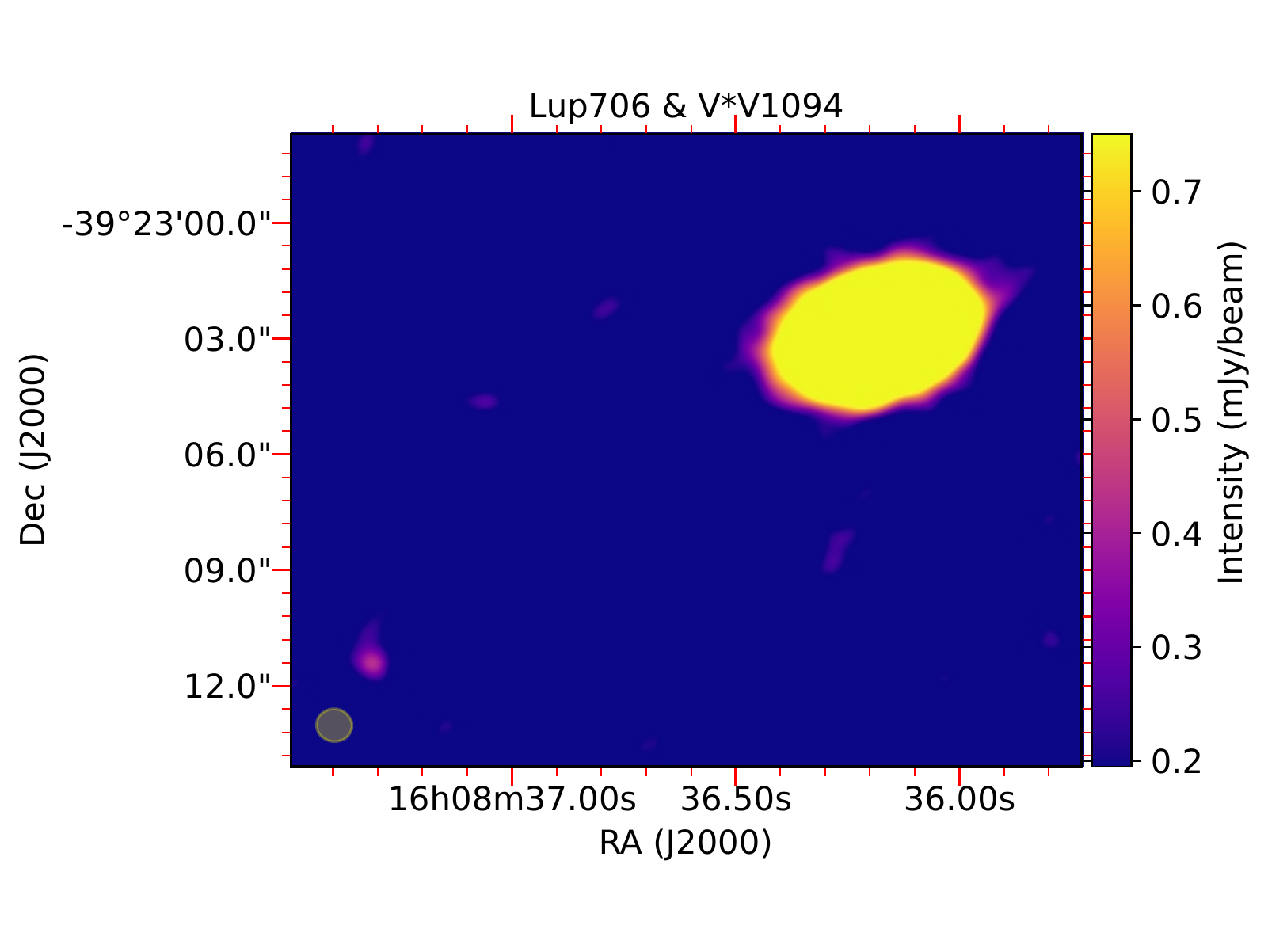}  
\caption[ALMA continuum images of the Lupus 1 and 3 sample with two sources inside the ALMA primary beam]{\label{dobles}ALMA 1.3 mm maps for fields that show two sources inside the ALMA primary beam. Grey ellipse at the bottom-left corner represents the synthesized beam. Individual sources can be seen in more detail in Figure \ref{panel_alma} and \ref{apendix_panel_alma}}
\end{figure*} 

\begin{figure*}
\subfloat{\includegraphics[width=0.46\textwidth]{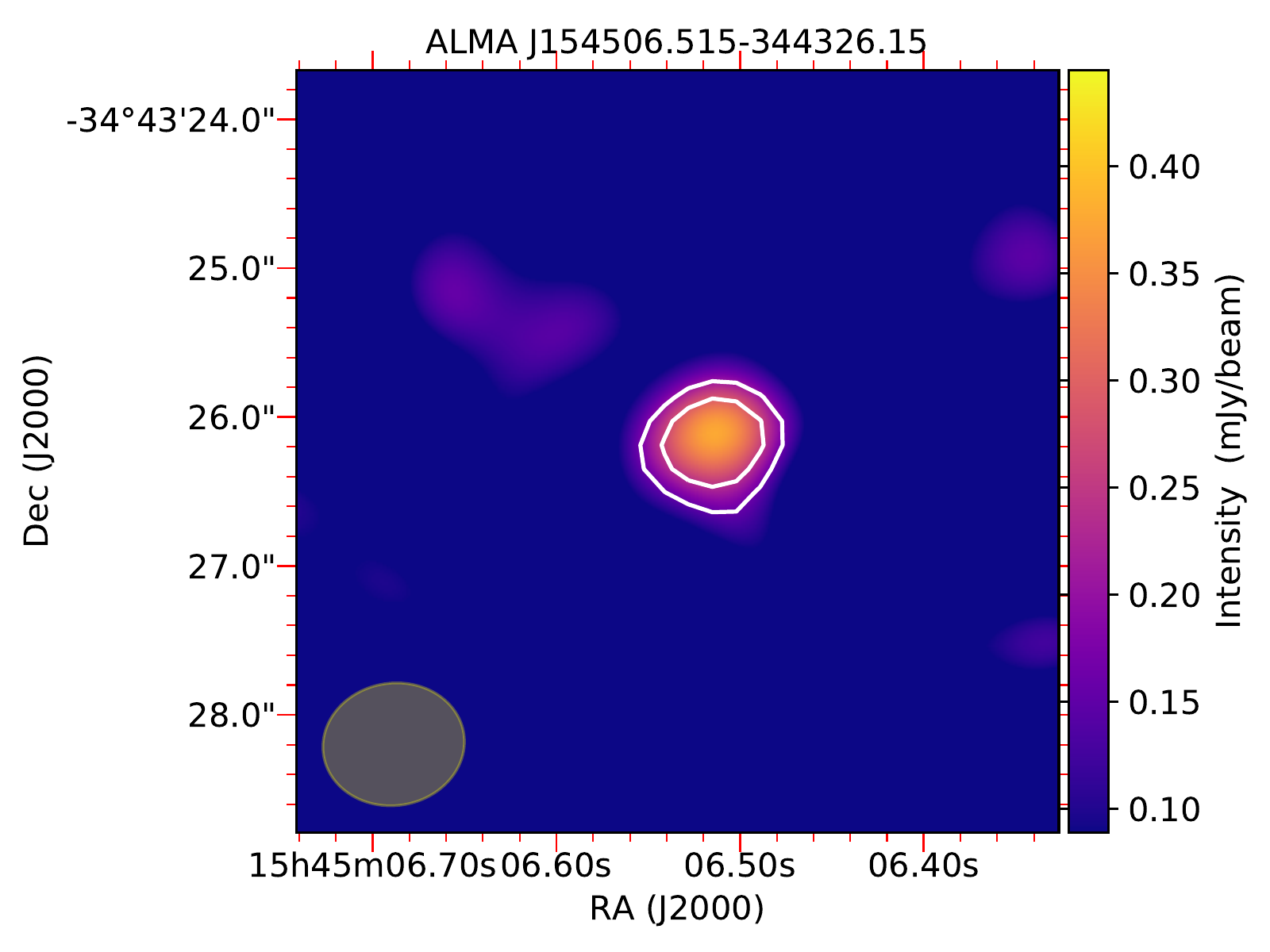}}
\subfloat{\includegraphics[width=0.46\textwidth]{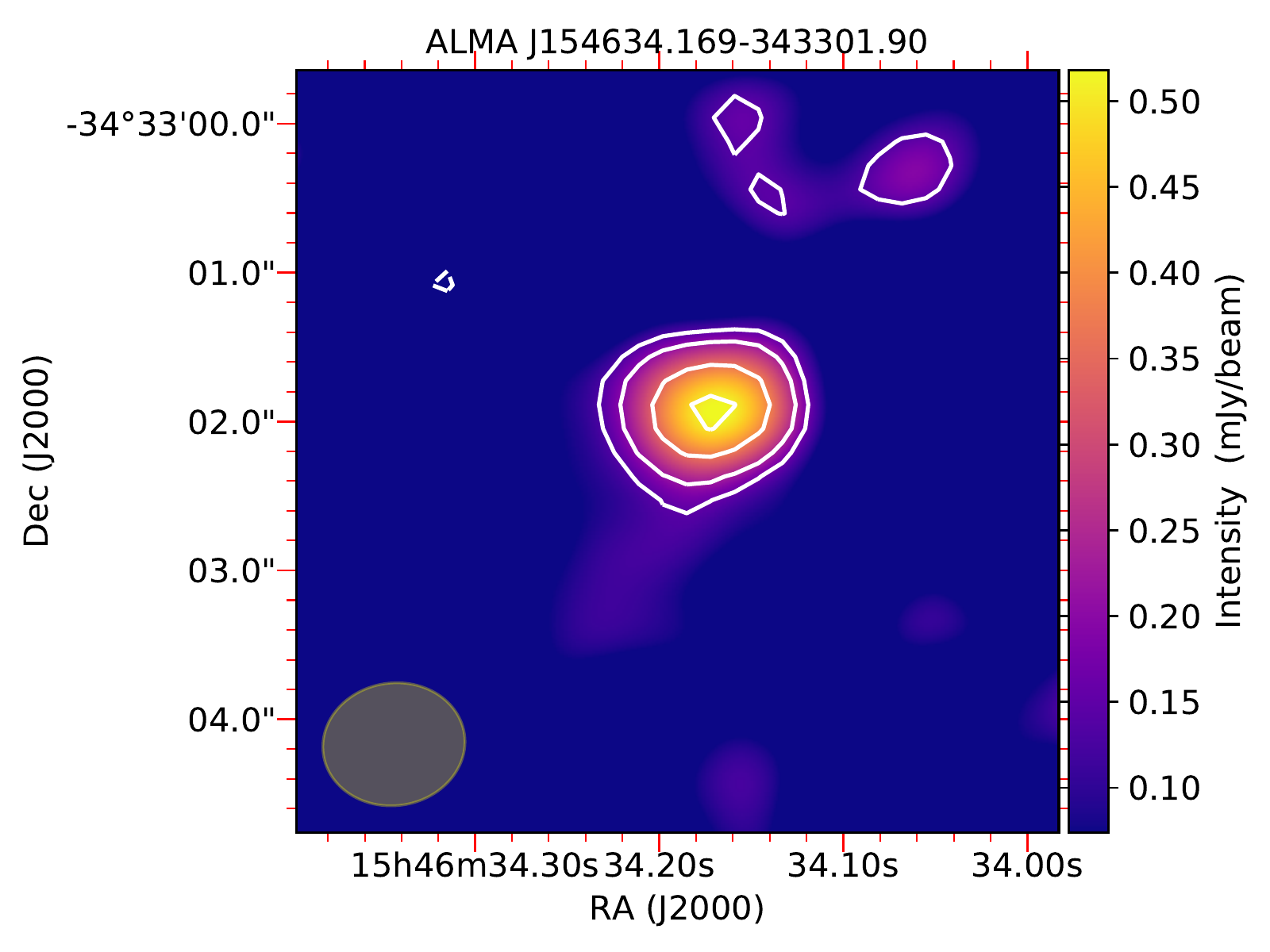}}
\qquad
\subfloat{\includegraphics[width=0.46\textwidth]{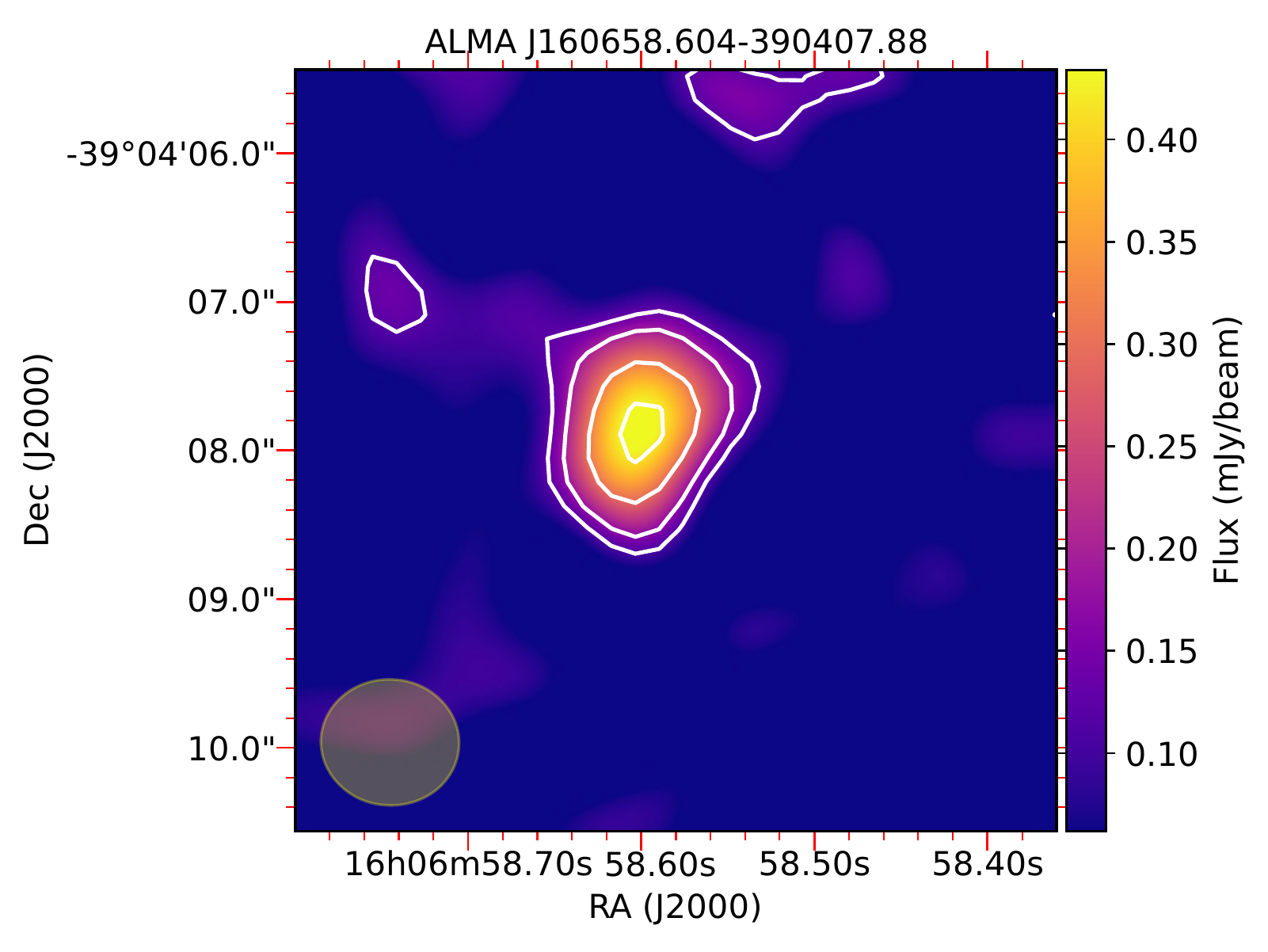}}
\subfloat{\includegraphics[width=0.46\textwidth]{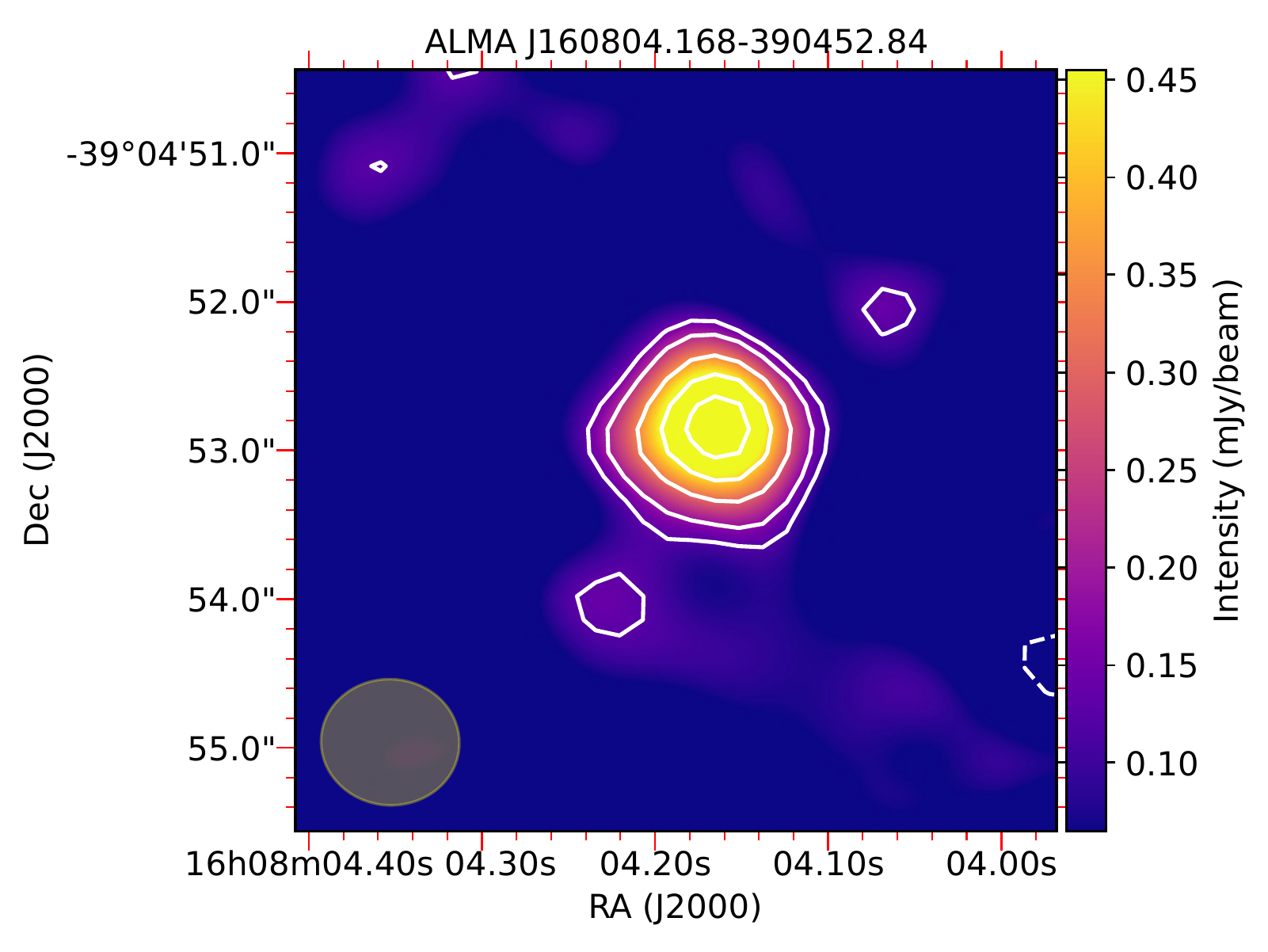}}
\qquad
\subfloat{\includegraphics[width=0.46\textwidth]{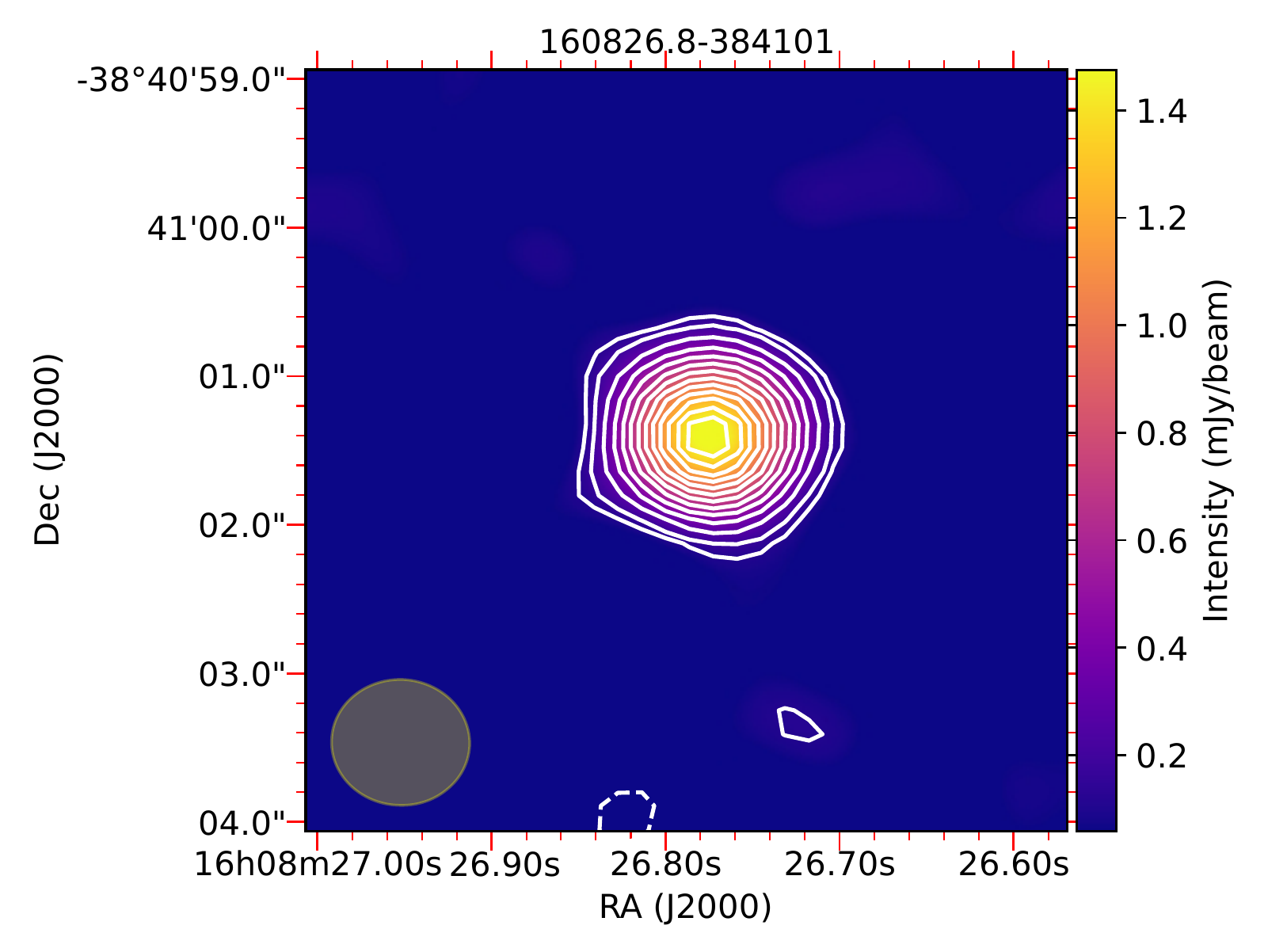}}
\subfloat{\includegraphics[width=0.46\textwidth]{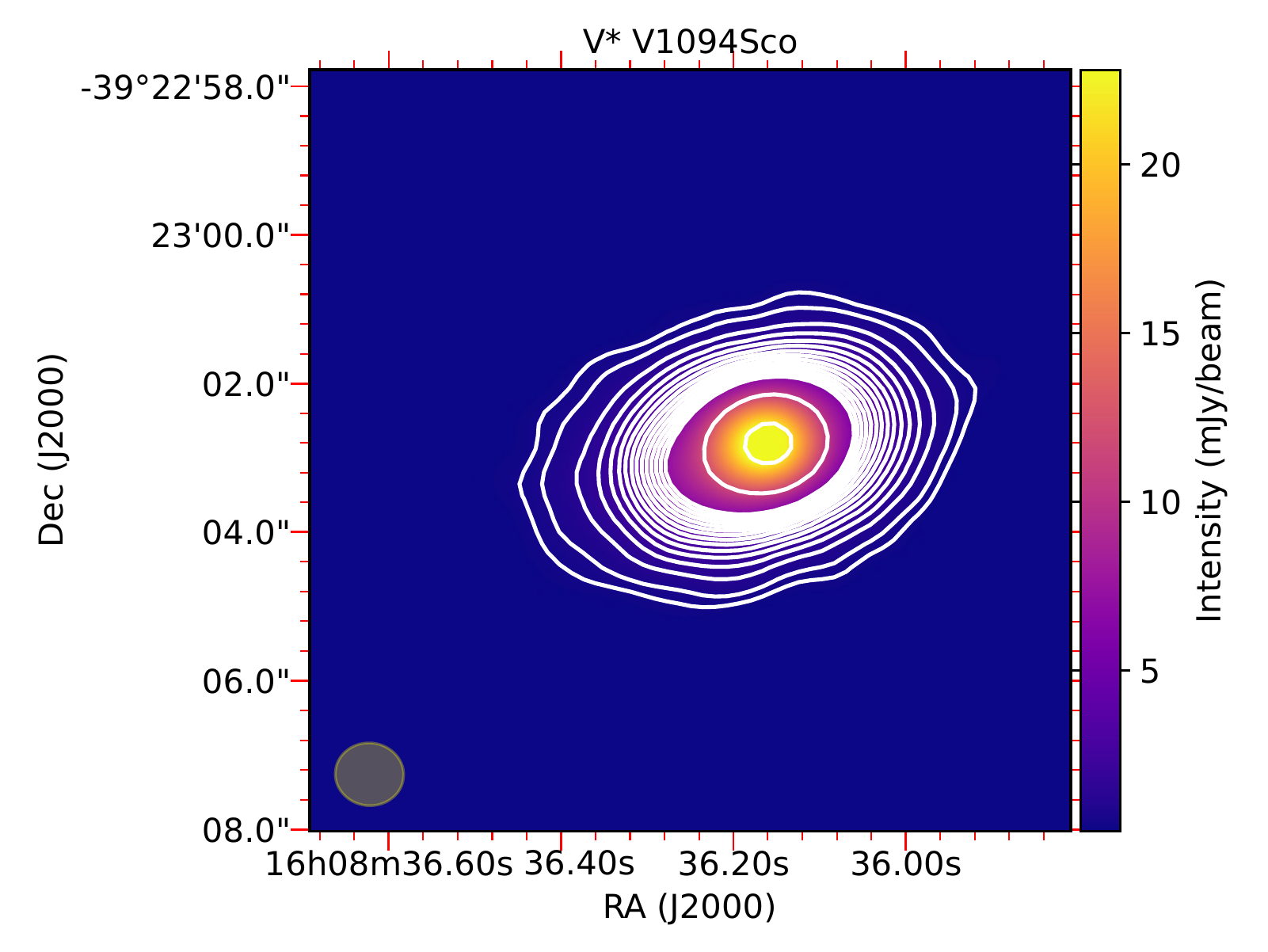}}
\caption{\label{apendix_panel_alma} continued from Figure \ref{panel_alma}}
\end{figure*}

\begin{figure*}
\subfloat{\includegraphics[width=0.46\textwidth]{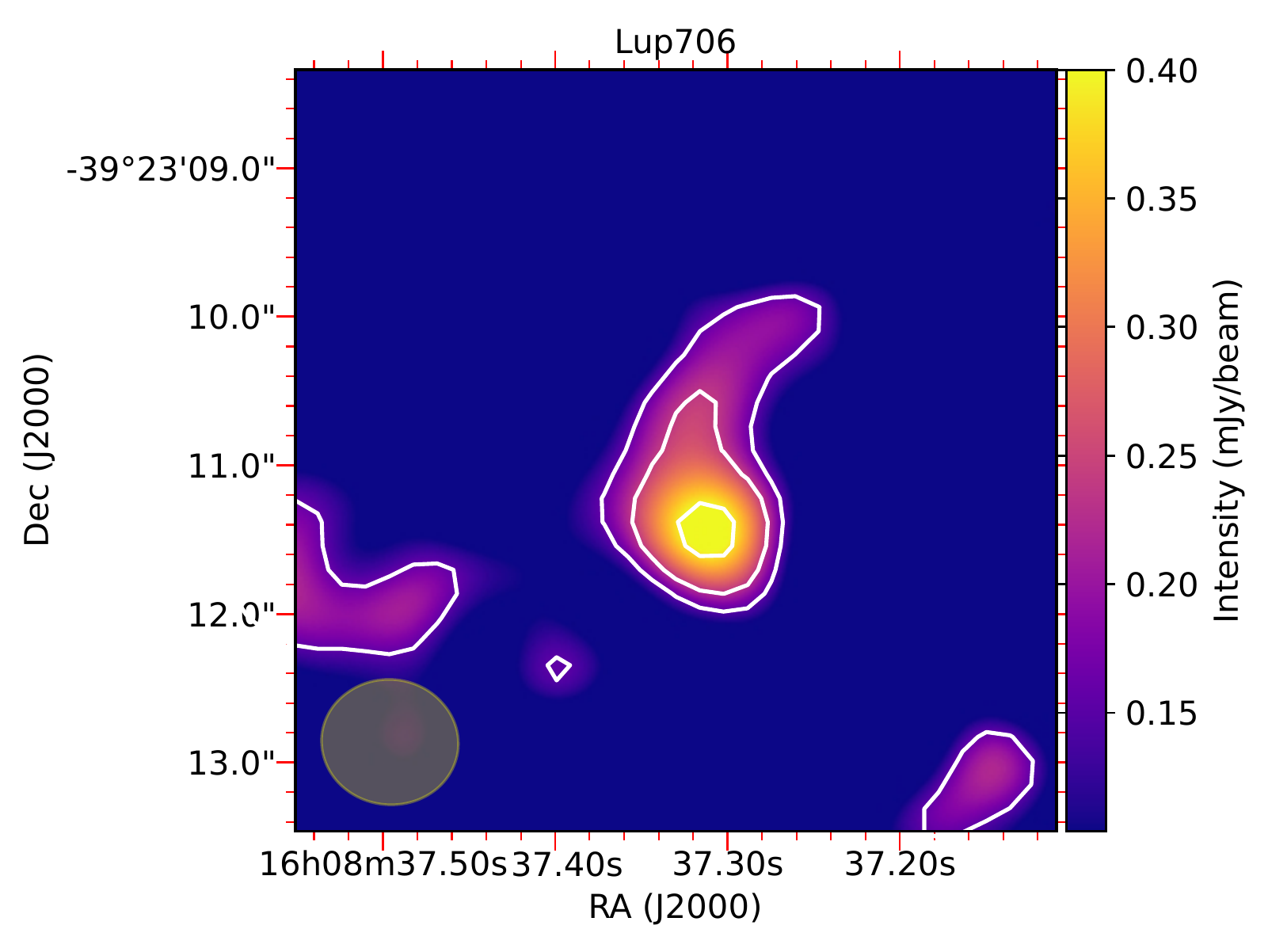}}
\subfloat{\includegraphics[width=0.46\textwidth]{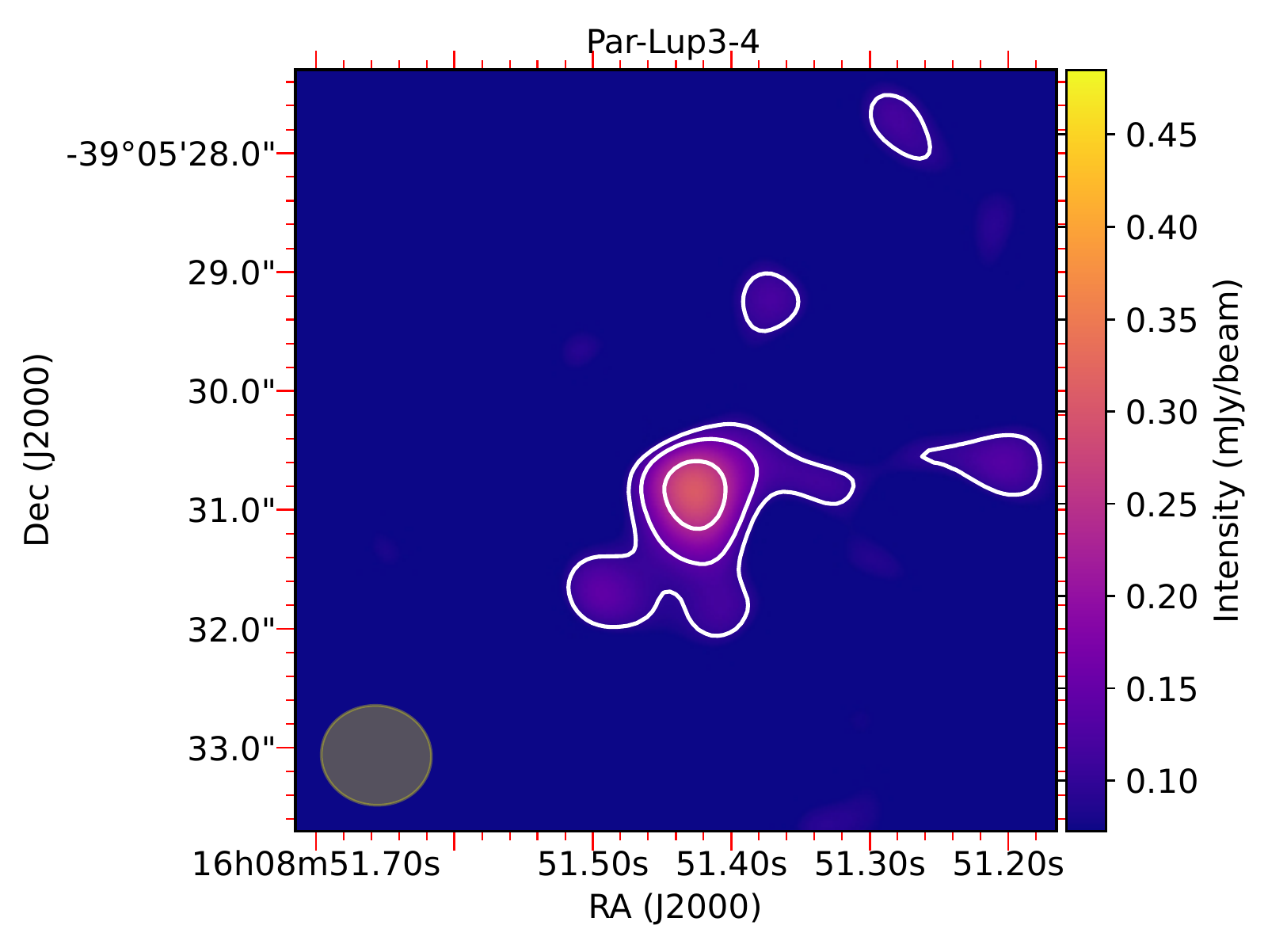}}
\qquad
\subfloat{\includegraphics[width=0.46\textwidth]{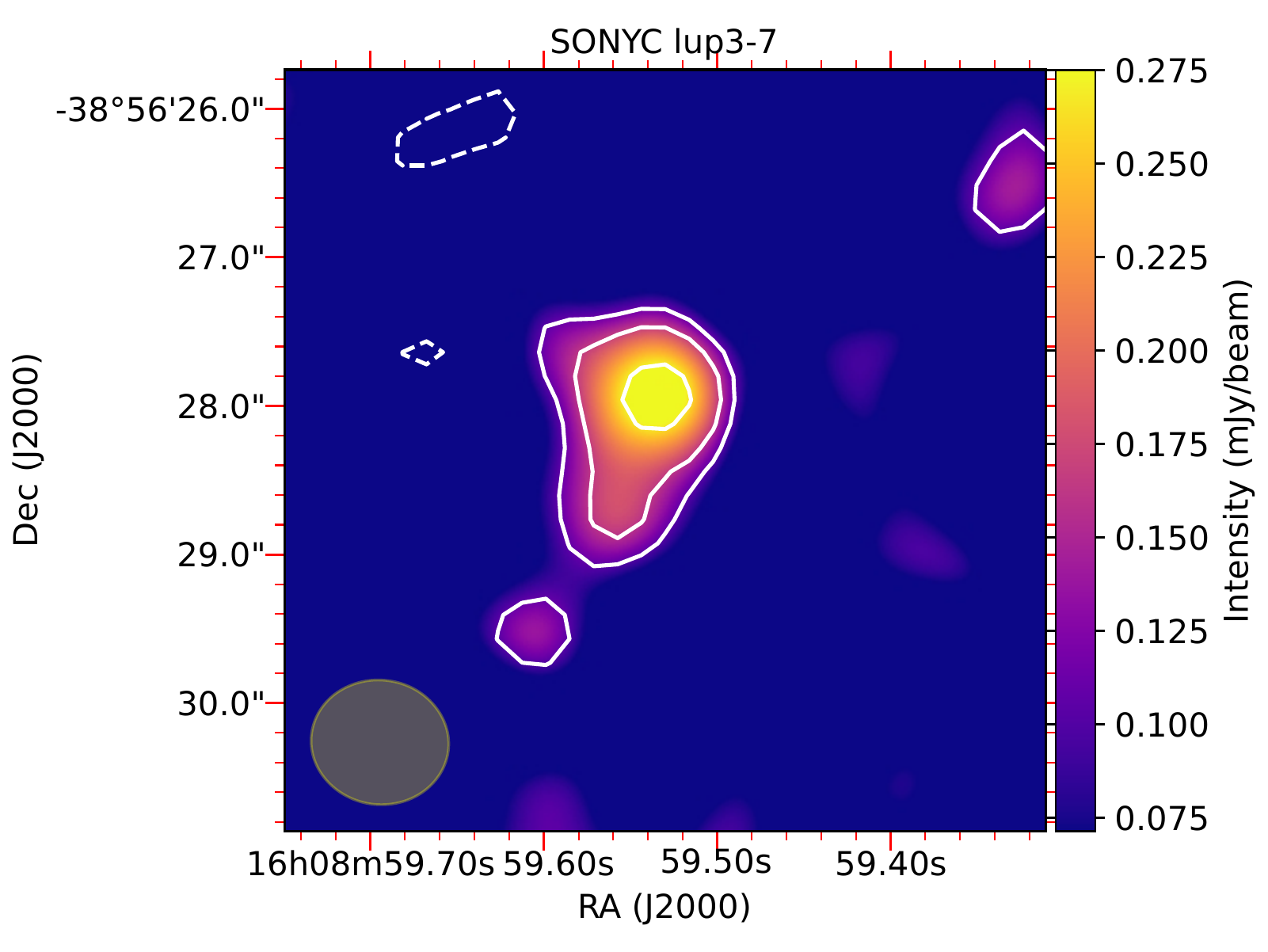}}
\subfloat{\includegraphics[width=0.46\textwidth]{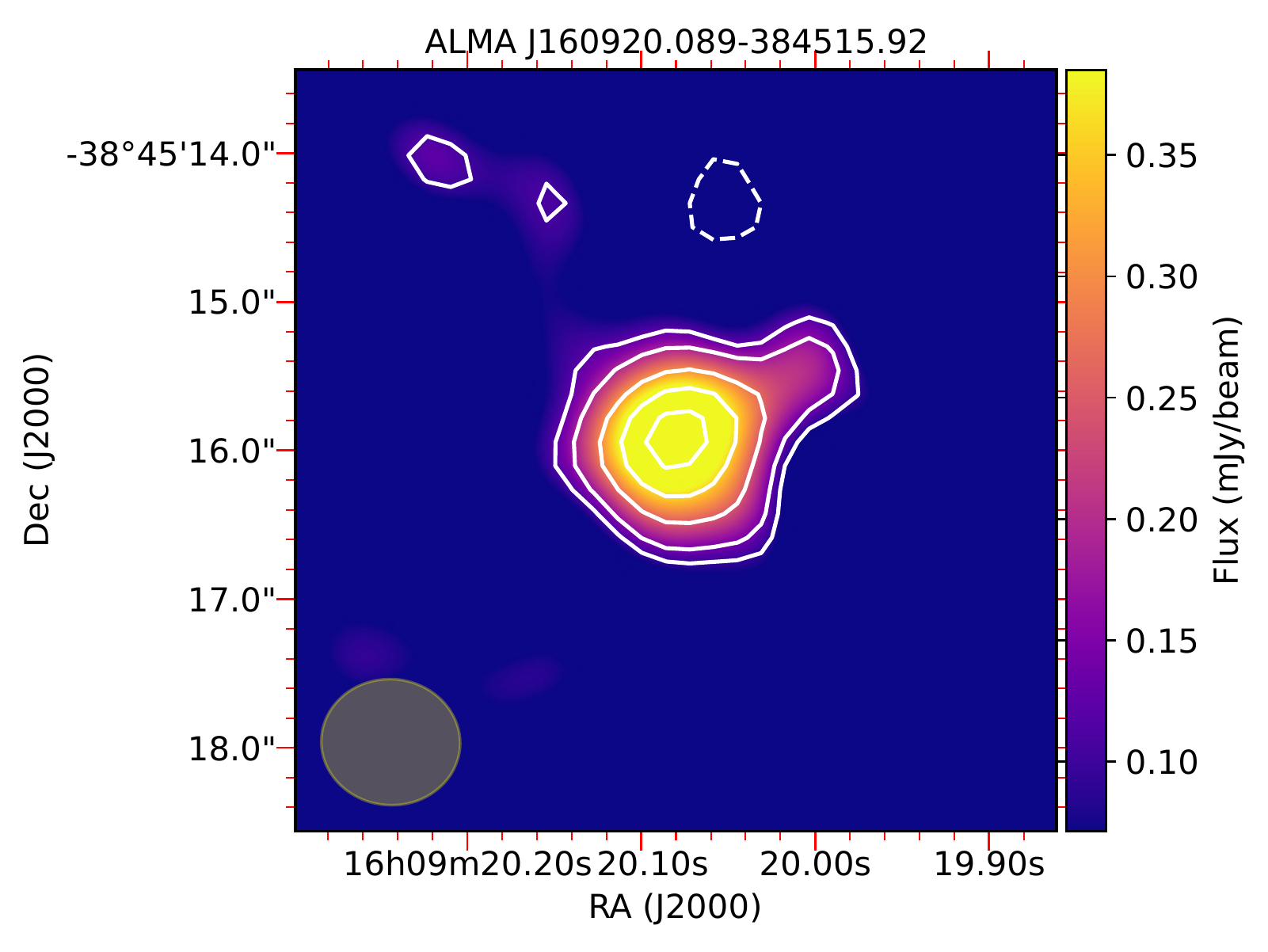}}
\qquad
\subfloat{\includegraphics[width=0.46\textwidth]{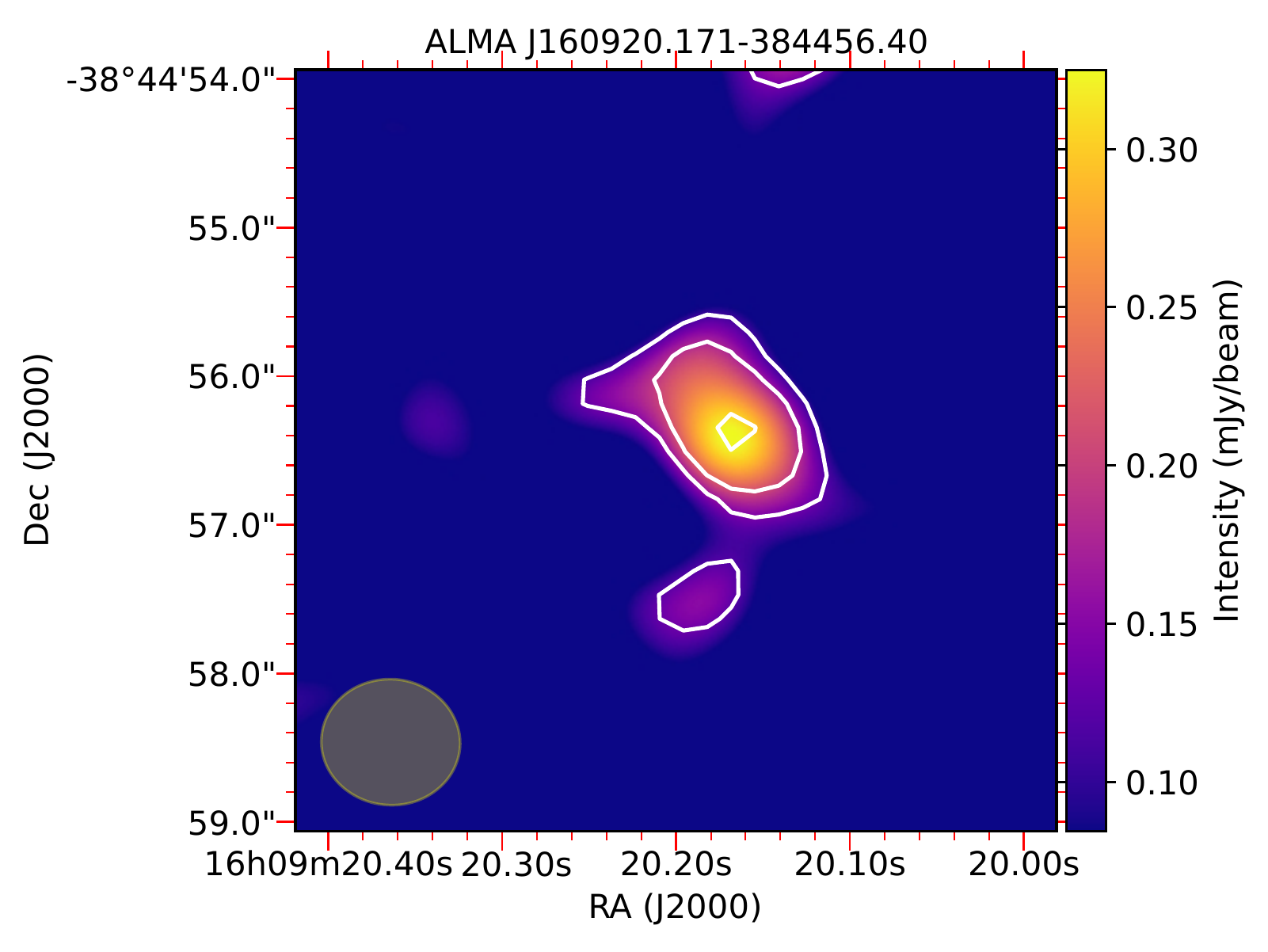}}
\subfloat{\includegraphics[width=0.46\textwidth]{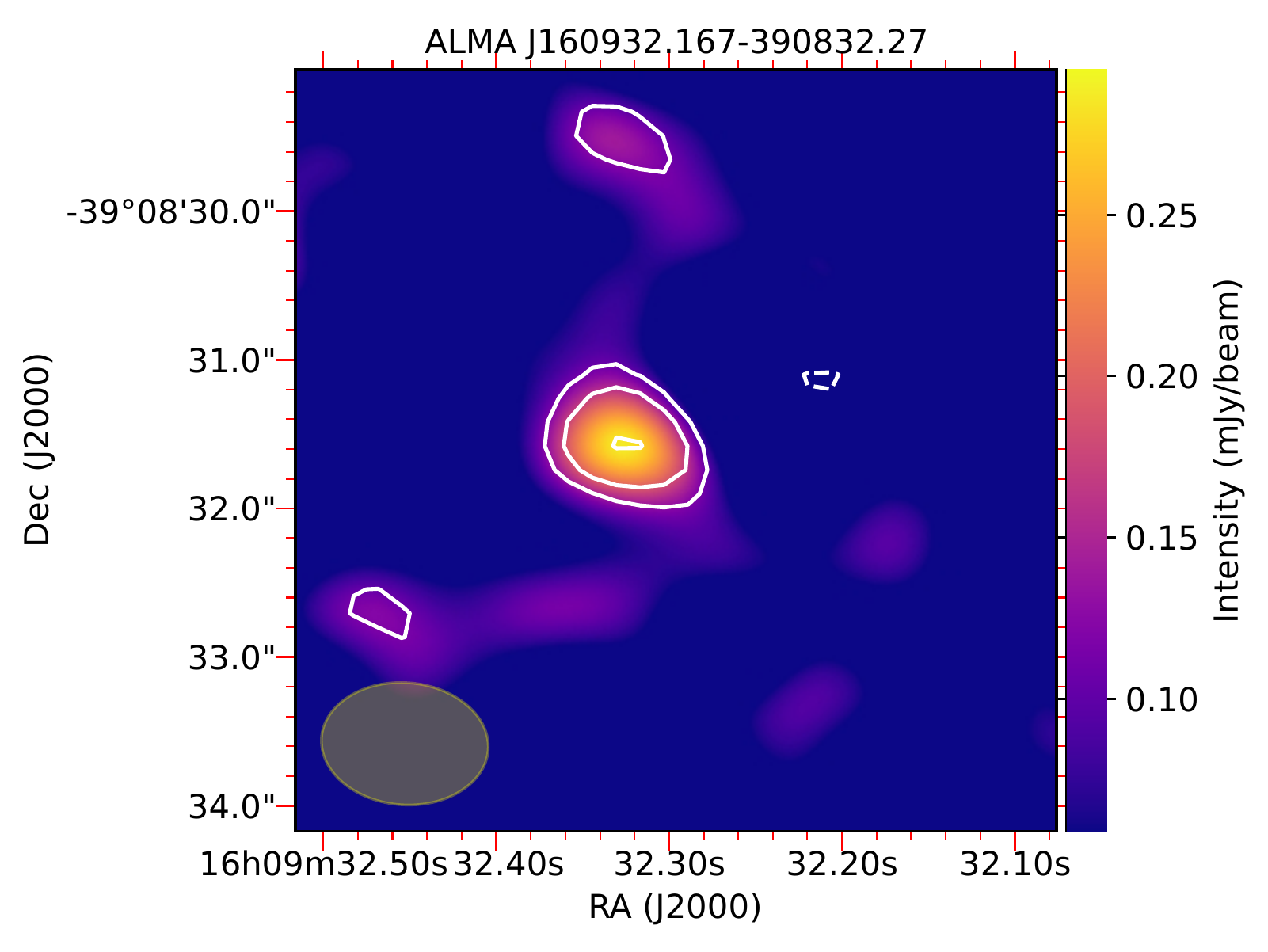}}
\caption*{(continued)}
\end{figure*}

\begin{figure*}
\subfloat{\includegraphics[width=0.46\textwidth]{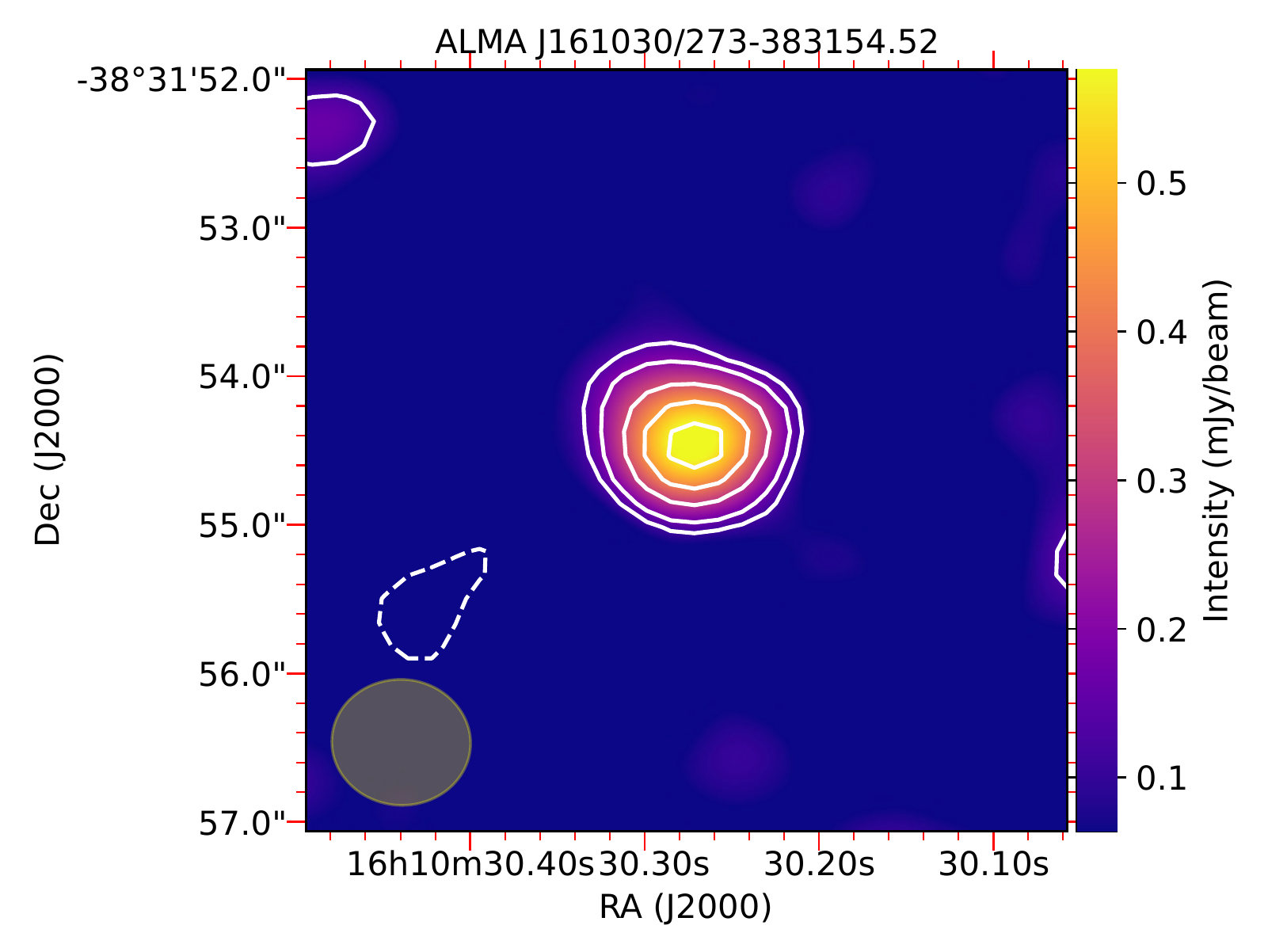}}
\caption*{(continued)}
\end{figure*}

\begin{figure*}
\includegraphics[height=1.01\textheight]{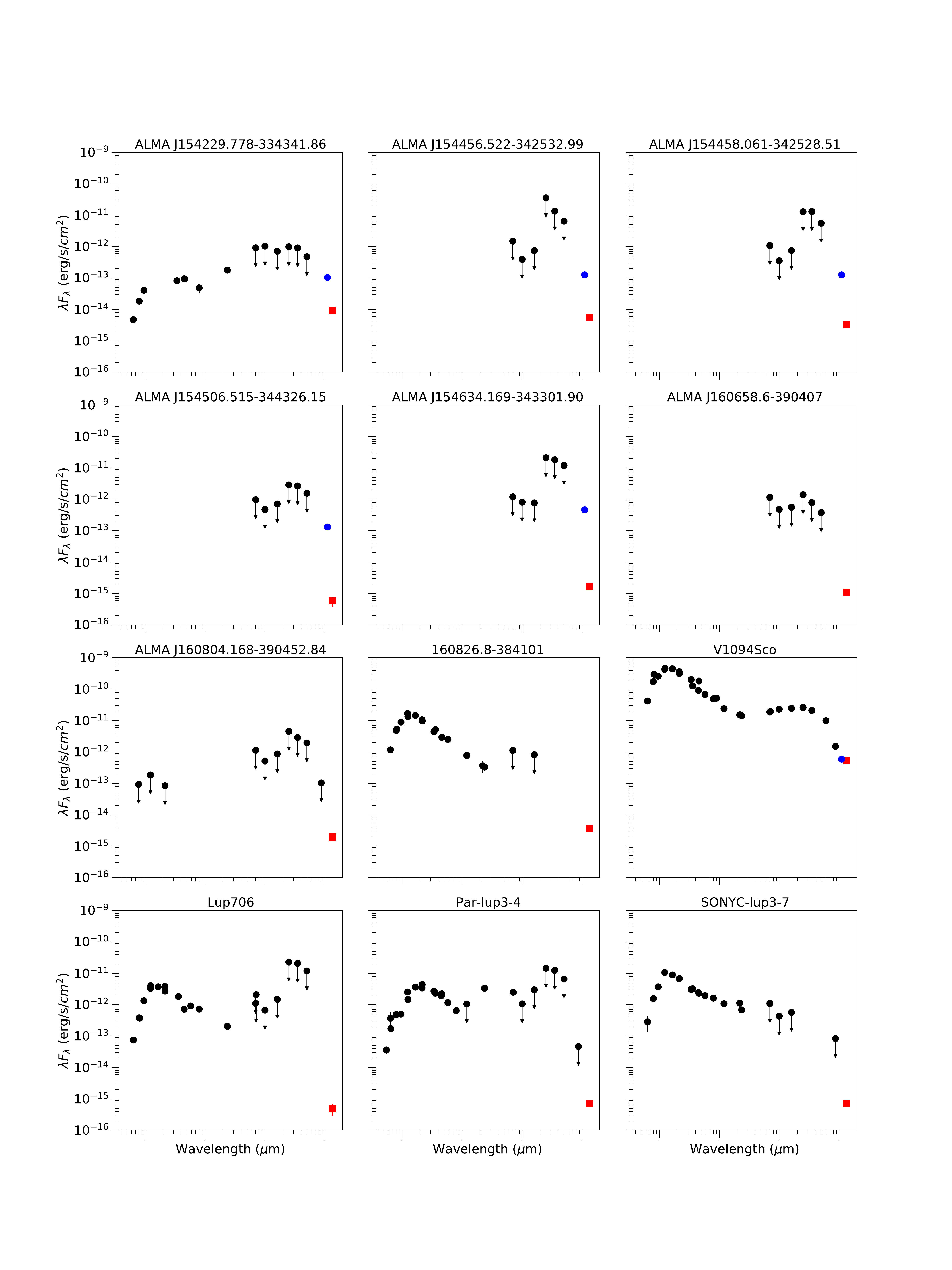}
\caption{\label{apendix_panel_sed}(continued from Figure \ref{panel_sed})}

\end{figure*} 

\begin{figure*}
\includegraphics[height=0.8\textheight]{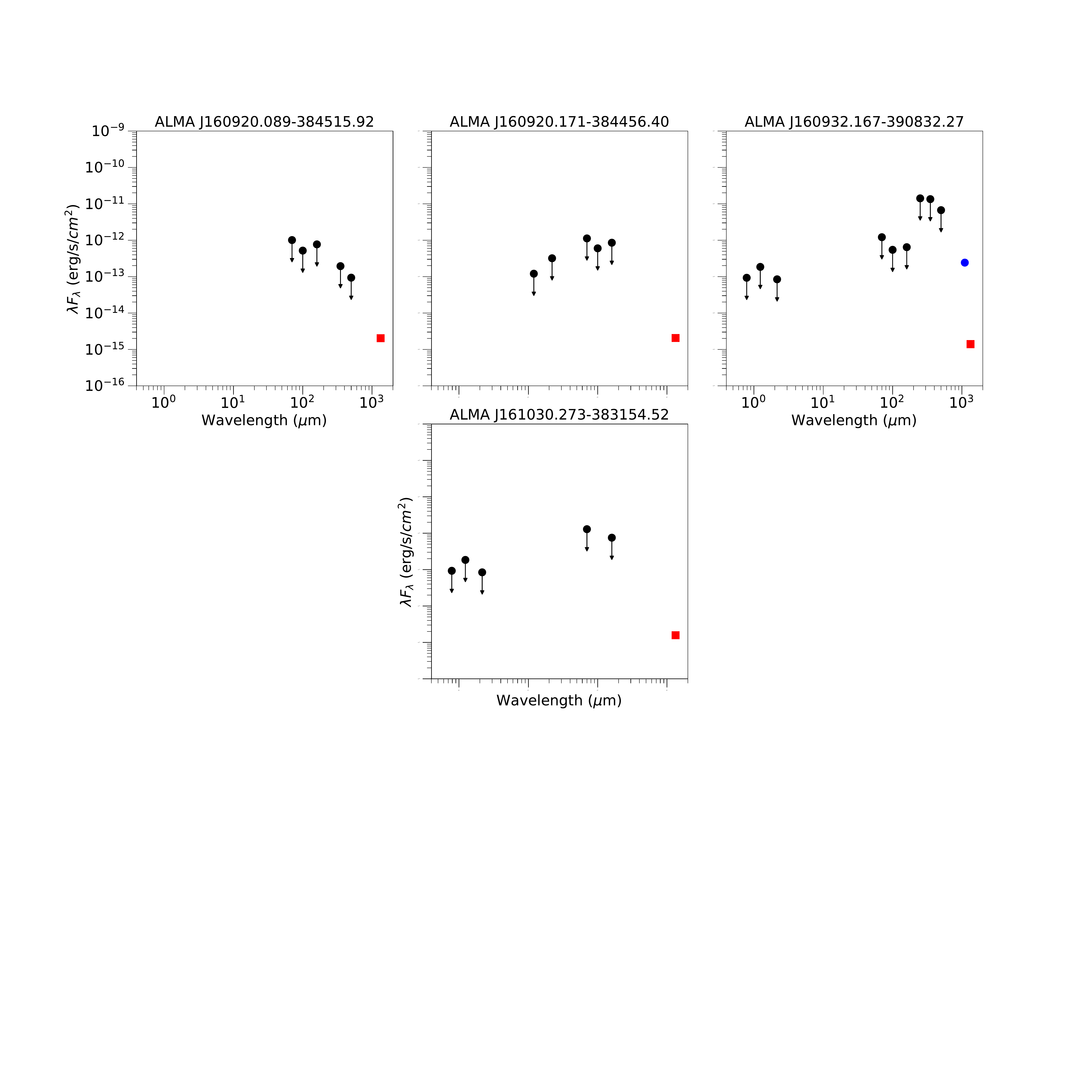}
\caption*{(continued from Figure \ref{panel_sed})}
\end{figure*} 
 
\begin{figure*}
\centering
\subfloat{\includegraphics[width=1.032 \textwidth]{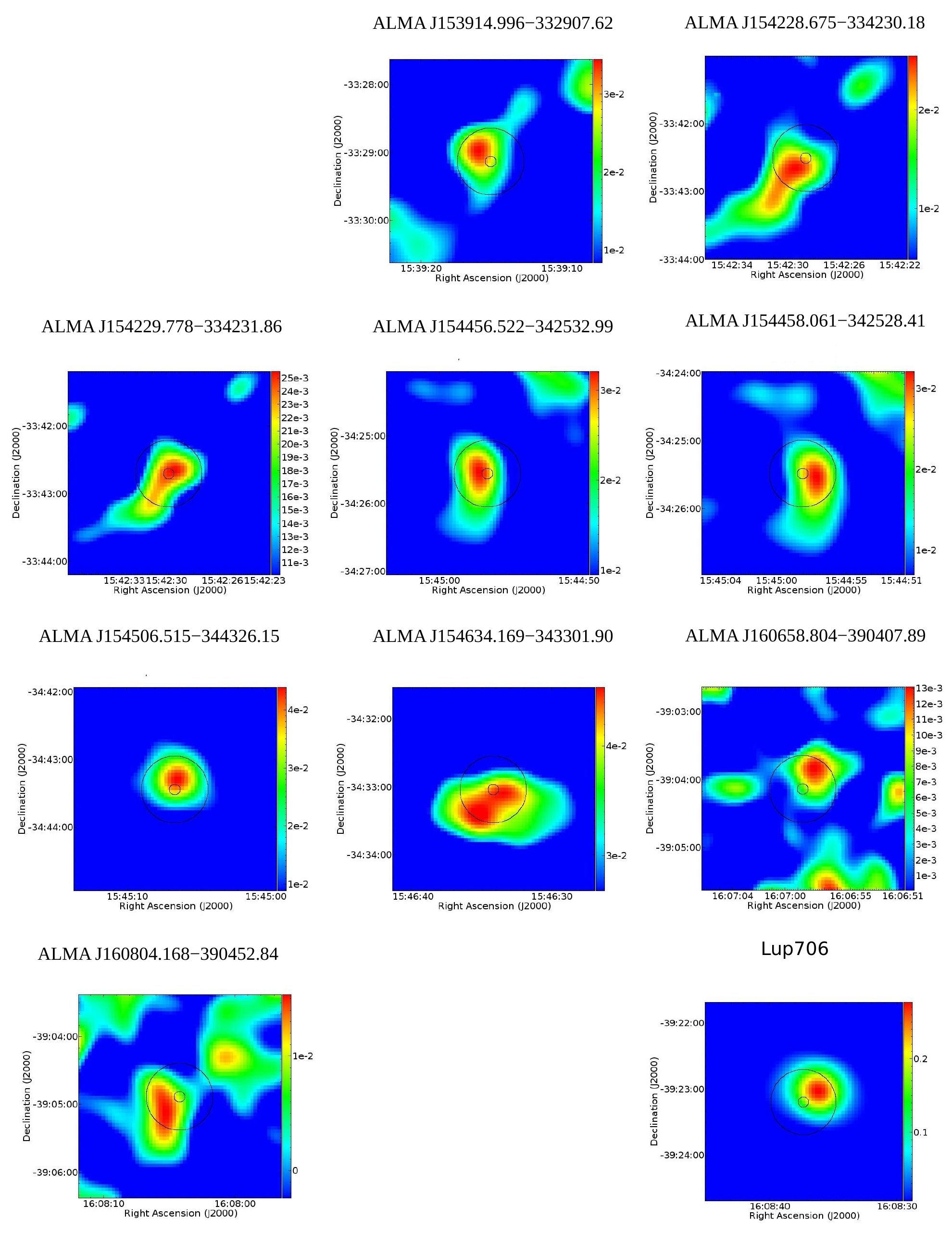}}
\caption{\label{aztec_maps} AzTEC maps centred at the position of the ALMA detections. The position is in RA and Dec (J2000). The wedge colour bar at the right shows the intensity (mJy/beam). The smaller circle is 5$\arcsec$ in diameter and the greater one 30$\arcsec$. SONYC-Lup3-7 and 161030.6-383151 have no images} 
\label{panel_aztec}

\qquad
\end{figure*}

\begin{figure*}
\centering
\subfloat{\includegraphics[width=1.032 \textwidth]{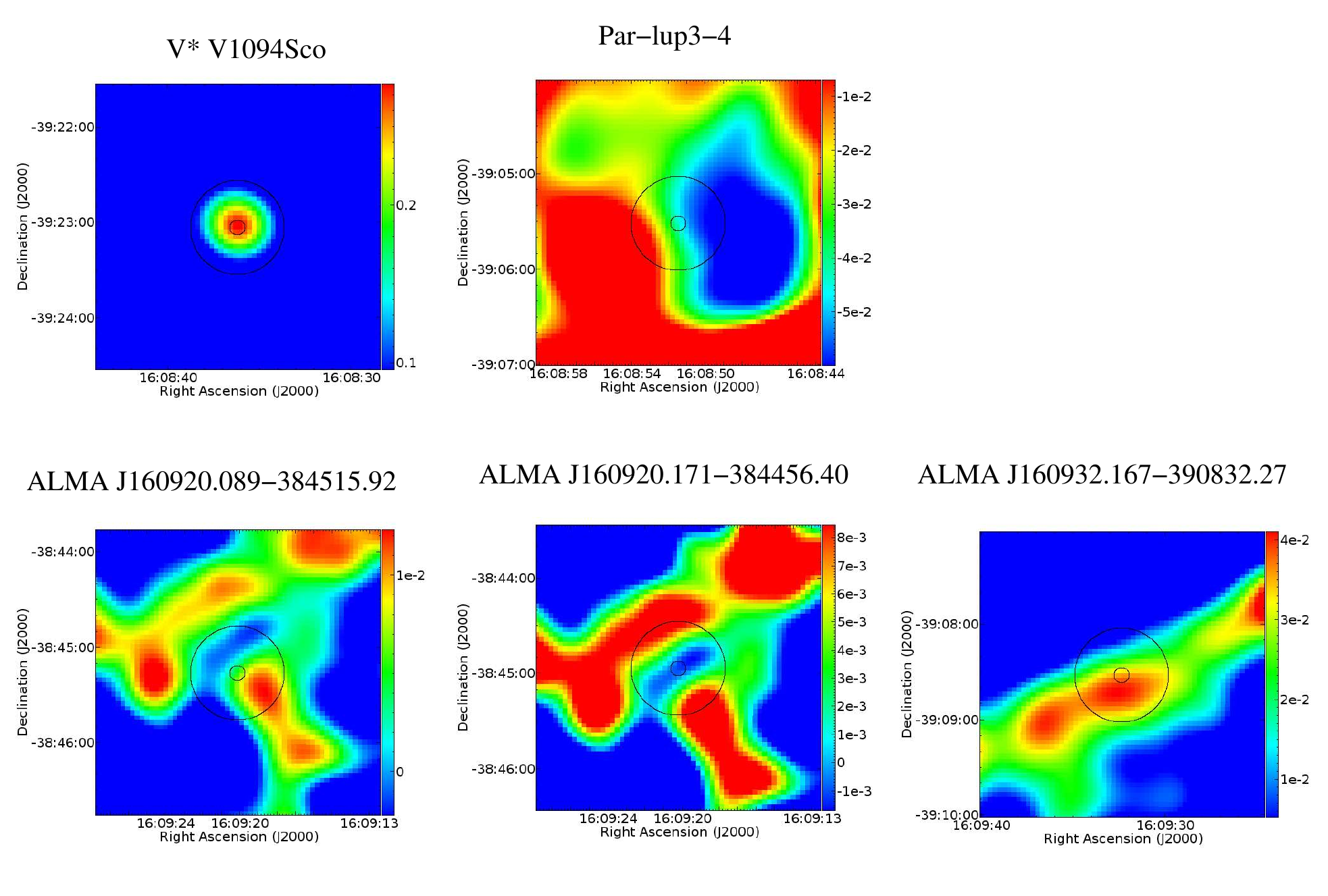}}
\caption*{continued}

\end{figure*}

\section{SEDs of the sources in this work}

\newgeometry{margin=1.5cm} 
\begin{landscape}
\begin{table*} 
\caption{Fluxes at each wavelength used to build the SEDs}\label{wavelength_fluxes_SED}
{\fontsize{7.5}{9.5} \selectfont 
\centering                                      
\begin{tabular}{|c|c|c|c|c|c|c|c|c|c|c|c|c|c|c|c|c|c|c|c|c|c|c|c|c|c|c|c|c|c|c|c|c|c|c}       
 
\hline  
Name & \multicolumn{2}{c}{Flux (mJy)} \\
\hline
&	0.55 $\micro$m &	0.64  $\micro$m &	0.65  $\micro$m&	0.77  $\micro$m &	0.79  $\micro$m&	0.80  $\micro$m &	0.82  $\micro$m &	0.96  $\micro$m &	1.23  $\micro$m &	1.25  $\micro$m &	1.66  $\micro$m\\ 
\hline  
ALMA J153702.653-331924.92 &	&	&	&	&	&	&	&	&	&	&	\\
\hline  
ALMA J153914.996-332907.62 &	&	0.0013 $\pm$ 0.0003 &	&	&	&	0.0071  $\pm$ 0.0013 &	&	0.017 $\pm$ 0.003 &	&	&	\\
\hline
ALMA J154228.675-334230.18 &	&	&	&	&	&	&	&	&	&	&	\\
\hline
ALMA J154229.778-334341.86 &	&	0.0010 $\pm$ 0.0002 &	&	&		0.0049 $\pm$ 0.0011 & &	&	0.013 $\pm$ 0.003 &	&	&	\\
\hline
ALMA J154456.522-342532.99 &	&	&	&	&	&	&	&	&	&	&	\\
\hline
ALMA J154458.061-342528.51 &	&	&	&	&	&	&	&	&	&	&	\\
\hline
ALMA J154506.515-344326.15 &	&	&	&	&	&	&	&	&	&	&	\\
\hline
ALMA J154634.169-343301.90 &	&	&	&	&	&	&	&	&	&	&	\\
\hline
ALMA J160658.604-390407.88 &	&	&	  &	&	&	&	 &	&	 &	 \\
\hline
ALMA J160804.168-390452.84 &	&	&	&	&	$\le$ 0.02 &	&	&	&	$\le$ 0.08 &	&	\\
\hline
160826.8-384101 &	&	0.250 $\pm$ 0.004 &	&	&	&	1.30 $\pm$ 0.02 &	1.48 $\pm$ 0.07 &	2.88 $\pm$ 0.04 &	6.95 $\pm$ 0.15 &	5.7 $\pm$ 0.5 &	8.06 $\pm$ 0.19 \\
\hline
V1094Sco &	&	8.97 $\pm$  0.03 &	&	&	&	46.50 $\pm$ 0.13 &	82 $\pm$ 2 &	82.91 $\pm$ 0.15 &	177 $\pm$ 4 &	193.0 $\pm$ 8.9 &	247 $\pm$ 5 \\
\hline
Lup706 &	&	0.016  $\pm$ 0.001 &	&	&	&	0.102 $\pm$ 0.005 &	0.10 $\pm$ 0.02 &	0.42 $\pm$ 0.01 &	1.36 $\pm$ 0.07 &	1.68 $\pm$ 0.20 &	2.06 $\pm$ 0.09 \\
\hline
Par-Lup3-4 &	0.007 $\pm$ 0.002 &	0.08 $\pm$ 0.04 &	0.038 $\pm$ 0.005 &	&	&	0.13 $\pm$ 0.01 &	&	0.16 $\pm$ 0.01 &	1.04 $\pm$ 0.06 &	0.61 $\pm$ 0.120 &	2.02 $\pm$ 0.07 \\
\hline
SONYC-Lup3-7  &	&	0.06 $\pm$ 0.03 &	&	&	&	0.42 $\pm$ 0.01 &	&	1.19 $\pm$ 0.02 &	4.38 $\pm$ 0.12 &	&	4.93 $\pm$ 0.14 \\
\hline
ALMA J160920.089-384515.92 &	&	 &	&	& &	 &	&	 &	 &	&	\\
\hline
ALMA J160920.171-384456.40 &	&	&	&	&	&	&	&	&	 &	&	 \\
\hline
ALMA J160932.167-390832.27 &	&	&	&	&	$\le$ 0.02 &	&	&	&	$\le$ 0.08 &	&	\\
\hline
ALMA J161030.273-383154.52 &	&	&	&	&	$\le$ 0.02 &	&	&	&	$\le$ 0.08 &	&	\\
\hline
\end{tabular}
}
\end{table*}

\begin{table*} 
{\fontsize{7.5}{9.5} \selectfont 
\centering                                      
\begin{tabular}{|c|c|c|c|c|c|c|c|c|c|c|c|c|c|c|c|c|c|c|c|c|c|c|c|c|c|c|c|c|c|c|c|c|c|c}  
\hline
& 2.15 $\micro$m &	2.16 $\micro$m&	3.4 $\micro$m&	3.6 $\micro$m&	4.5 $\micro$m&	4.6 $\micro$m&	5.8 $\micro$m&	8 $\micro$m&	9 $\micro$m&	12 $\micro$m&	22 $\micro$m\\
\hline 
ALMA J153702.653-331924.92 &	&	&	&	&	&	&	&	&	&	&	\\
\hline
ALMA J153914.996-332907.62 &	&	&	&	&	&	&	&	&	&	&	\\
\hline
ALMA J154228.675-334230.18 &	&	&	&	&	&	&	&	&	&	&	\\
\hline
ALMA J154229.778-334341.86 &	&	&	0.092 $\pm$ 0.007 &	&	0.14 $\pm$  0.01 &	0.142 $\pm$ 0.013 &	&	0.13 $\pm$ 0.04 &	&	&	$\le$ 4.7 \\
\hline
ALMA J154456.522-342532.99 &	&	&	&	&	&	&	&	&	&	&	\\
\hline
ALMA J154458.061-342528.51 &	&	&	&	&	&	&	&	&	&	&	\\
\hline
ALMA J154506.515-344326.15 &	&	&	&	&	&	&	&	&	&	&	\\
\hline
ALMA J154634.169-343301.90 &	&	&	&	&	&	&	&	&	&	&	\\
\hline
ALMA J160658.604-390407.88 &	 &	&	&	&	&	&	&	&	&	&	\\
\hline
ALMA J160804.168-390452.84 &	$\le$ 0.06 &	&	&	&	&	&	&	&	&	&	\\
\hline
160826.8-384101 &	7.6 $\pm$ 0.7 &	7.04 $\pm$ 0.19 &	5.04 $\pm$ 0.11 &	6.2 $\pm$ 0.4 &	&	4.50 $\pm$ 0.09 &	4.9 $\pm$ 0.27 &	&	&	3.11 $\pm$ 0.19 &	2.7 $\pm$ 1.1 \\
\hline
V1094Sco &	259 $\pm$ 12 &	230  $\pm$ 4 &	230 $\pm$ 5 &	153 $\pm$ 10 &	138 $\pm$ 8 &	280 $\pm$ 3 &	132 $\pm$ 6 &	132 $\pm$ 6 &	156 $\pm$ 24 &	96.1 $\pm$ 1.4 &	113 $\pm$ 3 \\
\hline
Lup706 &	2.7 $\pm$ 0.5 &	1.96 $\pm$ 0.08 &	&	2.18 $\pm$ 0.11 &	1.08 $\pm$ 0.11 &	&	1.77 $\pm$ 0.11 &	1.9 $\pm$ 0.1 &	&	&	\\
\hline
Par-Lup3-4 &	2.5 $\pm$ 0.4 &	3.18 $\pm$ 0.12 &	3.09 $\pm$ 0.07 &	2.80 $\pm$ 0.15 &	 2.88 $\pm$ 0.15 &	3.43 $\pm$ 0.07 &	2.24 $\pm$ 0.13 &	1.7 $\pm$ 0.1 &	&	$\le$ 4.22 &	\\
\hline
SONYC-Lup3-7  &	&	4.90 $\pm$ 0.14 &	3.49 $\pm$ 0.08 &	3.9 $\pm$ 0.2 &	3.69 $\pm$ 0.19 &	3.55 $\pm$ 0.07 &	3.8 $\pm$ 0.2 &	4.3 $\pm$ 0.2 &	&	4.3 $\pm$ 0.2 &	8.2 $\pm$ 1.2 \\
\hline
ALMA J160920.089-384515.92 &	& &	&	&	&	&	&	&	&	&	\\
\hline
ALMA J160920.171-384456.40 &	& &	 &	&	&	 &	&	&	&	$\le$ 0.5 &	$\le$ 2.4 \\
\hline
ALMA J160932.167-390832.27 &	&	$\le$ 0.06 &	&	&	&	&	&	&	&	&	\\
\hline
ALMA J161030.273-383154.52 &	&	$\le$ 0.06 &	&	&	&	&	&	&	&	&	\\
\hline

\end{tabular}
}
\end{table*}

\begin{table*} 
{\fontsize{7.5}{9.5} \selectfont 
\centering                                      
\begin{tabular}{|c|c|c|c|c|c|c|c|c|c|c|c|c|c|c|c|c|c|c|c|c|c|c|c|c|c|c|c|c|c|c|c|c|c|c}  
\hline  
&	23.67 $\micro$m&	70 $\micro$m&	71.42 $\micro$m&	100 $\micro$m&	160 $\micro$m&	250 $\micro$m&	350 $\micro$m&	500 $\micro$m&	869 $\micro$m&	1100 $\micro$m&	1330 $\micro$m\\
\hline
ALMA J153702.653-331924.92 &	&	$\le$20.9 &	&	&	$\le$37.7 &	$\le$135 &	$\le$123 &	$\le$122 &	& &	0.45 $\pm$ 0.10 \\
\hline
ALMA J153914.996-332907.62 &	&	$\le$ 27.1 &	&	$\le$ 17.8 &	$\le$ 47.4 &	$\le$ 1100 &	$\le$ 1266 &	$\le$ 820 &	&	40 $\pm$ 4 &	0.73 $\pm$ 0.09 \\
\hline
ALMA J154228.675-334230.18 &	&	$\le$21.6 &	&	$\le$14.1 &	$\le$35.8 &	$\le$110 &	$\le$178 &	$\le$ 114 &	&	25 $\pm$ 3 &	2.84 $\pm$ 0.14 \\
\hline
ALMA J154229.778-334341.86 &	1.4 $\pm$ 0.3 &	$\le$ 21.3 &	&	$\le$ 34.4 &	$\le$ 38.0 &	$\le$ 82   &	$\le$ 106  & $\le$ 79	&	&	38 $\pm$ 4 &	4.1 $\pm$ 0.1 \\
\hline
ALMA J154456.522-342532.99 &	&	$\le$ 34.8 &	&	$\le$ 13.1 &	$\le$ 39.6 &	$\le$ 2950 &	$\le$ 1570 &	$\le$ 1080 &	&	29 $\pm$ 3 &	2.50 $\pm$ 0.07 \\
\hline
ALMA J154458.061-342528.51 &	&	$\le$25.2 &	&	$\le$11.8 &	$\le$39.80 &	$\le$1070 &	$\le$1520 &	$\le$920 &	&	17 $\pm$ 2 &	1.42 $\pm$ 0.01 \\
\hline
ALMA J154506.515-344326.15 &	&	$\le$22.6 &	&	$\le$15.8 &	$\le$37.8 &	$\le$242 &	$\le$310 &	$\le$278 &	&	48 $\pm$ 5 &	0.26 $\pm$ 0.09 \\ 
\hline
ALMA J154634.169-343301.90 &	&	$\le$27.7 &	&	$\le$17.0 &	$\le$40.7 &	$\le$1750 &	$\le$2110 &	$\le$1990 &	&	170 $\pm$ 17 &	0.75 $\pm$ 0.08 \\
\hline
ALMA J160658.604-390407.88 &	&	$\le$ 26.9 &	&	$\le$ 15.9 &	$\le$ 29.8 &	$\le$ 116 &	$\le$ 91 &	$\le$ 63 &	&	&	0.48 $\pm$ 0.06 \\
\hline
ALMA J160804.168-390452.84 &	&	$\le$ 26.5 &	&	& $\le$ 17.2 &	  $\le$ 46.3 &	 $\le$ 378 &	$\le$ 338 &	$\le$ 324 &	 &	0.87 $\pm$ 0.09 \\
\hline
160826.8-384101 &	2.6 $\pm$ 0.3 &	$\le$  26.11 &	&	&	$\le$  43.5 &	&	&	&	&	 &	1.57 $\pm$ 0.06 \\
\hline
V1094Sco &	113 $\pm$ 10 &	448 $\pm$ 40 &	439 $\pm$ 48 &	766 $\pm$ 25 &	1316 $\pm$ 13 &	2160 $\pm$ 430 &	2470 $\pm$ 490 &	1989 $\pm$ 398 &	438 $\pm$ 9 &	218 $\pm$ 6 &	244.1 $\pm$ 0.2 \\
\hline
Lup706 &	1.6 $\pm$ 0.3 &	$\le$ 25.9 &	$\le$ 50 &	$\le$ 22.4 &	$\le$79.3 &	$\le$ 1905 &	$\le$ 2427 &	$\le$ 1983 &	&  &	0.22 $\pm$ 0.09 \\
\hline
Par-Lup3-4 &	27 $\pm$ 3 & &	59 $\pm$ 2 &		$\le$ 30.8 &	 $\le$ 159.2 &	$\le$ 1215 &	$\le$ 1455 &	$\le$ 1101 &	 $\le$ 14 &	 &	0.31 $\pm$ 0.05 \\
\hline
SONYC-Lup3-7  &	5.4 $\pm$ 0.6 &	$\le$25.6 &	&	$\le$ 14.4 &	 $\le$ 30.4 &	&	&	&	$\le$ 24 &	&	0.32 $\pm$ 0.06 \\
\hline
ALMA J160920.089-384515.92 &	&	$\le$ 23.6 &	&	$\le$ 17.3 &	$\le$ 41.1 &	&	$\le$ 23 &	$\le$ 15.62 &	&	&	0.90 $\pm$ 0.08 \\
\hline
ALMA J160920.171-384456.40 &	&	$\le$  26.2 &	&	$\le$  20.0 &	$\le$ 45.6 &	&	&	&	&	&	0.91 $\pm$ 0.18 \\
\hline
ALMA J160932.167-390832.27 &	&	$\le$ 28.3 &	&	$\le$ 18.2 &	$\le$ 34.41 &	$\le$ 1180 &	$\le$ 1572 &	$\le$ 1119 &	&	89 $\pm$ 9 &	0.62 $\pm$ 0.13 \\
\hline
ALMA J161030.273-383154.52 &	&	$\le$ 30.2 &	&	&	$\le$ 40.3 &	&	&	&	&	&	0.70 $\pm$ 0.06 \\
\hline 

\end{tabular}
}
\end{table*}
\end{landscape}
\restoregeometry

\end{appendix}

\end{document}